\documentclass[12pt]{article}
\usepackage{amssymb}
\usepackage{graphicx,psfrag}

\newcommand \bra[1]{\left< {#1} \,\right\vert}
\newcommand \ket[1]{\left\vert\, {#1} \, \right>}

\newcommand{\bea}{\begin{eqnarray}}
\newcommand{\eea}{\end{eqnarray}}
\newcommand{\simgt}{\hbox{ \raise3pt\hbox to 0pt{$>$}\raise-3pt\hbox{$\sim$} }}
\newcommand{\simlt}{\hbox{ \raise3pt\hbox to 0pt{$<$}\raise-3pt\hbox{$\sim$} }}

\newcommand{\clfn}{\setcounter{footnote}{0}}

\newcommand{\LQ}{\Lambda_{\rm QCD}}

\begin{document}

\begin{titlepage}
\title{\bf \LARGE
\vspace{28mm}
Static QCD Potential at \boldmath $r < \LQ^{-1}$:\vspace{5mm}\\
\Large
Perturbative expansion and operator-product expansion
\vspace{7mm}}
\author{
Y.~Sumino
\\ \\ \\ Department of Physics, Tohoku University\\
Sendai, 980-8578 Japan
}
\date{}
\maketitle
\thispagestyle{empty}
\vspace{-4.5truein}
\begin{flushright}
{\bf TU--743}\\
{\bf May 2005}
\end{flushright}
\vspace{4truein}
\begin{abstract}
\noindent
{\small
We analyze the static QCD potential $V_{\rm QCD}(r)$
in the distance region $0.1~{\rm fm}\simlt r 
\simlt 1~{\rm fm}$ using perturbative QCD and
operator-product expansion (OPE) as basic theoretical
tools.
We assemble theoretical developments up to
date and perform a solid and accurate analysis.
The analysis consists of 3 major steps:
(I) We study large-order behavior of the perturbative
series of $V_{\rm QCD}(r)$ analytically.
Higher-order terms are estimated by large-$\beta_0$
approximation or by renormalization group,
and the renormalization scale is varied around the
minimal-sensitivity scale.
A ``Coulomb''+linear potential can be identified
with the scale-independent and renormalon-free
part of the prediction
and can be separated from the renormalon-dominating
part.
(II) In the frame of OPE, we define two types of
renormalization schemes for the
leading Wilson coefficient.
One scheme belongs to the 
class of conventional factorization schemes.
The other scheme belongs to a new class, which is
independent of the factorization scale, derived from a 
generalization of the ``Coulomb''+linear potential
of (I).
The Wilson coefficient is free from
IR renormalons and IR divergences in both schemes.
We study properties of the Wilson coefficient and
of the corresponding non-perturbative contribution
$\delta E_{\rm US}(r)$ in each scheme.
(III) We compare numerically perturbative predictions of
the Wilson coefficient and lattice computations of
$V_{\rm QCD}(r)$ when $n_l=0$.
We confirm either correctness or consistency (within
uncertainties) of the theoretical predictions made in (II).
%We confirm that the theoretical predictions obtained in
%(II) are either correct or consistent within the present
%level of uncertainties.
Then we perform fits to simultaneously
determine $\delta E_{\rm US}(r)$ and 
$ r_0 \Lambda_{\overline{\rm MS}}^{\mbox{\scriptsize 3-loop}} $
(relation between lattice scale and $\Lambda_{\overline{\rm MS}}$).
As for the former quantity, 
we improve bounds
as compared to the previous determination;
as for the latter quantity, our analysis provides a new
method for its determination.
We find that (a) $\delta E_{\rm US}(r)=0$ is disfavored, and
(b) $ r_0 \Lambda_{\overline{\rm MS}}^{\mbox{\scriptsize 3-loop}} 
=0.574\pm 0.042$.
We elucidate the mechanism for the sensitivities and
examine sources of errors in detail.
}
\end{abstract}
\vfil

\end{titlepage}

%%%%%%%%%%%%%%%%%%%%%%%%%%%%%%%%%%%%%%%%%%%%%%%%%%%%%%%%%
\section{Introduction}
\label{s1}

In this article, we study the QCD potential
for a static quark-antiquark ($Q\bar{Q}$) pair,
in the distance region
$0.5~{\rm GeV}^{-1}\, (0.1~{\rm fm})\simlt r 
\simlt 5~{\rm GeV}^{-1}\, (1~{\rm fm})$.
%, or parametrically at $r \simlt \LQ^{-1}$.
This region is known to be relevant to the spectroscopy
of the heavy quarkonium states.
We use perturbative QCD and operator-product-expansion
(OPE) as basic theoretical tools, taking
advantage of dramatic theoretical developments that took place 
in the last decade.
In addition, we use recent accurate results of lattice computations of
the QCD potential.

For 30 years, the static QCD potential $V_{\rm QCD}(r)$ 
has been studied extensively
for the purpose of elucidating the nature of the interaction between heavy
quark and antiquark.
Generally, $V_{\rm QCD}(r)$ at short-distances can be computed accurately
by perturbative QCD.
On the other hand,
the potential shape at long-distances should be determined by
non-perturbative methods, such as
lattice simulations or phenomenological potential-model analyses;
in the latter approach phenomenological potentials are extracted
from experimental data for the heavy quarkonium spectra.

Computations of $V_{\rm QCD}(r)$ in perturbative QCD has a long
history.
At tree-level, $V_{\rm QCD}(r)$ is merely a Coulomb potential,
$-(4/3)(\alpha_S/r)$, arising from one-gluon-exchange diagram.
The 1-loop correction (with massless internal quarks) was already computed in
\cite{Appelquist:es,Fischler:1977yf}.
The 1-loop correction due to massive internal quarks was computed
in \cite{Billoire:1979ih}.
It took a rather long time before
the 2-loop correction (with massless internal quarks) was computed in 
\cite{Peter:1996ig}; part of this result was corrected
soon in \cite{Schroder:1998vy}.
The 2-loop correction due to massive internal quarks was computed
in \cite{Melles:2000dq,Melles:2000ey,Hoang:2000fm};
misprints in \cite{Melles:2000ey,Hoang:2000fm} were corrected
in \cite{Recksiegel:2001xq}.
The logarithmic correction at 3-loop
originating from the ultrasoft scale
was first pointed out in \cite{Appelquist:es} and
computed in \cite{Brambilla:1999qa,Kniehl:1999ud}.
A renormalization-group (RG) improvement of $V_{\rm QCD}(r)$ at
next-to-next-to-leading logarithmic order (NNLL), including the
ultrasoft logarithms, 
was performed in \cite{Pineda:2000gz}.
(There exist estimates of higher-order corrections to the perturbative
QCD potential 
in various methods 
\cite{Chishtie:2001mf,Pineda:2001zq,Cvetic:2003wk}.)\footnote{
Recently 2-loop correction to the octet QCD potential
has been computed \cite{Kniehl:2004rk}.
}

For a long time, the perturbative QCD predictions of $V_{\rm QCD}(r)$
were {\it not} successful 
in the distance region relevant to the bottomonium and charmonium
states, 
$0.5~{\rm GeV}^{-1} \simlt r \simlt
5~{\rm GeV}^{-1}$.
In fact, the perturbative series turned out to be very poorly convergent at
$r \simgt 0.5~{\rm GeV}^{-1}$; 
uncertainty of the series is so large that one could hardly obtain
meaningful prediction in this distance region.
Even if one tries to improve the perturbation series by
certain resummation prescriptions (such as RG improvement),
scheme dependence of the results turns out to be very large;
hence, one can neither
obtain accurate prediction of the potential in this distance region.
For instance, the QCD potential bends downwards at large $r$
as compared to the Coulomb potential if the $V$-scheme running
coupling constant is used, whereas the potential bends
upwards at large $r$ if the $F$-scheme running coupling constant
is used \cite{Grunberg:1989xf}.
(See e.g.\ Fig.~4 of \cite{Sumino:2001eh}.)
It was later pointed out that the large uncertainty of the perturbative 
QCD prediction can be understood as caused by
the ${\cal O}(\LQ)$ infrared (IR) renormalon contained
in $V_{\rm QCD}(r)$ \cite{Aglietti:1995tg}.

Empirically it has been known that phenomenological potentials
and lattice computations of $V_{\rm QCD}(r)$ are both
approximated well by the sum of a Coulomb potential and a linear
potential in the above range $0.5~{\rm GeV}^{-1} \simlt r \simlt
5~{\rm GeV}^{-1}$ \cite{Bali:2000gf}.
The linear behavior of $V_{\rm QCD}(r)$ at large distances
$r \gg \LQ^{-1}$, verified numerically by lattice simulations,
is consistent with the quark confinement picture.
For this reason,  and given the very poor predictability of perturbative
QCD,
it was often said that, while the ``Coulomb'' part of $V_{\rm QCD}(r)$  
(with logarithmic corrections at short-distances) is
contained in the perturbative QCD prediction, the linear part
is purely non-perturbative and absent in
the perturbative QCD prediction (even at $r < \LQ^{-1}$), and that
the linear potential needs to be added to the perturbative prediciton to obtain 
the full QCD potential.
Nevertheless, to the best of our knowledge, there was no firm theoretical basis
for this argument.

Since the discovery \cite{Pineda:id,Hoang:1998nz,Beneke:1998rk} of the 
cancellation of ${\cal O}(\Lambda_{\rm QCD})$
renormalons in the total energy of a static quark-antiquark pair
$E_{\rm tot}(r) \equiv V_{\rm QCD}(r) + 2m_{\rm pole}$,
convergence of the perturbative series for $E_{\rm tot}(r)$
improved drastically and
much more accurate perturbative predictions 
for the potential shape became available.
It was understood that a large uncertainty originating from
the ${\cal O}(\Lambda_{\rm QCD})$
renormalon in $V_{\rm QCD}(r)$ can be absorbed into
twice of the quark pole mass $2m_{\rm pole}$.
Once this is achieved, perturbative
uncertainty of $E_{\rm tot}(r)$ is 
estimated to be 
${\cal O}(\Lambda_{\rm QCD}^3 r^2)$
at $r \simlt \Lambda_{\rm QCD}^{-1}$ \cite{Aglietti:1995tg},
based on the renormalon dominance hypothesis.

On the other hand, OPE
of $V_{\rm QCD}(r)$ for $r \ll \Lambda_{\rm QCD}^{-1}$ 
was developed \cite{Brambilla:1999qa,Brambilla:1999xf} within an 
effective field theory  
``potential non-relativistic QCD'' (pNRQCD)
\cite{Pineda:1997bj}.
In this framework, $V_{\rm QCD}(r)$ is expanded in $r$
(multipole expansion).
At each order of this expansion, short-distance contributions are factorized
into Wilson coefficients (perturbatively computable)
and long-distance contributions into matrix elements of operators
(non-perturbative quantities).
The leading non-perturbative contribution to the potential is
contained in the ${\cal O}(r^2)$ term of the multipole expansion.
%to be (at most)
%${\cal O}(\Lambda_{\rm QCD}^3 r^2)$ \cite{Brambilla:1999qa} ???.

Subsequently, several studies 
\cite{Sumino:2001eh,Recksiegel:2001xq,Pineda:2002se,Recksiegel:2002um}
showed that perturbative
predictions for  $V_{\rm QCD}(r)$ agree well
with phonomenological potentials  
and lattice calculations of $V_{\rm QCD}(r)$, 
once the ${\cal O}(\Lambda_{\rm QCD})$ renormalon contained 
in $V_{\rm QCD}(r)$ is cancelled.
In particular, in the context of OPE, the leading Wilson coefficient was 
shown to be in agreement with
lattice computations of $V_{\rm QCD}(r)$, after the
subtraction of the ${\cal O}(\Lambda_{\rm QCD})$ 
renormalon \cite{Pineda:2002se}.
Ref.\ \cite{Lee:2002sn} showed that 
a Borel resummation of the perturbative series gives a potential shape
which agrees with lattice results, if the ${\cal O}(\Lambda_{\rm QCD})$ 
renormalon is properly taken into account.
In fact, these agreements hold within
uncertainties of  ${\cal O}(\Lambda_{\rm QCD}^3 r^2)$ 
estimated from the residual renormalon.
That is,
a linear potential of ${\cal O}(\Lambda_{\rm QCD}^2\, r)$ 
at $r \simlt \Lambda_{\rm QCD}^{-1}$
was ruled out numerically  in the differences 
between the perturbative predictions
and phenomenological potentials/lattice results.
These observations support the validity of renormalon dominance
hypothesis.

A crucial point is that,
once the ${\cal O}(\Lambda_{\rm QCD})$ renormalon is cancelled and
the perturbative prediction is made accurate,
the perturbative potential becomes steeper than
the Coulomb potential as $r$ increases.
This feature is understood, within perturbative QCD, 
as an effect of the {\it running} of the strong coupling constant 
\cite{Brambilla:2001fw,Sumino:2001eh}.

Soon after,
it was shown analytically \cite{Sumino:2003yp} that the
perturbative QCD potential approaches a ``Coulomb''+linear
form at large orders, up to an ${\cal O}(\Lambda_{\rm QCD}^3 r^2)$ uncertainty.
(Here and hereafter, the ``Coulomb'' potential {\it with quotes}
represents a Coulombic potential with logarithmic corrections at 
short distances.)
Higher-order terms were estimated by the large-$\beta_0$ approximation
or by RG equation and
a scale-fixing prescription based on renormalon dominance hypothesis was used.
The ``Coulomb''+linear potential can be computed systematically 
via RG; up to NNLL,
it shows a convergence towards lattice computations of $V_{\rm QCD}(r)$.
Furthermore, the ``Coulomb''+linear potential was shown to coincide
with the leading Wilson coefficient in the framework of OPE,
up to an ${\cal O}(r^2)$ difference \cite{Sumino:2004ht}.
\medbreak

In this paper, we perform a precise and solid analysis, on the basis of
our previous works \cite{Sumino:2003yp,Sumino:2004ht}.
This work extends our previous works in the following respects:
\begin{itemize}
\item
We incorporate a degree of freedom for varying renormalization scale
into the analysis of 
\cite{Sumino:2003yp}.
In this way, the ``Coulomb''+linear potential is identified with the
scale-independent part of the prediction.
Details of the derivation and formulas not delivered so far are also presented.
\item
We promote the ``Coulomb''+linear potential to the leading Wilson coefficient
in the framework of OPE, taking advantage of the result of \cite{Sumino:2004ht}.
We study properties of the Wilson coefficient and
the corresponding non-perturbative
correction $\delta E_{\rm US}(r)$.
\end{itemize}
In addition, we present the following analysis:
\begin{itemize}
\item
We determine the non-perturbative correction $\delta E_{\rm US}(r)$
using perturbative computations of the
Wilson coefficients and recent lattice data.
\item
As a byproduct, we determine the relation between lattice scale
(Sommer scale) and $\Lambda_{\overline{\rm MS}}$.
This provides a new method to determine this relation.
\end{itemize}
In this analysis, we 
assemble all the developments of perturbative computations and of
OPE up to date.

Organization of the paper is as follows.
Sec.~\ref{s2} is devoted to a review: we review the current status of the
perturbative QCD computations of $V_{\rm QCD}(r)$ (Sec.~\ref{s2.1}), 
convergence property of $E_{\rm tot}(r)$ up to ${\cal O}(\alpha_S^3)$ 
(Sec.~\ref{s2.2}),
large-order
behavior of the perturbative series
based on renormalon argument (Sec.~\ref{s2.3}),
and the predictions of OPE for $V_{\rm QCD}(r)$ (Sec.~\ref{s2.4}).
In Sec.~\ref{s3}, we analyze the large-order behavior of the perturbative
prediction of $V_{\rm QCD}(r)$ analytically:
After explaining the strategy in Sec.~\ref{s3.1}, we present
the results when the higher-order terms are estimated by 
the large-$\beta_0$
approximation and by RG in Secs.~\ref{s3.2} and \ref{s3.3},
respectively.
Details of the derivation are given through Secs.~\ref{s3.4} and \ref{s3.5}.
(The readers may as well skip these details in the first reading.)
Sec.~\ref{s4} defines two types of
renormalization schemes for the leading Wilson coefficient 
in the context of OPE (Secs.~\ref{s4.1} and \ref{s4.2}) and
discusses properties of the Wilson coefficient and of the 
corresponding non-perturbative contributions (Sec.~\ref{s4.3}).
In Sec.~\ref{s5} we compare the perturbative computations of the
Wilson coefficient with lattice compuations of $V_{\rm QCD}(r)$.
We first check consistency of theoretical predictions based on OPE
(Sec.~\ref{s5.1}).
Then we determine the 
non-perturbative contribution
in each scheme as well as the relation between lattice scale and
$\Lambda_{\overline{\rm MS}}$ (Secs.~\ref{s5.2} and \ref{s5.3}).
Summary and conclusions are given in Sec.~\ref{s6}.

App.~\ref{appA} collects the formulas necessary for the computation of
the perturbative series of the QCD potential.
In App.~\ref{appB}, we give a derivation of the one-parameter integral
representation of $[\alpha_{V}^{\rm PT}(q)]_\infty$.
In App.~\ref{appC}, we present the analytic formula for the linear potential
up to NNLL.
Methods for numerical evaluation of the Wilson coefficient
are given in App.~\ref{appD}.

\section{Perturbation Series and OPE of \boldmath $V_{\rm QCD}(r)$ (Review)}
\label{s2}
\clfn

\subsection{Definitions and conventions}
\label{s2.1}

Throughout this paper, color factors of QCD are denoted as
\bea
C_F = \frac{4}{3},
~~~~~
C_A = N_C = 3,
~~~~~
T_F = \frac{1}{2},
\eea
where $N_C$ is the number of color,
$C_F$ is the second Casimir operator of the fundamental 
representation,
$C_A$ is the second Casimir operator of the adjoint representation,
and $T_F$ is the trace normalization of the fundamental
representaion of the color $SU(3)$ group.
Furthermore, we denote the number of light quark flavors 
by $n_l$.
We assume that all light quarks are massless (except in Sec.~\ref{s2.2}).

The static QCD potential is defined from an expectation value
of the Wilson loop as
\bea
V_{\rm QCD}(r) 
&=& 
- \lim_{T \to \infty} \frac{1}{iT} \,
\log \frac{\bra{0} {\rm{Tr\, P}} \exp
\biggl[ i g_S \oint_{\cal P} dx^\mu \, A_\mu(x) \biggr]
\ket{0}}
{\bra{0} {\rm Tr} \, {\bf 1} \ket{0}}
\label{Wilson-loop}
\\
&=& 
\int \frac{d^{d}\vec{q}}{(2\pi)^{d}} \, e^{i \vec{q} \cdot \vec{r}}
\, \biggl[
-4 \pi  C_F \, \frac{\alpha_V(q)}{q^2}
\biggr]
~~~;~~~~~
q = |\vec{q}|
,
\eea
where ${\cal P}$ is a rectangular loop of spatial extent $r$ and
time extent $T$.
The second line defines the $V$-scheme coupling contant,
$\alpha_V(q)$, in momentum space.
In dimensional regularization, there are one temporal dimension
and $d = D-1 = 3-2\epsilon$ spatial dimensions.

In perturbative QCD, $\alpha_V(q)$ is calculable in series
expansion of the strong coupling constant.
We denote the perturbative evaluation of 
$\alpha_V(q)$ as
\bea
\alpha_V^{\rm PT}(q) 
&=& \alpha_S(\mu) \, \sum_{n=0}^{\infty} P_n(\log (\mu/q) ) \,
\biggl( \frac{\alpha_S(\mu)}{4\pi} \biggr)^n
\label{alfVPT}
\\
&=& 
\alpha_S(q) \, \sum_{n=0}^{\infty} P_n(0) \,
\biggl( \frac{\alpha_S(q)}{4\pi} \biggr)^n 
.
\label{alfVRG}
\eea
Here, $\alpha_S(\mu)$ denotes the strong coupling constant
renormalized at the renormalization scale $\mu$,
defined in the modified minimal subtraction ($\overline{\rm MS}$) scheme;
$P_n(\ell)$ denotes an $n$-th-degree polynomial of $\ell$.
In the second equality, we set $\mu = q$ using $\mu$-independence
of $\alpha_V^{\rm PT}(q)$.
Eq.~(\ref{alfVRG}) is reduced to eq.~(\ref{alfVPT}),
if we insert 
the series expansion of $\alpha_S(q)$ in terms of $\alpha_S(\mu)$.
This expansion is determined by the RG equation
\bea
q^2 \, \frac{d}{dq^2} \, \alpha_S(q) = \beta(\alpha_S(q)) =
- \alpha_S(q) \sum_{n=-1}^{\infty} \beta_n 
\biggl( \frac{\alpha_S(q)}{4\pi} \biggr)^{n+1},
\label{RGeq}
\eea
where $\beta_n$ represents the $(n+1)$-loop coefficient of the
beta function.\footnote{
In dimensional regularization and $\overline{\rm MS}$ scheme,
$\beta_{-1} \neq 0$ when the space-time dimension is different
from 4; see App.~\ref{appA}, eq.~(\ref{betafn-gen-dim}).
}
Eqs.~(\ref{alfVPT})(\ref{alfVRG}) show that, 
at each order of the expansion of $\alpha_V^{\rm PT}(q)$ in $\alpha_S(\mu)$, 
the only part of the polynomial $P_n(\log (\mu/q) )$ 
that is not determined by the RG equation is $P_n(0)$.

\begin{figure}[t]\centering
\includegraphics[width=8cm]{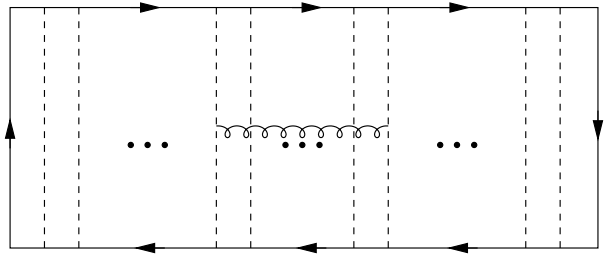}
\caption{\small 
Class of diagrams contributing to the QCD potential at
${\cal O}(\alpha_S^4\log\alpha_S)$.
Dashed lines represent Coulomb gluons;
curly line represents transverse gluon.
\label{Coulomb-resum}}
\end{figure}

It is known \cite{Appelquist:es} that 
$P_n(0)$ for $n \ge 3$ contain IR divergences.
Namely, the perturbative QCD potential is IR divergent 
and not well-defined at and beyond
${\cal O}(\alpha_S^4)$.
There are two ways to deal with this problem.
One way is to use OPE, in which
the QCD potential is factorized into Wilson coefficients
and matrix elements.
The Wilson coefficients include only ultraviolet
(UV) contributions, hence they
are computable in perturbative expansion in $\alpha_S$
free from IR divergences.
IR contributions are contained in the matrix elements which
are non-perturbative quantities.
Another way is to expand the QCD potential as a double
series in $\alpha_S$ and $\log \alpha_S$.
This is achieved by resummation of certain class of diagrams
(Fig.~\ref{Coulomb-resum}) as indicated by \cite{Appelquist:es}.
More systematically, this can be achieved
within pNRQCD framework \cite{Brambilla:1999qa,Kniehl:1999ud,Brambilla:1999xf}.
We will use both methods for regularization of IR divergences
through Secs.~\ref{s3}--\ref{s5}.\footnote{
In our analysis in Sec.~\ref{s3}, regularization
of IR divergences is rather a conceptual matter;
there, our practical analysis concerns only up to the orders where
IR finite terms are involved.
On the other hand, in 
Secs.~\ref{s4}--\ref{s5}, we include ${\cal O}(\alpha_S^4)$
term in our analysis, hence the regularization becomes
practically relevant.
}

Let us explain our terminology for the order counting.
When we state ``$\alpha_V^{\rm PT}(q)$
up to ${\cal O}(\alpha_S^N)$,'' we mean that we truncate the series
on the right-hand-side of eq.~(\ref{alfVPT}) and take the sum
for $0 \leq n \leq N\! -\! 1$.
We also improve the perturbation series using the RG
evolution of the $\overline{\rm MS}$ coupling.\footnote{
It is known that, up to NNLL, the RG-improved $\overline{\rm MS}$ 
running coupling
is more convergent than the RG-improved running coupling in the
$V$-scheme or $F$-scheme, hence the RG-improvement in the 
$\overline{\rm MS}$-scheme leads to a more stable prediction
of the potential shape; see \cite{Jezabek:1998pj} and Sec.~4 
of \cite{Sumino:2001eh}.
For this reason, we adopt the RG-improvement in the 
$\overline{\rm MS}$-scheme in this paper.
}
By $\alpha_V^{\rm PT}(q)$
up to LL, NLL, NNLL and NNNLL, 
we mean that we define $\alpha_V^{\rm PT}(q)$ by
eqs.~(\ref{alfVRG})(\ref{RGeq}) and take the
sums for $0 \leq n \leq 1$, 2, 3 and 4, respectively, in both equations
(i.e.\ 1-, 2-, 3- and 4-loop running coupling constants are 
used for $\alpha_S(q)$, respectively).
This procedure resums logarithms of the forms 
$\alpha_S(\mu) [\alpha_S(\mu) \log (\mu/q)]^n$,
$\cdots$, 
$\alpha_S(\mu)^4 [\alpha_S(\mu) \log (\mu/q)]^n$, respectively.

On the other hand, the IR divergences at ${\cal O}(\alpha_S^4)$
and beyond induce additional
powers of $\log (\mu/q)$ in $\alpha_V(q)$
at NNLL and beyond, 
which are not resummed
by the evolution of $\alpha_S(q)$ via eq.~(\ref{RGeq}).
Hence, at these orders, it is more consistent 
(with respect to naive power counting)
to resum these IR logarithms (referred usually
as ultrasoft logarithms)
as well, 
although physical origins of the logarithms are quite different.
The ultrasoft (US) logarithms at NNLL can be resummed
by replacing the $V$-scheme coupling
constant as \cite{Pineda:2000gz}
\bea
\alpha_V^{\rm PT}(q) \to \alpha_V^{\rm PT}(q) 
+ \frac{C_A^3}{6\beta_0} \, \alpha_S(q)^3 \,
\log \biggl[ \frac{\alpha_S(q)}{\alpha_S(\mu_f)} \biggr] ,
\label{NNLLUSlog}
\eea
where $\mu_f$ denotes the factorization scale.
We will examine the resummation of US logs separately.

For $n \leq 2$, we define $a_n \equiv P_n(0)$.
For $n \geq 3$, we include US logs into $a_n$ in addition.
Explicit expressions for
$P_n(\ell)$, $a_n$, $\beta_n$ up to $n = 3$
(except for the unknown part of $a_3$)
are listed in App.~\ref{appA}.
Furthermore, for convenience, we will denote
\bea
\delta = {\beta_1}/{\beta_0^2} 
\eea
in the following.
Other formulas, useful
for evaluation of $\alpha_V^{\rm PT}(q)$,
are collected in App.~\ref{appA} as well.

\subsection{Convergence and scale-dependences of
\boldmath $E_{\rm tot}(r)$ up to ${\cal O}(\alpha_S^3)$}
\label{s2.2}

Let us demonstrate the improvement of accuracy of the perturbative
prediction for the total energy
$E_{\rm tot}(r) = 2 m_{\rm pole} + V_{\rm QCD}(r)$ 
up to ${\cal O}(\alpha_S^3)$, when the cancellation of
${\cal O}(\LQ)$ renormalons is incorporated.
This is achieved (even without any knowledge of renormalons)
if one re-expresses the quark pole mass $m_{\rm pole}$ by
the $\overline{\rm MS}$ mass in series expansion 
in $\alpha_S(\mu)$.
Presently perturbation series of $V_{\rm QCD}(r)$ 
\cite{Peter:1996ig,Schroder:1998vy} and 
$m_{\rm pole}$ \cite{Melnikov:2000qh} 
are both known up to ${\cal O}(\alpha_S^3)$.

\begin{figure}[t]\centering
\psfrag{mu}{\hspace{0mm}$\mu$~[GeV]}
\psfrag{Etot}{\hspace{2mm}$E_{\rm tot}(r)$~~[GeV]}
\psfrag{r = 2.5}{\hspace{0mm}$r=2.5$~GeV$^{-1}$}
\psfrag{Pole-mass scheme}{\hspace{0mm}\raise-0pt\hbox{Pole-mass scheme}}
\psfrag{MSbar-mass scheme}{\hspace{0mm}$\overline{\rm MS}$-mass scheme}
\hspace{-3cm}
\includegraphics[width=8cm]{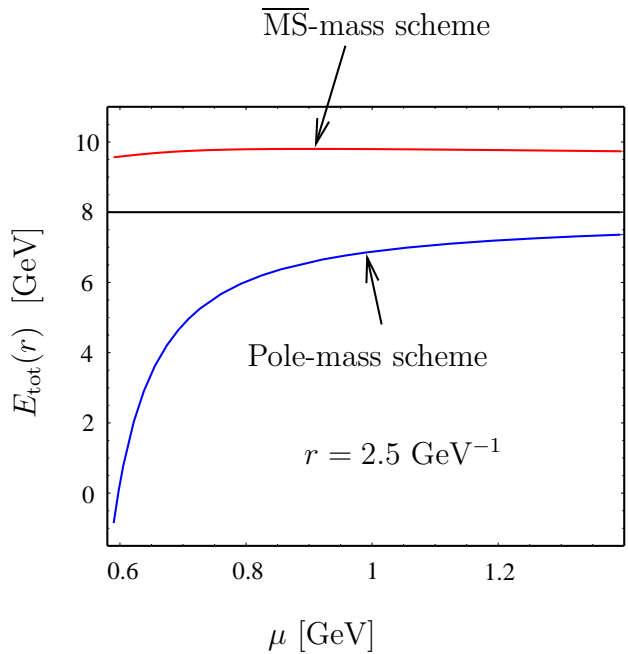}
\caption{\small 
Scale dependences of $E_{{\rm tot}}^{b\bar{b}}(r)$ 
up to ${\cal O}(\alpha_S^3)$ at 
$r =2.5~{\rm GeV}^{-1} \approx 0.5$~fm, in the pole-mass and 
$\overline{\rm MS}$-mass schemes.
A horizontal line at 8~GeV is shown for a guide.
\label{scale-dep}}
\end{figure}

As an example, we take the bottomonium case:\footnote{ 
$E_{{\rm tot}}^{b\bar{b}}(r) = 2m_{b,{\rm pole}}+V_{\rm QCD}(r)$ 
has been applied to computations of the 
bottomonium spectrum \cite{Recksiegel:2002za}.
}
We choose the $\overline{\rm MS}$
mass of the $b$-quark, renormalized at the  $b$-quark $\overline{\rm MS}$
mass, as
$\overline{m}_b \equiv 
m_b^{\overline{\rm MS}}(m_b^{\overline{\rm MS}})
= 4.190$~GeV;
in internal loops, four flavors of light quarks are included
with $\overline{m}_u=\overline{m}_d=\overline{m}_s=0$
and $\overline{m}_c=1.243$~GeV.
(See the formula for $E_{{\rm tot}}^{b\bar{b}}(r)$ in \cite{Recksiegel:2001xq}.)
In Fig.~\ref{scale-dep}, we fix $r=2.5~{\rm GeV}^{-1} \approx 0.5$~fm
(midst of the distance range of our interest)
and examine the renormalization scale ($\mu$) dependence of $E_{{\rm tot}}(r)$.
We see that $E_{{\rm tot}}(r)$ is much less scale dependent when we
use the $\overline{\rm MS}$ mass (after cancellation
of renormalons) than when we use the pole mass 
(before cancellation of renormalons).
This shows clearly that the perturbative prediction of $E_{{\rm tot}}(r)$
is much more
stable in the former scheme.

We also compare the convergence behaviors
of the perturbative series of $E_{\rm tot}(r)$ for the same $r$
and when $\mu$ is fixed to the minimal-sensitivity scale \cite{Stevenson:1981vj}
(the scale at which $E_{\rm tot}$
becomes least sensitive to variation of $\mu$)
in the 
$\overline{\rm MS}$-mass scheme.
At $r =2.5~{\rm GeV}^{-1}$, the minimal-sensitivity scale is
$\mu = 0.90$~GeV.
Convergence of the perturbation series turns out to be close to optimal
for this scale choice:\footnote{
In the pole-mass scheme, there exists no minimal-sensitivity scale within a
wide range of $\mu$, and the convergence behavior of the series is
qualitatively similar to eq.~(\ref{conv-polemass-scheme}) within this range.
}
\bea
E_{{\rm tot}}^{b\bar{b}}(r) &=& 
10.408 - 0.275 - 0.362 - 0.784 ~~{\rm GeV}
~~~~~
\mbox{(Pole-mass scheme)}
\label{conv-polemass-scheme}
\\
&=&
~\,8.380 + 1.560 -0.116 - 0.022~~{\rm GeV}
~~~~~
\mbox{($\overline{\rm MS}$-mass scheme)} .
\eea
The four numbers represent the ${\cal O}(\alpha_S^0)$,
${\cal O}(\alpha_S^1)$, ${\cal O}(\alpha_S^2)$ and 
${\cal O}(\alpha_S^3)$ terms of the series expansion in each
scheme.
The ${\cal O}(\alpha_S^0)$ terms represent the twice of
the pole mass and of the $\overline{\rm MS}$ mass, respectively.
As can be seen, if we use the pole mass, the series is not
converging beyond ${\cal O}(\alpha_S^1)$, whereas
in the $\overline{\rm MS}$-mass scheme, the series is converging.
One may further verify that, 
when the series is converging ($\overline{\rm MS}$-mass scheme),
$\mu$-dependence of $E_{\rm tot}(r)$ decreases
as we include more terms of the perturbative series,
whereas when the series is diverging
(pole-mass scheme), $\mu$-dependence does not decrease with
increasing order.
(See e.g.\ \cite{Kiyo:2002rr}.)

We observe qualitatively the same features at
different $r$ and for different number of light quark
flavors $n_l$, or even if we change values 
of the masses $\overline{m}_b$, $\overline{m}_c$.
Generally, at smaller $r$,
$E_{\rm tot}(r)$ becomes less
$\mu$-dependent and more convergent, 
due to the asymptotic freedom of QCD \cite{Recksiegel:2001xq}.

The stability against scale
variation and convergence of the 
perturbative series are closely connected with each other.
Formally, scale dependence vanishes 
at all order of perturbation series.
This means that,
for a truncated perturbative series up to ${\cal O}(\alpha_S^N)$,
scale dependence is of ${\cal O}(\alpha_S^{N+1})$.
Hence, the scale dependence decreases for larger $N$
as long as the series is converging.
Thus, the truncated perturbative series is expected to become 
less $\mu$-dependent with increasing order
when the series is converging.
It also follows that the series
is expected to be most convergent 
when $\mu$ is close to
the minimal-sensitivity scale.
This observation, supported by the above numerical verification
up to ${\cal O}(\alpha_S^3)$, 
forms a basis of our analysis in Sec.~\ref{s3}.
\medbreak

As already mentioned in the Introduction,
once $E_{\rm tot}(r)$ is expressed in terms of
the $\overline{\rm MS}$ mass
and an accurate prediction is obtained,
it agrees well with phenomenological potentials and lattice computations
of the QCD potential in the range of $r$ of our interest.
As more terms of the series expansion are included, $E_{\rm tot}(r)$ 
becomes steeper in this range.
This behavior originates from an increase of the interquark force due to the
running of the strong coupling constant \cite{Sumino:2001eh}.\footnote{
See \cite{Brambilla:2001fw,Sumino:2001nd} 
for a more microscopic explanation of this feature.
}
$E_{\rm tot}(r)$ up to a finite order in perturbative expansion
has a functional
form $1/r \times (\mbox{Polynomial of $\log r$})$,
apart from an $r$-independent constant;
cf.\ App.~\ref{appA}, eq.~(\ref{expand-alfVbar}).
On the other hand, we 
see a tendency that, as we increase the order, 
$E_{\rm tot}(r)$
approaches phenomenological potentials/lattice results, 
which are typically
represented by a Coulomb+linear potential.
This observation motivates us to examine 
the perturbative prediction for $E_{\rm tot}(r)$ at large orders,
which will be given in Sec.~\ref{s3}.
For that analysis, we need to know
large-order behaviors of the perturbative
series of $E_{\rm tot}(r)$.

\subsection{Large-order behaviors and IR renormalons}
\label{s2.3}
\clfn

The nature of the perturbative series of
$V_{\rm QCD}(r)$ and $E_{\rm tot}(r)$ at large orders, 
including their
uncertainties, can be 
understood within the argument based on renormalons.
The argument gives certain estimates of higher-order terms, 
and empirically it
gives good estimates even at relatively low orders of perturbative
series.

Before starting any argument on large-order behaviors, 
one may be perplexed because the
perturbative expansion of $V_{\rm QCD}(r)$ contains IR divergences
beyond ${\cal O}(\alpha_S^3)$.
For definiteness,
let us assume (conceptually) that we regularize the IR divergences by
expanding $V_{\rm QCD}(r)$ in double series in
$\alpha_S$ and $\log \alpha_S$;
then we identify ${\cal O}(\alpha_S^n)$ term with the sum of
${\cal O}(\alpha_S^n \log^k \! \alpha_S)$ terms for all $k$.\footnote{
There exists evidence that renormalon dominance may be 
valid in such an expansion \cite{Kiyo:2002rr}.
}

\begin{figure}[t]\centering
  \psfrag{Etotn}{$|E_{\rm tot}^{(n)}(r)|$}
  \psfrag{Pole-mass scheme}{Pole-mass scheme}
  \psfrag{MSbar-mass scheme}{$\overline{\rm MS}$-mass scheme}
  \psfrag{NLO}{\raise-3mm\hbox{$
                  N_1 = \frac{6\pi}{\beta_0\alpha_S(\mu)}$}}
  \psfrag{LO}{\hspace{-8mm}\raise-3mm\hbox{$
                  N_0 = \frac{2\pi}{\beta_0\alpha_S(\mu)}$}}
  \psfrag{n}{$n$}
\vspace{7mm}
  \includegraphics[width=8.5cm]{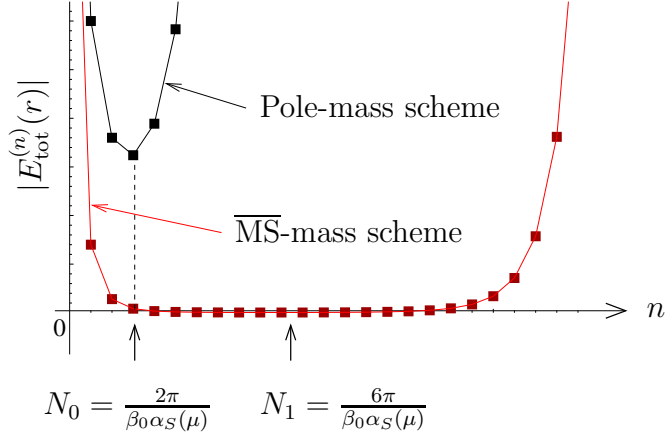}\vspace*{7mm}
\caption{\small
Diagram showing the $n$-dependence of 
$|V_{\rm QCD}^{(n)}(r)|$ (or
$|E_{\rm tot}^{(n)}(r)|$ in the pole-mass scheme) [black squares]
and that of $|E_{\rm tot}^{(n)}(r)|$ in the $\overline{\rm MS}$-mass 
scheme [red squares],
based on renormalon estimates.
\label{Vn}}
\end{figure}

Let us denote the ${\cal O}(\alpha_S^{n+1})$ term %of $V_{\rm QCD}(r)$
as $V^{(n)}_{\rm QCD}(r)$.
According to the renormalon argument, 
the leading behavior of $V^{(n)}_{\rm QCD}(r)$
at large orders $n \gg 1$ is given by
\bea
V^{(n)}_{\rm QCD}(r) \sim
{ {\rm const.}}\times n! \, \left(
\frac{\beta_0 \alpha_S(\mu)}{2\pi}\right)^n \, n^{\delta/2} ,
\label{LO-renormalon}
\eea 
up to a relative correction of ${\cal O}(1/n)$ \cite{Beneke:1998ui}.
It follows that $|V^{(n)}_{\rm QCD}(r)|$ becomes minimal at order
$n \approx N_0 \equiv 2\pi/(\beta_0 \alpha_S(\mu))$, while
$|V^{(n)}_{\rm QCD}(r)|$ scarcely changes in the range 
$N_0 - \sqrt{N_0} \ll n \ll N_0 + \sqrt{N_0}$.
For $n \gg N_0 + \sqrt{N_0}$, the series diverges rapidly.
(See Fig.~\ref{Vn}, black squares.)
Due to the divergence
(the series is an asymptotic series), 
there is a limitation to the
achievable accuracy of the perturbative prediction for $V_{\rm QCD}(r)$.
An uncertainty of the asymptotic series may be estimated by
the size of the terms around the minimum, 
$\sqrt{N_0} \times |V_{\rm QCD}^{(N_0)}(r)|$, which gives an uncertainty of 
${\cal O}(\Lambda_{\rm QCD})$ \cite{Aglietti:1995tg}.

The perturbative series
of $E_{\rm tot}(r)$ in the pole-mass scheme is the same
as that of $V_{\rm QCD}(r)$ except for the 
${\cal O}(\alpha_S^0)$ term.
If we re-express $E_{\rm tot}(r)$ in terms of the $\overline{\rm MS}$ mass,
the leading behavior of $V^{(n)}_{\rm QCD}(r)$ is cancelled against that of
the perturbative series of $2m_{\rm pole}$.\footnote{
In order to realize the cancellation of the leading behavior of
the perturbative series at each order of the expansion,
one needs to expand $V_{\rm QCD}(r)$ and $m_{\rm pole}$ in the
{\it same} coupling constant $\alpha_S(\mu)$.
This is somewhat involved technically, since usually 
$V_{\rm QCD}(r)$ and $m_{\rm pole}$
are expressed in terms of different coupling constants;
see \cite{Sumino:2001eh,Recksiegel:2001xq,Recksiegel:2002um}.
}
Then the large-order behavior of $E_{\rm tot}(r)$ becomes 
\bea
E^{(n)}_{\rm tot}(r) \sim
{\rm const}.\times r^2 \, n! \, 
\left(\frac{\beta_0 \alpha_S(\mu)}{6\pi}\right)^n \, n^{3\delta/2} .
\label{NLO-renormalon}
\eea
$|E^{(n)}_{\rm tot}(r)|$ becomes minimal at
$n \approx N_1 \equiv 6\pi/(\beta_0 \alpha_S(\mu))$
and its size scarcely changes for 
$N_1 - \sqrt{N_1} \ll n \ll N_1 + \sqrt{N_1}$.
As compared to $E_{\rm tot}(r)$ in the pole-mass scheme,
the series converges faster and up to a larger order,
but beyond order $\alpha_S^{N_1}$ again the series 
diverges.
(See Fig.~\ref{Vn}, red squares.)
An uncertainty of the perturbative prediction for $E_{\rm tot}(r)$
can be estimated similarly as
$\sqrt{N_1} \times |E_{\rm tot}^{(N_1)}(r)|
\sim {\cal O}(\Lambda_{\rm QCD}^3r^2)$ \cite{Aglietti:1995tg}.

We note that each term of the perturbation series 
($V_{\rm QCD}^{(n)}$, $E^{(n)}_{\rm tot}$) 
is dependent on
the scale $\mu$.
Hence, its large-order behavior, including
the order at which its size becomes minimal 
[$N_0,N_1\propto 1/\alpha_S(\mu)$], is
also dependent on $\mu$.
The estimated uncertainty ($\sqrt{N_0} \times |V_{\rm QCD}^{(N_0)}(r)|$,
$\sqrt{N_1} \times |E_{\rm tot}^{(N_1)}(r)|$),
however, is independent of $\mu$.

These estimates of large-order behaviors,
according to renormalons, follow primarily from analyses of
IR sensitivities of certain classes of Feynman diagrams; 
then the estimates are improved and reinforced via consistency
with RG equation \cite{Beneke:1998ui}.
The ${\cal O}(\LQ^{2k})$ IR renormalon,
corresponding to the perturbative series
\bea
c^{ren}_n(k) \sim {\rm const.}\times n! \,
\left(\frac{\beta_0 \alpha_S(\mu)}{4k\pi}
\right)^n \, n^{k\delta} ,
\label{cren}
\eea
originates typically from an
integral of the form
\bea
\int_0^{\mu_f} dq \, q^{2k-1} \, \alpha_S(q) =
\sum_n c^{ren}_n(k) ,
\eea
where $\mu_f \gg \LQ$ is a UV cutoff.
Nevertheless, in general contributions originate also from more complicated
loop integrals.

\subsection{OPE of \boldmath $V_{\rm QCD}(r)$}
\label{s2.4}
\clfn

A most solid way to separate perturbative and non-perturbative
contributions to the QCD potential is to use OPE.
OPE of the QCD potential
was developed \cite{Brambilla:1999qa,Brambilla:1999xf} within 
pNRQCD \cite{Pineda:1997bj}, which is
an effective field theory (EFT) tailored to describe dynamics
of ultrasoft gluons coupled to a quark-antiquark ($Q\bar{Q}$) system,
when the distance $r$ between $Q$ and $\bar{Q}$
is small, and when the motions of $Q$ and $\bar{Q}$ are non-relativistic. 
(In the case of the QCD potential, they are static.)
Within this EFT, the QCD potential is expanded in
$r$ (multipole expansion),
when the following hierarchy of scales exists:
\bea
\Lambda_{\rm QCD} \ll \mu_f \ll \frac{1}{r} .
\label{hierarchy}
\eea
Here, $\mu_f$ denotes the factorization scale.
Non-perturbative contributions to the QCD potential are factorized into
matrix elements of operators, while
short-distance contributions are factorized into
potentials, which are in fact Wilson coefficients.
Conceptually, physics from IR region $q<\mu_f$ is contained
in the former, while physics from UV region $q>\mu_f$ is
contained in the latter.

Explicitly, the QCD potential
is given by \cite{Brambilla:1999xf}
\bea
&&
V_{\rm QCD}(r) = V_S(r) + \delta E_{\rm US}(r),
\label{OPE}
\\ &&
\delta E_{\rm US}(r)=
- i g_S^2 \frac{T_F}{N_C}
\int_0^\infty \! \! dt \, e^{-i \, \Delta V(r)\, t} \,
\bra{0} \vec{r}\cdot\vec{E}^a(t) \varphi_{\rm adj}(t,0)^{ab}
\vec{r}\cdot\vec{E}^b(0) \ket{0}
~+~
{\cal O}(r^3) .
\label{deltaEUS}
\eea
The leading short-distance contribution to $V_{\rm QCD}(r)$
is given by the singlet potential
$V_S(r)$.
It is a Wilson coefficient, which represents 
the potential between the static $Q\bar{Q}$ pair in
the color singlet state.
The leading long-distance contribution
is contained in the matrix element in
eq.~(\ref{deltaEUS}).
It is ${\cal O}(r^2)$ in multipole expansion.
$\Delta V(r) = V_O(r) - V_S(r)$ denotes the
difference between the octet and singlet potentials;
$\vec{E}^a$ denotes the color electric
field at the center of gravity of the $Q\bar{Q}$ system.
See \cite{Brambilla:1999xf} for details.

\begin{figure}[t]\centering
\psfrag{singlet}{\small singlet}
\psfrag{octet}{\small octet}
\psfrag{k}{$q<\mu_f$}
\psfrag{vtx}{$g_s \, \vec{r}\cdot \vec{E}^a$}
\includegraphics[width=5cm]{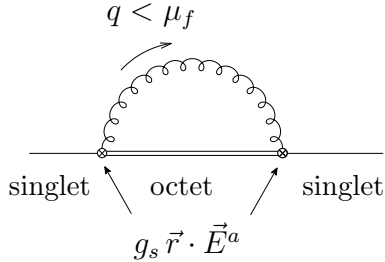}
\caption{\small 
Leading contribution of US gluon to $\delta E_{\rm US}(r)$
in pNRQCD.
\label{USgluon}}
\end{figure}

Intuitively we may understand why the leading non-perturbative 
matrix element is 
${\cal O}(r^2)$
as follows.
As well known,
the leading interaction (in expansion in $r$) between soft gluons and a 
color-singlet 
$Q\bar{Q}$ state
of size $r$ is given by the dipole interaction
$\vec{r}\cdot \vec{E}^a$.
It turns the color singlet $Q\bar{Q}$ state into a color octet $Q\bar{Q}$ state
by emission of soft gluon(s).
To return to the color singlet $Q\bar{Q}$ state, 
the color octet state needs to reabsorb the
soft gluon(s), which requires an additional dipole interaction.
Thus, the leading contribution of soft gluons to the 
total energy is ${\cal O}(r^2)$.
See Fig.~\ref{USgluon}.

Although $\delta E_{\rm US}(r)$ is ${\cal O}(r^2)$ in terms 
of the expansion
of operators, it has an additional dependence on $r$ through 
the Wilson coefficient $\Delta V(r)$.
After all, we would like to know how $\delta E_{\rm US}(r)$
depends on $r$ in the region of our interest.
The leading power of $r$ can be determined in some cases.
Since, however, the argument depends on the renormalization 
of the
singlet potential within pNRQCD, let us discuss this
issue first.

The Wilson coefficient
$V_S(r)$ can be computed in perturbative expansion in $\alpha_S$
by matching pNRQCD to QCD.
It turns out that
$V_S(r)$ thus computed
coincides with the perturbative expansion of $V_{\rm QCD}(r)$
(in dimensional regularization);
in particular, this means that $V_S(r)$
includes IR divergences beyond ${\cal O}(\alpha_S^3)$.
This result follows from a simple argument:
%The IR contribution 
Formally,
$\delta E_{\rm US}(r)$ can be computed also
in series expansion in $\alpha_S$.
This expansion, 
in dimensional regularization,
vanishes to all orders, since 
all diagrams are given by scaleless integrals.\footnote{
We neglect the masses of quarks in internal loops.
}

On the other hand,
$\delta E_{\rm US}(r)$ is expected to be non-zero beyond naive perturbation
theory.
For instance, this can be verified by computing $\delta E_{\rm US}(r)$
in pNRQCD when $\alpha_S(1/r) \ll 1$.
According to the concept of the EFT, $V_S(r)$ and $\Delta V(r)$
should be expanded in $\alpha_S$ only after all
loop integrations are carried out.
Since this theory is
assumed to correctly describe physics at energy scales much below $1/r$,
$\Delta V(r)$ ($ \ll 1/r$) should be kept in the denominator
of the propagator $[E-\Delta V(r)]^{-1}$.\footnote{
This situation is similar to the case, where one should not expand
the electron propagator by the electron mass if one wants to describe
the physics of collinear photon emission in the region
$E\theta \ll m_e$.
}
Thus, if we expand all factors except $\Delta V(r)$ in $\alpha_S$ in
eq.~(\ref{deltaEUS}), $\delta E_{\rm US}(r)$ becomes non-zero
since $\Delta V(r)$ acts as an IR regulator.
(One may expand $\Delta V(r)$ in $\alpha_S$ only after all the integrations
are performed.
Then $\log \alpha_S$ appears, in contrast to the formal expansion in $\alpha_S$,
where everything is expanded before integrations.)
In this case, $\delta E_{\rm US}(r)$ contains
UV divergences.
In dimensional regularization ($D=4-2\epsilon$), they are
given as poles in $\epsilon$, which exactly cancel 
the poles corresponding to the IR divergences in $V_S(r)$.
Consequently, in the sum eq.~(\ref{OPE}), $V_{\rm QCD}(r)$ becomes
finite as $\epsilon \to 0$.

These divergences in $V_S(r)$ and $\delta E_{\rm US}(r)$, respectively,
can be regarded as artefacts of dimensional regularization, where the
integral regions of virtual momenta extend from 0 to $\infty$.
If we introduce a hard cutoff to each momentum integration, corresponding
to the factorization scale $\mu_f$, 
$V_S(r)$ ($q>\mu_f$) and $\delta E_{\rm US}(r)$ ($q<\mu_f$),
respectively, would become finite and dependent on $\mu_f$.
This observation 
calls for renormalization of $V_S(r)$ and $\delta E_{\rm US}(r)$ within pNRQCD
also in dimensional regularization.
For example, $V_S(r)$ can be made finite
by multiplicative renormalization,
i.e.\ by adding
a counter term $(Z_S-1) V_S(r)$.

With respect to the spirit of factorization in OPE,
it is natural to subtract IR renormalons from $V_S(r)$ in a similar manner.
In \cite{Pineda:2001zq,Pineda:2002se}, this was advocated and
in practice subtraction of (only) the 
${\cal O}(\Lambda_{\rm QCD})$ renormalon was carried out explicitly.
The known IR renormalons of the bare $V_S(r)$
are contained in the integral
\cite{Beneke:1998ui}\footnote{
Here, we neglect the contributions of the instanton-anti-instanton-induced 
singularities
\cite{Beneke:1998ui}
on the positive real axis in the Borel
plane.
These contributions are known to be rather small.
}
\bea
&&
\int_0^{\mu_f} \! dq \, \frac{\sin (qr)}{qr} \, \alpha_V^{\rm PT}(q) 
=
\int_0^{\mu_f} \! dq \,
\biggl( { 1 \! - \! \frac{q^2r^2}{6} \!+\! \cdots} \biggr)
\nonumber\\
&&~~~~~~~~~~~~~~~~
\times
\biggl[ \alpha_S(q) + a_1 \Bigl( \frac{\alpha_S(q)}{4\pi} \Bigr)^2
+ \cdots \biggr]
.
\label{int-renormalon}
\eea
[Note that the perturbative expansion of 
the bare $V_S(r)$ coincides with
that of $V_{\rm QCD}(r)$.]
As for the ${\cal O}(\LQ^3r^2)$ renormalon, it was shown
that the IR renormalon contained in the bare
$V_S(r)$ and the UV renormalon contained in the
bare $\delta E_{\rm US}(r)$
cancel in dimensional regularization \cite{Sumino:2004ht}. 
In a hard cutoff renormalization
scheme, contributions of gluons to $\delta E_{\rm US}(r)$ close to
the UV cutoff region $q \sim \mu_f$ can be analyzed
using perturbative expansion in $\alpha_S$
within pNRQCD, due to
the hierarchy (\ref{hierarchy}).
It has exactly the structure suitable to absorb the ${\cal O}(\LQ^3r^2)$
renormalon contained in eq.~(\ref{int-renormalon}).
Namely, in a hard cutoff 
scheme,
the ${\cal O}(\LQ^3r^2)$ renormalon is subtracted from $V_S(r)$ and 
absorbed into $\delta E_{\rm US}(r)$.
The $\mu_f$-dependences that enter as a consequence
cancel between
the renormalized $V_S(r)$ and $\delta E_{\rm US}(r)$ \cite{Brambilla:1999xf}.
Hence, everything holds in parallel with the case of
IR divergences discussed above.
Therefore, 
it is appropriate to subtract from $V_S(r)$ the IR renormalons,
e.g.\ in the form of
eq.~(\ref{int-renormalon}), in addition to subtracting IR divergences, 
and to define a renormalized
singlet potential.
(We will give explicit renormalization prescriptions in Sec.~\ref{s4}.)

More generally, it is known that, 
in a wide class of physical observables (whenever OPE is available),
IR renormalons in perturbation series
are deeply connected with OPE of the corresponding
physical observables.
As we have seen, renormalon uncertainties have power-like behaviors in the ratio of
a large scale and $\LQ$ [in our case $ (r\,\LQ)^k \times \LQ$].
In OPE, non-perturbative contributions (matrix elements of operators)
have the same power-like structures.
Therefore, in an appropriate renormalization prescription, 
IR renormalons contained in 
perturbative series can be subtracted from Wilson coefficients and
absorbed into matrix elements in OPE, thereby leaving
Wilson coefficients free from IR renormalons.
It means that (in principle) Wilson coefficients can be computed to
arbitrary accuracy by perturbative expansion.
At the same time, renormalon ambiguities are replaced by
matrix elements of operators
(condensates), the values of 
which can be determined by comparing to various experimental data
or results of lattice simulations.

Now we return to the discussion on
the $r$-dependence of $\delta E_{\rm US}(r)$
when $r \ll \LQ^{-1}$ \cite{Brambilla:1999xf}. 
We assume that $V_S(r)$ and $\delta E_{\rm US}(r)$ are
renormalized in a hard cutoff scheme, according to the above 
discussion.
One can derive the $r$-dependence of $\delta E_{\rm US}(r)$
clearly when 
$\Delta V(r) \approx C_A\alpha_S/r \gg \mu_f $ $(\gg \LQ)$.
Since, in this case, the exponential factor in eq.~(\ref{deltaEUS})
is rapidly oscillating, we can expand the matrix element in $t$.
Then the matrix element reduces to a local gluon condensate, and
from purely dimensional analysis, $\delta E_{\rm US}(r)$
becomes ${\cal O}(\mu_f^4 r^3)$.\footnote{
Note that we may ignore $\LQ$ in comparison
to $\mu_f$, since $\mu_f \gg \LQ$.
An alternative derivation is to compute contributions of gluons
from the region $\LQ \ll q \simlt \mu_f$
using perturbative expansion in $\alpha_S$.
}
The condition $\Delta V(r) \gg \mu_f \gg \LQ$
is satisfied at sufficiently short 
distances.

Another case, in which $r$-dependence of 
$\delta E_{\rm US}(r)$ is known, is when
$\mu_f \gg \Delta V(r)$ is satisfied, in addition to the
hierarchy (\ref{hierarchy}).
This condition is expected to hold at $r \ll \LQ^{-1}$
but not for too small $r$.
Under this condition, $\delta E_{\rm US}(r)$
is dominated by contributions of gluons
from the region $\LQ, \Delta V(r) \ll q \simlt \mu_f$,
which can be computed in perturbative expansion
in $\alpha_S$.
This leads to
$\delta E_{\rm US}(r) \sim {\cal O}(\mu_f^3 r^2)$.

Let us discuss the case where $\mu_f $ is reduced and
taken close to $\LQ$.
This case violates the conventional hierarchy
condition (\ref{hierarchy}).
If $\Delta V(r) \gg \LQ$, the
matrix element can still be reduced to the local gluon condensate,
and $\delta E_{\rm US}(r) \sim {\cal O}(\LQ^4 r^3)$.
On the other hand, if $\Delta V(r) \sim \LQ$,
there is no way to predict the $r$-depndence of 
$\delta E_{\rm US}(r)$ in a model-independent way.
If  $\Delta V(r) \ll \LQ$, we can expand the
exponential factor in $\Delta V(r)$
in eq.~(\ref{deltaEUS})  and find 
$\delta E_{\rm US}(r) \sim {\cal O}(\LQ^3 r^2)$.\footnote{
In this paper, we do not consider the possibility $\Delta V(r) \ll \LQ$
henceforth,
since such large $r$ seem to lie beyond the applicable range of our analysis.
}

In the distance range of our interest,
$0.1~{\rm fm} \simlt r \simlt 1~{\rm fm}$,
the relation between $\LQ$ and $\Delta V(r)$ is
not very clear.
A rough estimate shows that, at small $r$ within
this range (perhaps $r < 0.3$~fm), $\Delta V(r) \gg \LQ$, whereas
at larger $r$ (perhaps $r > 0.3$~fm), $\Delta V(r) \sim \LQ$.
However, of course, this depends on a precise definition of $\LQ$ and
accurate knowledge of $\Delta V(r)$.
It is quite probable that 
there exists no $\mu_f$ within the above range of $r$ such that
$\Delta V(r) \gg \mu_f \gg \LQ$ can be satisfied.
Therefore, if we choose $\mu_f \gg \LQ$, we would expect
$\delta E_{\rm US}(r) \sim {\cal O}(\mu_f^3 r^2)$
in the entire range $0.1~{\rm fm} \simlt r \simlt 1~{\rm fm}$.
On the other hand,
if we choose $\mu_f \sim \LQ$, we conjecture that
at small distances
 (perhaps $0.1~{\rm fm} \simlt r \simlt 0.3$~fm), 
$\delta E_{\rm US}(r) \sim {\cal O}(\LQ^4 r^3)$,
whereas at larger distances, we cannot predict
the $r$-dependence of $\delta E_{\rm US}(r)$ in
a model-independent way.
\medbreak

To end this subsection, let us discuss what is indicated by
OPE of the QCD potential as given above.
Suppose we consider an expansion of $V_{\rm QCD}(r)$ at 
$r \simlt \LQ^{-1}$ (in the distance range of our interest):
\bea
V_{\rm QCD}(r) \approx  \frac{c_{-1}}{r} + c_0 + c_1\, r + c_2\, r^2 + \cdots .
\eea
This is (at best) only a qualitative argument, since we know that
there are logarithmic corrections to the Coulomb potential at
short-distances, and for this reason, 
$V_{\rm QCD}(r)$ cannot be expanded in Laurent series.
Nevertheless, empirically the above expansion 
is a good one, since many phenomenological
potentials have been successfully determined, by fitting them
to the experimental data of heavy quarkonium spectra,
assuming Coulomb+linear forms.
So, suppose that one may decompose $V_{\rm QCD}(r)$ as above
qualitatively.
Then, since the non-perturbative contribution $\delta E_{\rm US}(r)$
is expected to be ${\cal O}(r^2)$ (assuming $\mu_f \gg \Delta V, \LQ$),
the $c_2 \, r^2$ term (and beyond) would come from both 
$V_S(r)$ and $\delta E_{\rm US}(r)$, and their relative contributions
change as we vary the factorization scale $\mu_f$.
On the other hand, the Coulomb, constant, and linear terms, 
${c_{-1}}/{r} + c_0 + c_1\, r $,
should originate only from the perturbative prediction of $V_S(r)$,
that is, from the perturbative prediction of $V_{\rm QCD}(r)$.
(The constant term becomes predictable perturbatively only 
when the pole masses are added to $V_{\rm QCD}(r)$
and rewritten
in terms of a short-distance mass such as the 
$\overline{\rm MS}$-mass.)
In particular, they should be predictable independently of $\mu_f$.

\section{Perturbative QCD Potential at Large Orders}
\label{s3}
\clfn

In this section, we
present an analysis of the QCD potential 
at large orders of perturbative expansion.
We separate the perturbative prediction of 
the QCD potential at large orders
into a scale-independent
(prescription-independent) part
and scale-dependent (prescription-dependent) part, when
higher order terms are estimated via large-$\beta_0$ approximation
or via RG, and when the renormalization scale $\mu$
is varied around the minimal-sensitivity
scale.

\subsection{Strategy and general assumptions of the analysis}
\label{s3.1}

We consider the perturbative QCD potential up to ${\cal O}(\alpha_S^N)$:
\bea
V_N(r) \equiv [ V_{\rm QCD}(r) ]_N
= -4 \pi  C_F \,
\int \frac{d^3\vec{q}}{(2\pi)^3} \, \frac{e^{i \vec{q} \cdot \vec{r}}}{q^2}
\, [ {\alpha_V^{\rm PT}(q)} ]_N .
\label{defVN}
\eea
Here and hereafter, $[X]_N$ denotes the series expansion of $X$ in $\alpha_S(\mu)$
truncated at  ${\cal O}(\alpha_S(\mu)^N)$.
We examine $V_N(r)$ for $N \gg 1$.
For this analysis, we need (a) an
estimate for the all order terms of $V_N(r)$,
and (b) a scale-fixing prescription.

In the following subsections,
we estimate the higher-order terms of $V_N(r)$ 
using large-$\beta_0$ approximation (Sec.~\ref{s3.2}) and
using RG (Sec.~\ref{s3.3}).
There is a caveat:
The former estimate does not contain IR divergences at all, and in the
latter estimate, IR divergences appear only beyond NNLL; 
hence, in
most of our argument, we will discard IR divergences.
Since the true higher-order terms contain IR divergences beyond 
${\cal O}(\alpha_S^3)$,
we have to clarify what we mean by our estimates of higher-order terms.
Conceptually, we assume that we have
removed ambiguities related to IR divergences, while
keeping IR renormalons in the perturbative expansion of the potential.
This seems to be possible, since, up to our current best knowledge, 
IR divergences \cite{Appelquist:es} and IR renormalons \cite{Aglietti:1995tg} 
contained in the
perturbative QCD potential stem from quite
different physical origins.
As an explicit example to realize such a situation, 
we may assume that we analyze 
the singlet potential $V_S(r)$ instead of $V_{\rm QCD}(r)$,
after subtracting IR divergences (but not IR renormalons) via renormalization.
Alternatively, we may assume that we have regularized 
IR divergences by expanding $V_{\rm QCD}(r)$ in double series in
$\alpha_S$ and $\log \alpha_S$.

Let us explain our scale-fixing prescription (b).
Since within our estimates the perturbative series turns out to be
an asymptotic series, there exists a certain arbitrariness in making a
prediction from large-order analysis of the series.
We will give a prediction by choosing a
reasonable scale $\mu$ for each given $N$ and then taking 
the limit $N \to \infty$.
(Later we will justify our prescription by comparing 
the prediction with that in OPE.)
Perturbative QCD in itself does not provide any
scale-fixing procedure.
In practice, 
whenever a perturbative expansion up to some finite order is given,
one chooses a reasonable (range of) scale $\mu$,
as we have seen in Sec.~\ref{s2.2}.
We would like to fix the scale in a similar manner
in our large-order analysis.
According to the argument given in Secs.~\ref{s2.2} and \ref{s2.3},
if we choose a scale $\mu$ such that 
$\alpha_S(\mu) =  {6\pi}/({\beta_0N})$ is satisfied, around this scale,
$V_N(r)$ (after cancelling the leading-order renomalon) would become
least $\mu$-dependent and the perturbative series 
would become most convergent; cf.\ Fig.~\ref{Vn}.
In view of this property, we fix $\mu$ such that\footnote{
Here, we generalize the prescription of \cite{Sumino:2003yp} by
introducing an additional parameter $\xi$, cf.\ \cite{VanAcoleyen:2003gc}.
}
\bea
N  = \frac{6\pi}{\beta_0\alpha_S(\mu)} \, \xi
= N_1 \, \xi \, .
~~~~~~~
(\xi \sim 1)
\label{xi}
\eea
$\xi =1$ corresponds to an optimal choice;
by varying the parameter $\xi$, we may change the scale $\mu$ for
a given $N$.
Then we consider $V_N(r)$ for $N \gg 1$ while keeping
$\Lambda_{\overline{\rm MS}}$ finite.
(Here, we relate our scale-fixing prescription
to that of principle of minimal sensitivity \cite{Stevenson:1981vj} only weakly,
as argued above.
A close examination of the relation can be found 
in \cite{VanAcoleyen:2003gc}.)

An alternative way to regard this prescription is as follows.
Suppose we know the perturbative expansion of
$V_{\rm QCD}(r)$ up to all orders, according to a certain estimate.
When the expansion is asymptotic for any choice of $\alpha_S(\mu)$, 
we cannot sum all the terms.
Instead, following a standard prescription to deal with asymptotic
series,  we may truncate the series around the order where the term is
close to minimal.
This gives the truncated series
$V_N(r)$ with $N$ given by the relation eq.~(\ref{xi}).

The motivation for considering the large $N$ limit is that it corresponds
to the limits where the perturbative expansion becomes well-behaved 
(small expansion parameter) and where the estimate of $V^{(n)}_{\rm QCD}(r)$
by renormalon contribution becomes a better
approximation around $n \sim N$.
Note that large $N$ corresponds to small $\alpha_S(\mu)$ and
large $\mu$ due to the above relation.

Let us further
comment on some details concerning the relation (\ref{xi}).
(i) The relation (\ref{xi}) follows from the asymptotic form, 
eq.~(\ref{NLO-renormalon}),  
of the series independently of its
overall coefficient.
Although the overall coefficient is not known exactly,\footnote{
See \cite{Lee:1999ws} for a method for systematically estimating the overall coefficient.
}
other parts of eq.~(\ref{NLO-renormalon}) or (\ref{cren}) are
considered to be solid, based on consistency with RG equation.
Hence, the relation (\ref{xi}) is based on a solid 
part of the renormalon estimate.
(ii) The scale $\mu$ fixed by the relation (\ref{xi})
is independent of $r$.
Usually it is considered that
a natural choice of the scale is related to a physical scale,
typically $\mu \sim 1/r$, at low orders of perturbative expansion.
Moreover, the minimal-sensitivity scales corresponding to low orders
of perturbative expansion, as in the cases of
Sec.~\ref{s2.2},  are known to be strongly 
dependent on $r$ \cite{Sumino:2001eh,Recksiegel:2001xq}. 
This is, however, not expected to be the case at large orders.
It is because, in eq.~(\ref{int-renormalon}), contributions from $q<1/r$ are
dominant on the left-hand-side at low orders,
whereas at large orders, the term proportional to $-q^2r^2/6$
dominates on the right-hand-side of eq.~(\ref{int-renormalon}),
hence,  $r^2$ factors out as an overall 
coefficient; cf.~eq.~(\ref{NLO-renormalon}).
(iii) Based on the argument in Sec.~\ref{s2.3}, we may consider that an optimal 
choice of $\mu$ or $\xi$ corresponds to the
range $N_1 - \sqrt{N_1} \simlt N=N_1 \xi \simlt N_1+\sqrt{N_1}$
in the relation (\ref{xi}).
Then, $\xi \to 1$ as $N \to \infty$.

\subsection{\boldmath $V_N(r)$ at large orders: large-$\beta_0$ approximation}
\label{s3.2}
\clfn

The large--$\beta_0$ approximation \cite{Beneke:1994qe}
is an empirically successful method
for estimating higher-order corrections in perturbative QCD
calculations; 
see e.g.\ 
\cite{Beneke:1998ui,Melnikov:2000qh,Broadhurst:1994se,Lovett-Turner:1995ti}.
In general, 
the large-$\beta_0$ approximation of a 
physical quantity, at a given order of
perturbative expansion in $\alpha_S$, is defined in the following way.
We first compute the leading order contribution in an expansion in
$1/n_l$, which
comes from so-called bubble chain diagrams.
Then we transform this large $n_l$ result by a simplistic replacement
$n_l \to n_l - 33/2 = -(3/2)\beta_0$.

For the QCD potential, the large--$\beta_0$ approximation 
corresponds to setting 
$a_n=(5\beta_0/3)^n$ in eq.~(\ref{alfVRG})
and all $\beta_n=0$ except $\beta_0$ in eq.~(\ref{RGeq}).
Hence, it includes only the one-loop running of $\alpha_S(q)$.
In this subsection, with these estimates of the all-order terms of 
$\alpha_V^{\rm PT}(q)$, we examine 
$V_N(r)$, defined above, for $N \gg 1$.
The reasons for examining the large--$\beta_0$ approximation are
as follows.
First, because this approximation leads to
the renormalon dominance picture; in fact, the renormalon dominance picture 
has often been discussed in this approximation.
Secondly, the running of
the strong coupling constant makes the potential steeper at large distances
as compared to the Coulomb potential;
hence, we would like to see if the potential can be written in 
a ``Coulomb''+linear form when only
the one-loop running is incorporated as a simplest case.

We define 
$\widetilde{\Lambda} = e^{5/6} \,
\Lambda_{\overline{\rm MS}}^{\mbox{\scriptsize 1-loop}}$,
where 
\bea
\Lambda_{\overline{\rm MS}}^{\mbox{\scriptsize 1-loop}}
= \mu \exp \left[ - \frac{2\pi}{\beta_0 \alpha_S(\mu)} \right] .
\eea
In the following, we assume 
\bea
\widetilde{\Lambda}^{-1}\exp\Bigl(-\frac{N}{3\xi}\Bigr) \ll 
r \ll \widetilde{\Lambda}^{-1}\exp\Bigl(\frac{N}{3\xi}\Bigr),
\label{limits}
\eea
when we consider the double limits
$r \to 0$, $N \to \infty$ or
$r \to \infty$, $N \to \infty$.
Note that, as $N \to \infty$,
the lower bound ($\widetilde{\Lambda}^{-1}e^{-N/(3\xi)}$) and the upper bound
($\widetilde{\Lambda}^{-1}e^{N/(3\xi)}$) of $r$
go to 0 and $\infty$, respectively.

First we present the result
and discuss some properties when 
$\xi = 1$, which corresponds to an optimal choice of
scale $\mu$.
(Derivation will be given in Sec.~\ref{s3.4}.)
\vspace{3mm}\\
{\it Result for $\xi = 1$}\vspace{2mm}

$V_N(r)$ for $\xi = 1$ and
$N \gg 1$ within the large--$\beta_0$ approximation 
can be decomposed into four parts corresponding to
\{$r^{-1}$, $r^0$, $r^1$, $r^2$\} terms
(with logarithmic corrections in the $r^{-1}$ and $r^2$ terms):
\bea
&&
V_N^{(\beta_0)}(r)\Bigr|_{\xi = 1} = 
\frac{4C_F}{\beta_0} \, \widetilde{\Lambda}
\,\, v(\widetilde{\Lambda}r,N) ,
\label{VNbeta0}
\rule[-6mm]{0mm}{6mm}
\\ &&
v(\rho,N)=v_C(\rho)+B(N)+C\rho+D(\rho,N)
+(\mbox{terms that vanish as $N\to\infty$}).
\label{vrhoN}
\eea
(i)``Coulomb'' part:
\bea
v_C(\rho)=-\frac{\pi}{\rho} + \frac{1}{\rho}
\int_0^\infty dx \, e^{-x} \, \arctan 
\biggl[ \frac{\pi/2}{\log(\rho/x)} \biggr]
,
\eea
where $\arctan x \in [0,\pi)$.
The asymptotic forms are given by
\bea
\left\{
\begin{array}{ll}
\displaystyle
\rule[-7mm]{0mm}{6mm}
v_C(\rho) \sim - \, \frac{\pi}{2\rho\log(1/\rho)}, ~~~& \rho \to 0 \\
\displaystyle
v_C(\rho) \sim - \, \frac{\pi}{\rho}, & \rho \to \infty \\
\end{array}
\right.
\label{asym-vC}
\eea
and both asymptotic forms are smoothly interpolated in the
intermediate region.
The short-distance behavior is consistent with the
one-loop RG equation for the QCD potential.

\noindent
(ii) constant part\footnote{
The ${\cal O}(1/N)$ and ${\cal O}(1/N^2)$ terms in eq.~(\ref{BN})
are irrelevant for $N \to \infty$.
We keep these terms in $B(N)$ for convenience in examining $V_N^{(\beta_0)}(r)$
at finite $N$; see Fig.~\ref{VN-CplusL} below.
}:
\bea
B(N)= - \int_0^\infty dt \, \frac{e^{-t}}{t} \,
\biggl[ \Bigl( 1 + \frac{3}{N}\, t\, \Bigr)^N - 1 \biggr]
- \log 2 - \frac{9}{8N} + \frac{99}{64N^2} .
\label{BN}
\eea
The first term (integral) diverges rapidly for $N \to \infty$ as 
${\displaystyle - \, \frac{3}{2}\sqrt{\frac{2\pi}{N}}
\biggl( \frac{3}{e^{2/3}} \biggr)^N \,
\bigl[ \, 1 + {\cal O}(1/N) \bigr]}$.\footnote{
The integral can be expressed in terms of confluent hypergeometric
function.
}

\noindent
(iii) linear part:
\bea
C = \frac{\pi}{2} .
\eea

\noindent
(iv) quadratic part:
\bea
&&
D(\rho,N)=\rho^2 \, \biggl[ \frac{1}{12}  \log N + d(\rho) \biggr],
\label{DrhoN}
\\ &&
d(\rho)= -
\int_0^\infty dx \, \,
\frac{e^{-x}-\Bigl[1-x+\frac{1}{2}x^2-\frac{1}{6}x^3 \, \theta(1-x) \Bigr]}
{x^4} \,\,
\frac{\log (\rho/x)}{\log^2 (\rho/x) + \pi^2/4}
\nonumber \\ && ~~~~~ ~~~
- \frac{1}{12}\, 
\biggl[ \log \Bigl( \log^2 \! \rho + \frac{\pi^2}{4} \Bigr) + 
\log \frac{9}{2}
+ \gamma_E \biggr],
\label{drho}
\eea
where $\theta(x)$ is the unit step function and $\gamma_E = 0.5772...$
is the Euler constant.
The asymptotic forms of $d(\rho)$ are given by
\bea
\left\{
\begin{array}{ll}
\displaystyle
\rule[-7mm]{0mm}{6mm}
d(\rho) \sim - \frac{1}{12}\, \Bigl[ 2 \log \log (1/\rho) + 
\log \frac{9}{2} + \gamma_E \Bigr] ,~~~
& \rho \to 0 \\
\displaystyle
d(\rho) \sim - \frac{1}{12}\, \Bigl[ 2 \log \log {\rho} + 
\log \frac{9}{2} + \gamma_E \Bigr] , & \rho \to \infty \\
\end{array}
\right.
\eea
and in the intermediate region both asymptotic forms are smoothly interpolated.

\begin{figure}
\begin{center}
\psfrag{linear}{\hspace{-17mm}$\xi=1~~~~~~C\,\rho$}
\psfrag{Bottom title}{$\rho = \widetilde{\Lambda}\, r$}
\psfrag{Coulomb}{$v_C(\rho)$} 
\psfrag{1/Lambda}{$r=1/\widetilde{\Lambda}$}
\psfrag{D10}{$D(\rho,10)$}
\psfrag{D30}{$D(\rho,30)$}
\psfrag{D100}{$D(\rho,100)$}
\includegraphics[width=10cm]{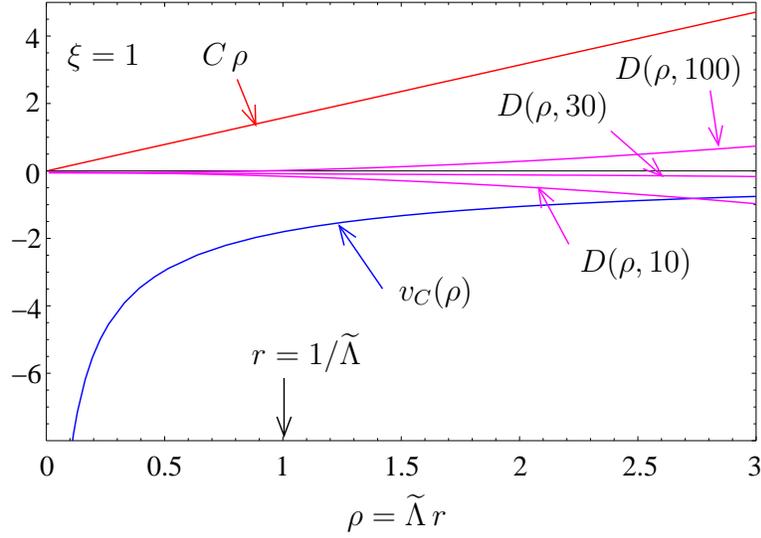}
\end{center}
\vspace*{-.5cm}
\caption{\small
$v_C(\rho)$, $C\rho$ and $D(\rho,N)$ ($N=10,30,100$) vs.\ $\rho$ for $\xi=1$.
\label{large-beta0-decomp}}
\end{figure}
The ``Coulomb'' [$v_C(\rho)$], linear [$C\rho$] and quadratic 
[$D(\rho,N)$] parts are shown in Fig.~\ref{large-beta0-decomp}.
The truncated potential $v(\rho,N)$ is compared with the
``Coulomb''+linear potential $v_C(\rho)+C\rho$ after
the constant $B(N)$ is subtracted, for
$N=10$, 30, 100, in Fig.~\ref{VN-CplusL}.\footnote{
One can find formulas convenient for computing $V_N(r)$ 
for a finite but large $N$ in App.~\ref{appA}.
}
In order to show how quickly $v(\rho,N)$ approaches 
$v_C(\rho)+B(N)+C\rho+D(\rho,N)$ as $N$ increases, we show
their differences for several values of $N$ in Fig.~\ref{conv-largeN}.
One sees that convergence is quite good at $\rho \simlt 1$ 
($r \simlt \widetilde{\Lambda}^{-1}$) for 
$N \ge 10$.
(For the purpose of separating different lines visibly, we
plot potentials up to fairly large distances in this section.
Nevertheless, we stress that, in most cases, our interests are 
in the region
$r \simlt \widetilde{\Lambda}^{-1}$.\footnote{
Roughly speaking, one may regard
$\widetilde{\Lambda}^{-1} \sim 1$~fm.
})
\medbreak
\begin{figure}
\begin{center}
\psfrag{rho=Lambda*r}{\hspace{10mm}$\rho = \widetilde{\Lambda}\, r$} 
\psfrag{vc(rho)+C*rho}{$v_C(\rho)+C\rho$} 
\psfrag{lefttitle}
{\hspace{1mm}$v(\rho,N)\! - \! B(N)$ and $v_C(\rho) \! + \! C\rho$} 
\psfrag{N=10}{$N=10$}
\psfrag{N=30}{\hspace{-2mm}$\! \! N=30$}
\psfrag{N=100}{\hspace{-35mm}$\xi=1~~~~~~~~~~~~~~~~N=100$}
\psfrag{r=1/Lambda}{$r=1/\widetilde{\Lambda}$}
\psfrag{r=1/LambdaMS}
{$r=1/\Lambda_{\overline{\rm MS}}^{\mbox{\scriptsize 1-loop}}$}
\includegraphics[width=10cm]{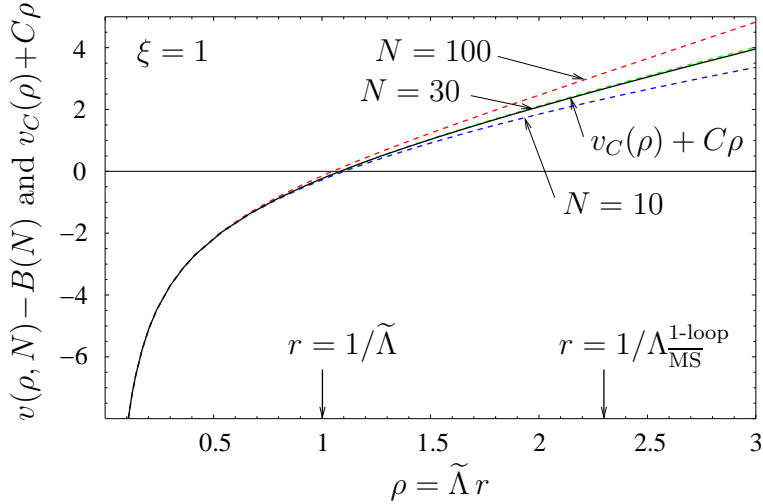}
\end{center}
\vspace*{-.5cm}
\caption{\small
Truncated potential after the constant term is subtracted,
$v(\rho,N)-B(N)$, (dashed) vs.\ $\rho$ for 
$N=10$, 30, 100 and $\xi=1$.
``Coulomb''+linear potential, 
$v_C(\rho)+C\rho$, (solid black)
is also plotted, which is hardly distinguishable from the 
$N=30$ curve.
\label{VN-CplusL}}
\end{figure}
\begin{figure}
\begin{center}
\psfrag{xi=1}{$\xi=1$}
\psfrag{rho=Lambda*r}{\hspace{0mm}$\rho = \widetilde{\Lambda}\, r$} 
\psfrag{vc(rho)+C*rho}{$v_C(\rho)+C\rho$} 
\psfrag{lefttitle}{}
%{\hspace{-32mm}$v(\rho,N)- [ v_C(\rho)+ B(N)+ C\rho + D(\rho,N)]$} 
\psfrag{N=3}{\hspace{-3mm}$N=3$}
\psfrag{N=10}{\hspace{-6mm}$N=10$}
\psfrag{N=30}{\hspace{-2mm}$N=30$}
\psfrag{N=100}{\hspace{0mm}$N=100$}
\psfrag{N=300}{\hspace{0mm}$N=300$}
\includegraphics[width=10cm]{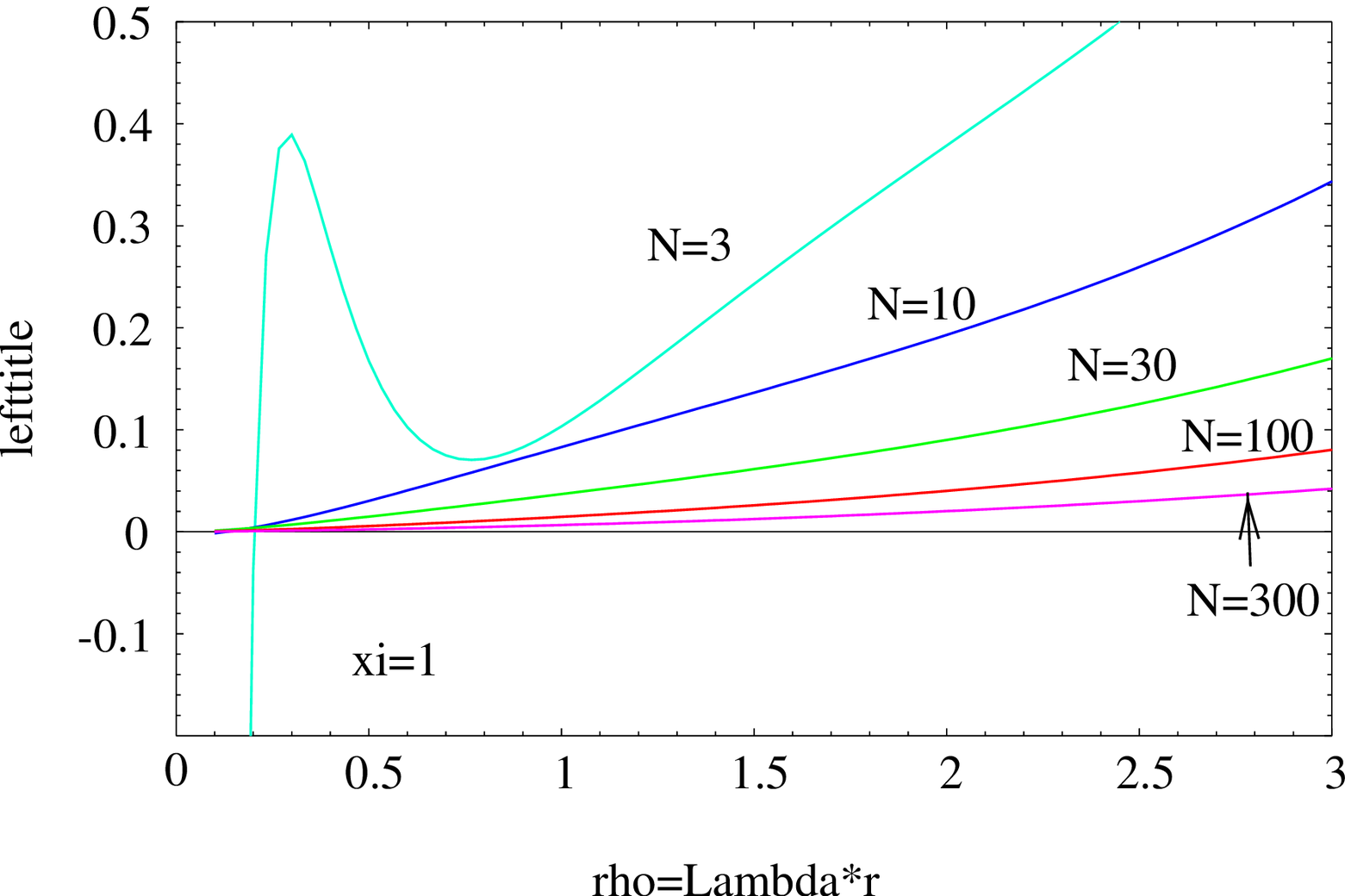}
\end{center}
\vspace*{-.5cm}
\caption{\small
Plots for
$v(\rho,N)- [ v_C(\rho)+ B(N)+ C\rho + D(\rho,N)]$ vs.\ $\rho$, 
showing 
convergence as $N$ increases ($\xi=1$).
Note that vertical scale is magnified widely as compared
to Figs.~\ref{large-beta0-decomp}, \ref{VN-CplusL}
for display purposes.
\label{conv-largeN}}
\end{figure}

Although the constant part of $V_N^{(\beta_0)}(r)\bigr|_{\xi=1}$ 
diverges rapidly as $N \to \infty$,
the divergence can be absorbed into the quark masses
in the computation of the total energy $E_{\rm tot}(r)$
(or the heavy quarkonium spectrum).
Therefore, in our analysis, 
we will not be concerned with the constant part of the potential
but only with the $r$-dependent terms.

The quadratic part of $V_N^{(\beta_0)}(r)\bigr|_{\xi=1}$  diverges slowly as
$\widetilde{\Lambda}^3 r^2 \log N  \sim 
\widetilde{\Lambda}^3 r^2 \log \log (\mu/\widetilde{\Lambda})$.
The dependence of $V_N^{(\beta_0)}(r)$ on $N$ is mild
(after the constant part is subtracted); 
for instance, as shown in Fig.~\ref{VN-CplusL},
the variation of $v(\rho,N)-B(N)$ is small
in the range $r \simlt \widetilde{\Lambda}^{-1}$ as we vary $N$ from
10 to 100; 
it corresponds to a variation of 
$\mu/\Lambda_{\overline{\rm MS}}^{\mbox{\scriptsize 1-loop}}$
from 30 to $3\times 10^{14}$.

The ``Coulomb'' part and the linear part are finite
as $N \to \infty$.
In Fig.~\ref{VN-CplusL}, we see that
$V_N^{(\beta_0)}(r)$ is approximated fairly well by the sum of the ``Coulomb''
part and the linear part (up to an $r$-independent constant) in the region 
$r \simlt \widetilde{\Lambda}^{-1}$ when we vary $N$ between
10 and 100.
Moreover, as long as $\frac{1}{12}\log N \simlt {\cal O}(1)$, the 
difference between $V_N^{(\beta_0)}(r)\bigr|_{\xi=1}$ and the ``Coulomb''+linear
potential remains at or below ${\cal O}(\widetilde{\Lambda}^3 r^2)$
in the {\it entire range} of $r$.
Note that $v(\rho,N)$ in this figure have
the form of $1/r\times (\mbox{Polynomial of }\log r)$, 
and a priori it is not obvious at all
that they approximate a
``Coulomb''+linear potential.
\vspace{3mm}\\
{\it Results for $\xi \ne 1$}\vspace{2mm}

\begin{figure}
\begin{center}
\psfrag{xi=0.9}{\hspace{1mm}$\xi=0.9$}
\psfrag{]}{\hspace{2mm}$\left.\rule[0mm]{0mm}{8mm} \right\}$}
\psfrag{N=30, xi=1.1}{\hspace{-10mm}$N=30,~\xi=1.1$}
\psfrag{rho=Lambda*r}{\hspace{5mm}$\rho = \widetilde{\Lambda}\, r$} 
\psfrag{vc+C*rho}{\hspace{-10mm}$v_C(\rho)+C\rho$} 
\psfrag{lefttitle}
{\hspace{-5mm}$v(\rho,N)$ and $v_C(\rho)+C\,\rho$} 
\psfrag{N=infty}{\hspace{0mm}$N=\infty$}
\psfrag{N=10}{\hspace{0mm}$N=10$}
\psfrag{N=100}{\hspace{0mm}$N=100$}
\includegraphics[width=10cm]{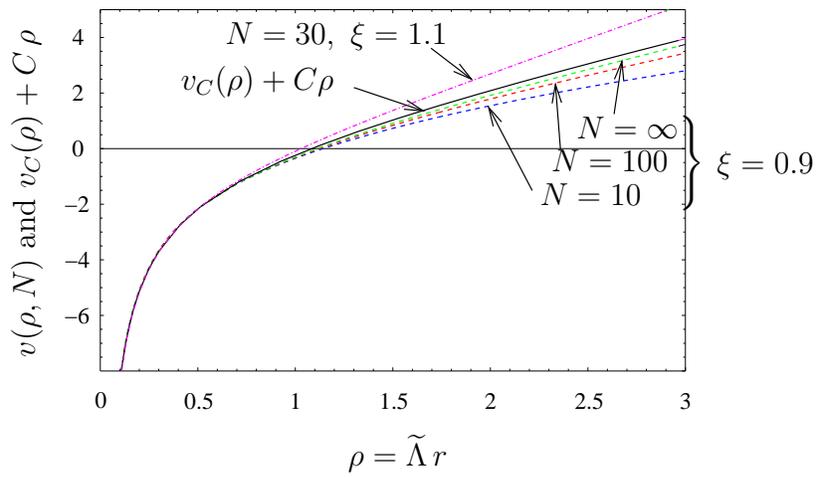}
\end{center}
\vspace*{-.5cm}
\caption{\small
$v(\rho,N)$ for different values of $\xi$ and $N$.
(Dashed lines for $\xi=0.9$ and dot-dahed line for
$\xi=1.1$.)
For comparison, the ``Coulomb''+linear potential
$v_C(\rho)+C\rho$ is also shown (solid line).
Constants have been added to 
$v(\rho,N)$ to make them coincide with $v_C(\rho)+C\rho$
at $\rho=0.5$.
\label{VN-vary-xi-N}}
\end{figure}

We vary $\xi$ in the scale-fixing prescription eq.~(\ref{xi})
and decompose $V_N^{(\beta_0)}(r)$ as in eqs.~(\ref{VNbeta0})
and (\ref{vrhoN}).
As a salient feature,  we obtain
the same ``Coulomb''+linear potential, $v_C(\rho)+C\rho$,
as in the $\xi=1$ case.
On the other hand, the constant $B(N)$ and 
$D(\rho,N)$ change.
The latter no longer takes a quadratic form.
Let us list how $D(\rho,N)$ change with $\xi$.
(See Fig.~\ref{VN-vary-xi-N}.)
\begin{itemize}
\item
$2/3 < \xi < 1$,  \vspace{3mm}\\
$D(\rho,N)$ is finite as $N\to\infty$:
\bea
&&
D(\rho,\infty) = - \rho^{3\xi -1} 
\int_0^\infty dx \, \,
\frac{e^{-x}-\Bigl(1-x+\frac{1}{2}x^2 \Bigr)}
{x^{1+3\xi}} \,\,
{\rm Im}\left[ \frac{e^{-3\pi i \xi/2}}{\log (\rho/x) - i \pi/2} \right] .
\label{Dxi}
\eea
Its asymptotic forms are given by
\bea
D(\rho,\infty) \sim
\rho^{3\xi -1}\,\frac{1}{\log \rho} \times \Gamma(-3\xi) 
\sin \Bigl( \frac{3}{2}\pi  \xi \Bigr) ,
~~~~~
\mbox{$\rho \to 0$~~or~~$\rho \to \infty$}.
\label{Dxi-asym}
\eea
The asymptotic forms at $\rho \to 0$ and $\rho \to \infty$ have
opposite signs.
In the intermediate region $D(\rho,\infty)$ changes sign once.
$D(\rho,\infty)$ for $\xi=0.85$, 0.9, 0.95 are plotted in 
Fig.~\ref{D-vary-xi}. 
\begin{figure}
\begin{center}
\psfrag{xi=0.85}{\hspace{1mm}$\xi=0.85$}
\psfrag{xi=0.9}{\hspace{-2mm}$\xi=0.9$}
\psfrag{xi=0.95}{\hspace{1mm}$\xi=0.95$}
\psfrag{rho}{\hspace{2mm}$\rho$} 
\psfrag{lefttitle}
{\hspace{0mm}$D(\rho,\infty)$} 
\includegraphics[width=10cm]{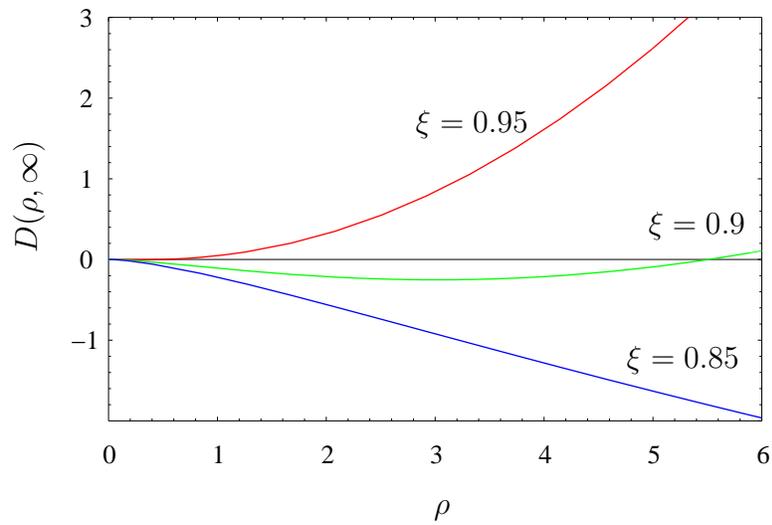}
\end{center}
\vspace*{-.5cm}
\caption{\small
$D(\rho,\infty)$ for different values of $\xi(<1)$.
\label{D-vary-xi}}
\end{figure}
\item
$\xi > 1$~ %($3\xi \ne \mbox{integer}$),
\vspace{3mm}\\
Even powers of $\rho$, corresponding to IR renormalons, become more
divergent as we increase $\xi$:
\bea
D(\rho,N)=d_2(N)\, \rho^2 + d_4(N)\, \rho^4 + \cdots + d_{\nu}(N)\, \rho^\nu + 
\mbox{(finite term as $N\to\infty$)} ,
\eea
where $\nu$ is the largest even integer satisfying $\nu \leq 3\xi -1$.
$d_i (N)$ diverges at least logarithmically 
(typically exponentially)
as $N \to \infty$.
It diverges more rapidly for larger $\xi$ and smaller $i$.
The asymptotic form of the finite ($N$-independent)
term as $\rho \to 0$ or $\rho \to \infty$
is $\rho^{3\xi-1} \times (\mbox{log correction})$.
\item
$\xi < 2/3$, \vspace{3mm}\\
$D(\rho,N)$ becomes more dominant than
the linear potential $C \rho$ at short-distances.
We do not consider this possibility henceforth.
($\xi =2/3$ is marginal; the asymptotic form eq.~(\ref{Dxi-asym})
is valid at $\rho \to 0$ but not at $\rho \to \infty$.)
\end{itemize}
Dependence of $B(N)$ on $\xi$ is similar:
It diverges more rapidly as $N \to \infty$ for larger $\xi$,
while it becomes finite when $\xi < 1/3$.

Thus, $B(N)$ and 
$D(\rho,N)$ are dependent on $\xi$, i.e.\ on the choice 
of scale via eq.~(\ref{xi});
they are also divergent as $N \to \infty$ for a sufficiently large $\xi$.
Namely, $B(N)$ and 
$D(\rho,N)$ are dependent on the prescription we adopted to
define our prediction.
It is natural to consider the prescription dependence as indicating 
uncertainties of our prediction.
In fact,  $B(N)$ and 
$D(\rho,N)$ are associated, respectively, with the
${\cal O}(\LQ)$ IR renormalon and ${\cal O}(\LQ^3r^2)$ IR renormalon
(and beyond) in
$V_{\rm QCD}(r)$.
We have already seen that 
these renormalons
induce uncertainties.
On the other hand, the ``Coulomb''+linear part
[$v_C(\rho)+C\rho$] are independent of $\xi$ and $N$.
Hence, $v_C(\rho)+C\rho$ 
can be regarded as a genuine part of the prediction.
In this regard, we remind the reader that there are no IR renormalons 
associated with the
$1/r$ and $r$ terms in the QCD potential \cite{Aglietti:1995tg}.

One may associate the 
${\cal O}(\LQ^3 r^2)$ renormalon with
$D(\rho,N)$ through following observations.
(1) When $\xi = 1$, the quadratic part of $V_N^{(\beta_0)}(r)$ 
diverges as $\widetilde{\Lambda}^3 r^2 \log N$.
If the series expansion of
$m_{\rm pole}(m_{\overline{\rm MS}},\alpha_S)$ or $V_{\rm QCD}(r)$
is truncated at the order corresponding to the
minimal term of the LO renormalon contribution,
i.e.\ at order $N_0 = 2\pi/(\beta_0 \alpha_S)$, 
$[m_{\rm pole}]_{N_0}$ or $[V_{\rm QCD}(r)]_{N_0}$ diverges as 
$\widetilde{\Lambda} \log N_0$
within the large-$\beta_0$ approximation.
We may compare $\widetilde{\Lambda} \log N_0$ with the usual interpretation 
that $m_{\rm pole}$ and $V_{\rm QCD}(r)$
contain ${\cal O}(\LQ)$ perturbative uncertainties
due to the LO renormalons.
(2) An argument similar to (1) applies for $\xi \ne 1$.
(3) As we will see in the next subsection, 
even if we estimate higher-order terms
using RG equation and incorporate effects of the 
two-loop running and beyond,
$D(\rho,N)$ has a similar behavior to that in the
large-$\beta_0$ approximation.

Let us further discuss questions concerning the strategy and results of
the analysis given above.

Naively, one would expect that scale-dependence decreases
as more terms of the perturbative expansion are summed, as long as
the series is converging.
Is this realized by our results?
In fixed-order perturbation theory, it is a common practice to
vary the scale $\mu$, say, by factor two, and examine the stability of
the prediction.
It may be more natural to vary $\mu$
such that $\xi$ changes by order $\sqrt{1/N}$, 
as we argued at the end of Sec.~\ref{s3.1}.
In either case, if we fix $N$ in eq.~(\ref{xi}),
the variation in $\xi$ vanishes in the large $N$ limit.
A closer examination shows that
the variation of $v_C(\rho,N)-B(N)$, corresponding to these changes of $\mu$,
also vanishes in the large $N$ limit,
as long as $\xi$ is close to 1.
In this sense, our prediction becomes stable against scale variation 
at large orders.

There exists an argument that the linear potential
cannot emerge in perturbative QCD:
From dimensional analysis, the coefficient of a linear 
potential should be non-analytic in $\alpha_S$, i.e.\
of order 
$(\Lambda_{\overline{\rm MS}}^{\mbox{\scriptsize 1-loop}})^2
 = \mu^2 \exp[-4\pi/(\beta_0\alpha_S)]$;
therefore, it should vanish at any order of perturbative expansion.
Within our large-order analysis, this argument is circumvented
as follows.
$r \,V_N(r)$ includes terms of the form
$T_n=\{ \frac{\beta_0\alpha_S(\mu)}{2\pi} \log (\mu r) \}^n$
for $0\le n\le N$.
If we substitute the relation (\ref{xi}) and take the limit 
$N\to \infty$ while fixing $n/N$ finite,
it is easy to see that 
$T_n \to(\Lambda_{\overline{\rm MS}}^{\mbox{\scriptsize 1-loop}}
 r)^{3\xi n/N}$.
Thus, 
perturbative terms converge to
$(\Lambda_{\overline{\rm MS}}^{\mbox{\scriptsize 1-loop}}
 r)^{P}$
with positive powers $0<P<3\xi$.
In fact, the power $P$ has a continuous distribution.
Our result shows that the continuous
distribution can be decomposed into
a sum of $\{r^{0},r^1,r^2,r^{3\xi}\}$ terms,
up to logarithmic corrections (for $2/3\leq\xi \leq 1$).
Non-analyticity in $\alpha_S$ enters through the relation (\ref{xi}).

Thus, the characteristic feature of our large-order analysis is the 
prescription eq.~(\ref{xi}).
We may consider that an additional input has been incorporated
through the relation (\ref{xi})
beyond a simple large-order analysis within perturbative QCD.
Here, we emphasize that 
the number of parameters has not decreased
from that of the original perturbative expansion
($\alpha_S$, $\mu$, $r$ and $N$)
apart from $N$.
(We fix $\Lambda_{\overline{\rm MS}}$, $r$ and
$\xi$ finite when sending 
the truncation order $N \to \infty$.)
The ``Coulomb''+linear part,
$v_C(\rho)+C\rho$, emerges independently of $\xi$ and $N$
in this limit.
In this sense, we consider $v_C(\rho)+C\rho$
a genuine prediction of perturbative QCD
at large orders, within our estimate of the higher-order terms.

\subsection{\boldmath $V_N(r)$ at large orders: RG estimates}
\label{s3.3}
\clfn

In this subsection we examine $V_N(r)$ for large $N$
using RG estimates of
the all-order terms of $V_{\rm QCD}(r)$.
We examine three cases, corresponding
to the estimates of 
$\alpha_V^{\rm PT}(q)$ up to LL, NLL and NNLL 
in eqs.~(\ref{alfVRG}) and (\ref{RGeq})
[note that $a_n = P_n(0)$ for $n \leq 2$]:
\begin{itemize}
\item[(a)] 
[LL]
$\beta_0$, $a_0$: exact values, $\beta_n = P_n(0) =0$ ($n \geq 1$);
\item[(b)] 
[NLL]
$\beta_0$, $\beta_1$, $a_0$, $a_1$: exact values, 
$\beta_n = P_n(0) =0$ ($n \geq 2$);
\item[(c)] 
[NNLL]
$\beta_0$, $\beta_1$, $\beta_2$, $a_0$, $a_1$, $a_2$: 
exact values, 
$\beta_n = P_n(0) =0$ ($n \geq 3$).
\end{itemize}
Namely, cases (a),(b),(c), respectively, correspond to taking the sum up to
$n=0$,1,2 in eq.~(\ref{alfVRG}) and
reexpanding in $\alpha_S(\mu)$.
From naive power counting of logarithms, one should also include 
US logarithms at NNLL.
We examine them separately:
\begin{itemize}
\item[(c$'$)] 
[NNLL$'$]
Resummation of US logs is included via
eq.~(\ref{NNLLUSlog}), in addition to (c).
\end{itemize}
We assume $\beta_0,\, \beta_1, \, \beta_2, a_0, a_1, a_2 
\,({\rm exact}) > 0$.\footnote{
This is the case when the number of active
quark flavors is less than 6 and all the quarks are massless.
}
In the standard 1-, 2-, and 3-loop RG improvement of 
the QCD potential (in $\overline{\rm MS}$ scheme), 
the same all-order terms as above are resummed; 
the difference of our treatment is that
the perturbative series are truncated at ${\cal O}(\alpha_S(\mu)^N)$.
We note that the estimate of higher-order behavior based on
renormalon dominance hypothesis, as given in Sec.~\ref{s2.3}, 
is consistent with
the above estimates, or more generally, with
the RG analysis \cite{Beneke:1998ui}.
All the results for case (a) can be obtained 
from the results of the large--$\beta_0$ approximation given
in the previous subsection,
if we replace $\widetilde{\Lambda}$ by
$\Lambda_{\overline{\rm MS}}^{\mbox{\scriptsize 1-loop}}$.

\begin{figure}
\begin{center}
\begin{tabular}{cc}
\psfrag{C1}{$C_1$} 
\psfrag{q}{$q$} 
\psfrag{q*}{$q_*$} 
\psfrag{0}{\hspace{-1mm}\raise-1mm\hbox{$0$}}
\psfrag{infty}{$\infty$}
\includegraphics[width=8cm]{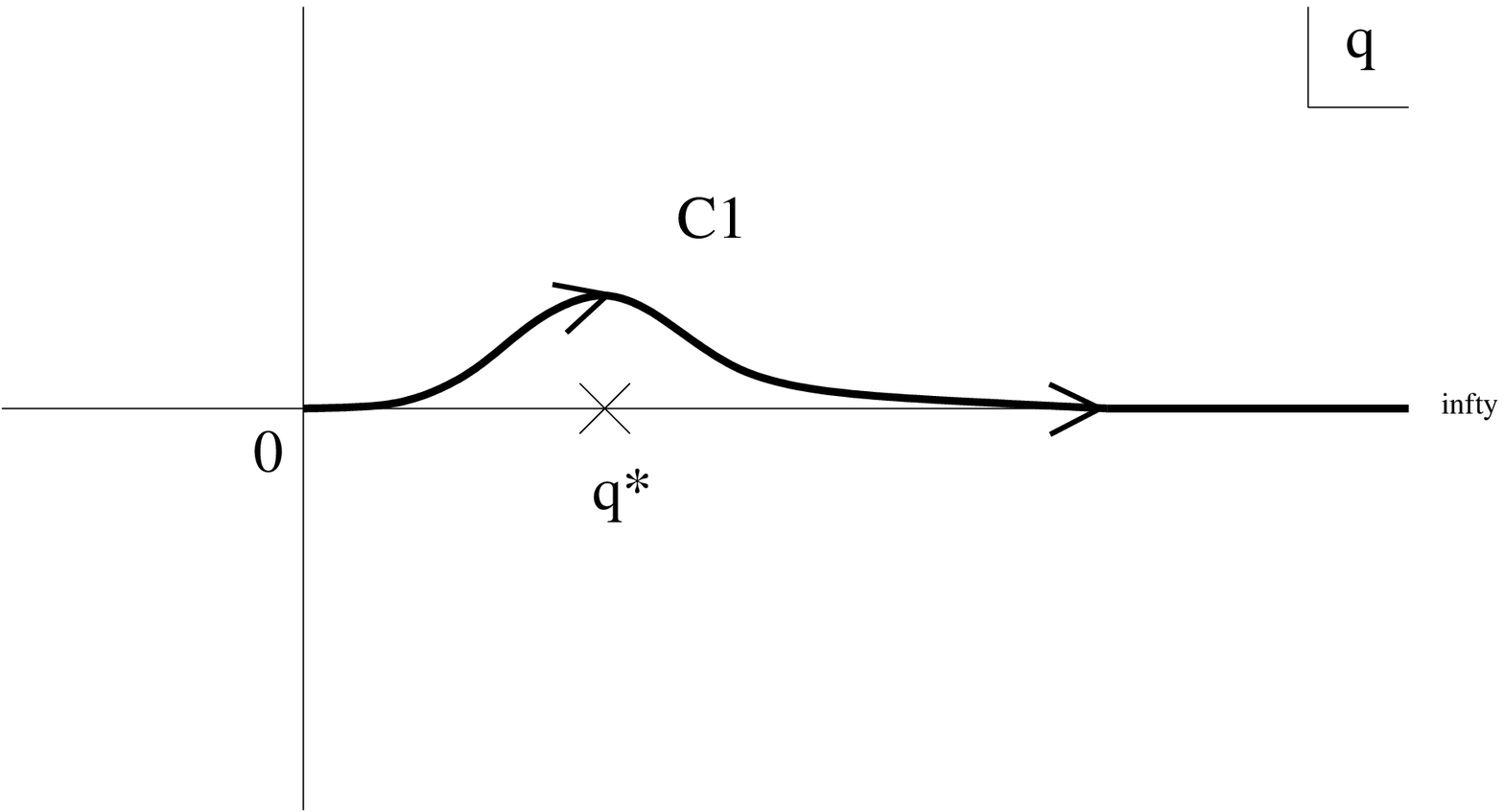} 
& \hspace{20mm}
\psfrag{C2}{$C_2$} 
\psfrag{q}{$q$} 
\psfrag{q*}{$q_*$} 
\psfrag{0}{\hspace{-1mm}\raise-1mm\hbox{$0$}}
\includegraphics[width=6cm]{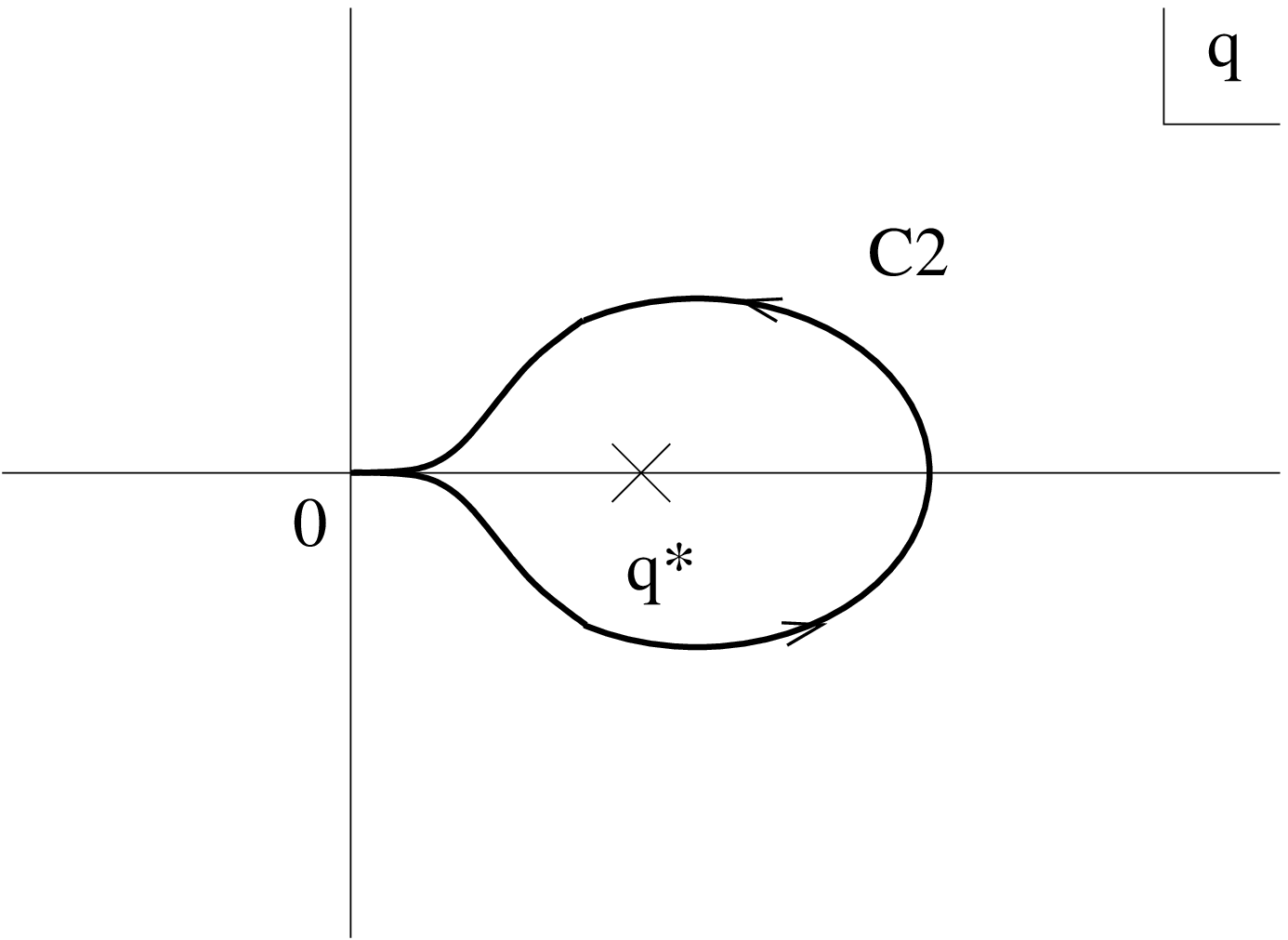}\\
(i) & \hspace{22mm}(ii)
\end{tabular}
\end{center}
\vspace*{-.5cm}
\caption{\small Integral contours $C_1$ and $C_2$ on the complex $q$-plane.
$q_*$ denotes the Landau singularity of $\alpha_S(q)$.
For 1-loop running, $q_*$ is a pole; for 2- and 3-loop running,
$q_*$ is a branch point.
In the latter case, branch cut is on the real axis starting from $q_*$
to $-\infty$.
\label{path}}
\end{figure}

Below we summarize our results. 
(See Sec.~\ref{s3.4} for derivation.)
Similarly to the previous subsection, we
can decompose $V_N(r)$ into four parts:
\bea
V_N(r) = V_C(r) + {\cal B}(N,\xi) + {\cal C} \, r + {\cal D}(r,N,\xi)
+ (\mbox{terms that vanish as $N\to\infty$}) ,
\label{decgen}
\eea
%The ``Coulomb'' and linear parts have finite limits as
%$N \to \infty$, whereas the constant and quadratic parts are divergent:
where 
\bea
&&
V_C(r) = - \frac{4\pi C_F}{\beta_0 r} 
- \frac{2C_F}{\pi} \, {\rm Im}
\int_{C_1}\! dq \, \frac{e^{iqr}}{qr} \, \alpha_V^{\rm PT}(q)
,
\label{genform1}
\\ &&
{\cal B}(N,\xi) = \lim_{r\to 0} \, \frac{2C_F}{\pi} \,{\rm Re}
\int_{C_1} \! dq \, e^{iqr} \, 
\Bigl\{ \alpha_V^{\rm PT}(q) - [ \alpha_V^{\rm PT}(q) ]_N  \Bigr\},
\\ &&
{\cal C} = \frac{C_F}{2\pi i} \int_{C_2}\! dq \, q \, \alpha_V^{\rm PT}(q) ,
\label{genform3}
\\ &&
{\cal D}(r,N,\xi) = V_N(r) - [V_C(r) + {\cal B}(N,\xi) + {\cal C} \, r] .
\label{genform4}
\eea
The integral contours $C_1$ and $C_2$
on the complex $q$-plane are displayed 
in Figs.~\ref{path}(i),(ii).
From the above equations, one can see that
the ``Coulomb'' and linear parts, $V_C(r)$ and ${\cal C}\, r$,
are independent of $\xi$ and $N$,
since $\alpha_V^{\rm PT}(q)$ is independent of $\xi$ and $N$.

The asymptotic behaviors of $V_C(r)$ for $r \to 0$ are same as those of
$V_{\rm QCD}(r)$ in the respective cases, as determined 
by RG equations;
the asymptotic behaviors of $V_C(r)$ for $r \to \infty$ are given
by the first term of eq.~(\ref{genform1}) in all the cases.
Namely,
\bea
&&
V_C(r)  \sim 
-\frac{2\pi C_F}{\beta_0 }\,
\frac{1}{r |\log (\Lambda_{\rm \overline{MS}}\,r)|}
\left[
1-\frac{\delta}{2} \, \frac{\log | \log (\Lambda_{\rm \overline{MS}}\,r) |}
{|\log (\Lambda_{\rm \overline{MS}}\,r)|}
\right],
~~~~~~~~~~
r \to 0 ,
\label{asym-VC-1}
\\
&&
V_C(r)  \sim  - \frac{4\pi C_F}{\beta_0 r} ,
~~~~~~~~~~
~~~~~~~~~~
~~~~~~~~~~
~~~~~~~~~~
~~~~~~~~~~
~~~~~~~~~~~
r \to \infty ,
\label{asym-VC-2}
\eea
where $\delta=0$ in case (a).
In the intermediate region both asymptotic forms are smoothly interpolated.

Evaluating the integral eq.~(\ref{genform3}),
the coefficient of the linear potential
can be expressed analytically in cases (a)--(c):
\bea
&&
{\cal C}^{\rm (a)}= \frac{2\pi C_F}{\beta_0} \,
\Bigl( \Lambda_{\overline{\rm MS}}^{\mbox{\scriptsize 1-loop}} \Bigr)^2 ,
\\ &&
{\cal C}^{\rm (b)}= \frac{2\pi C_F}{\beta_0} \,
\Bigl( \Lambda_{\overline{\rm MS}}^{\mbox{\scriptsize 2-loop}} \Bigr)^2
\, \frac{e^{-\delta}}{\Gamma (1+\delta)} \,
\biggl[ 1 + \frac{a_1}{\beta_0} \, \delta^{-1-\delta} \, e^\delta \,
\gamma (1+\delta,\delta) \biggr] ,
\eea
where $\gamma (x,\tau) \equiv \int_0^\tau dt \, t^{x-1} \, e^{-t}$
represents the incomplete gamma function;
see e.g.~\cite{Chetyrkin:1997sg} for definitions of
$\Lambda_{\overline{\rm MS}}^{\mbox{\scriptsize $n$-loop}}$.
In case (c), the expression for ${\cal C}$ is lengthy and
is given in App.~\ref{appC}.
Numerical values of ${\cal C}/\Lambda_{\overline{\rm MS}}^2$
for various $n_l$ are shown in Tab.~\ref{tab-str-tension}.
\begin{table}
\begin{center}
\begin{tabular}{l|cccccc}
\hline
$n_l$ & 0 & 1 & 2 & 3 & 4 & 5 \\
\hline
${\cal C}^{\rm (a)}/
\Bigl( \Lambda_{\overline{\rm MS}}^{\mbox{\scriptsize 1-loop}} \Bigr)^2$
& 0.762 & 0.811 & 0.867 & 0.931 & 1.005 & 1.093 \\
\hline
${\cal C}^{\rm (b)}/
\Bigl( \Lambda_{\overline{\rm MS}}^{\mbox{\scriptsize 2-loop}} \Bigr)^2$
& 0.591 & 0.622 & 0.664 & 0.722 & 0.807 & 0.935 \\
\hline
${\cal C}^{\rm (c)}/
\Bigl( \Lambda_{\overline{\rm MS}}^{\mbox{\scriptsize 3-loop}} \Bigr)^2$
& 1.261 & 1.317 & 1.385 & 1.465 & 1.556 & 1.644 \\
\hline
\end{tabular}
\end{center}
\caption{
Coefficients of the linear potential 
normalized by the Lambda parameter in
$\overline{\rm MS}$ scheme, for different values of $n_l$.
\label{tab-str-tension}}
\end{table}

${\cal B}(N,\xi)$ and ${\cal D}(r,N,\xi)$ depend on $\xi$ and diverge as
$N \to \infty$ if $\xi$ is sufficiently large.
In fact, apart from the overall normalization (and some details),
behaviors of
${\cal B}(N,\xi)$ and ${\cal D}(r,N,\xi)$ are similar to those presented in
the previous subsection.
We give two examples.
%\vspace{3mm}\\
\begin{itemize}
\item
${\cal D}(r,N,\xi)$ 
in case (b) with $a_1=0$ and $\xi=1$:
\bea
&&
{\cal D}^{\rm (b)}(r,N,\xi) 
\Bigr|_{\xi=1,a_1=0} = \frac{4C_F}{\beta_0 } \,
\bigl( \Lambda_{\overline{\rm MS}}^{\mbox{\scriptsize 2-loop}} \bigr)^3\, r^2
\,
\biggl[
\frac{1}{12} \Bigl( \frac{3}{2e} \Bigr)^{3\delta/2} 
\frac{1}{\Gamma(1+{\scriptstyle \frac{3}{2}}\delta)}
\, \log N + { d^{\rm (b)}}
(r \,\Lambda_{\overline{\rm MS}}^{\mbox{\scriptsize 2-loop}})
\biggr] ,
\nonumber\\
\label{logNinRG}
\\ &&
d^{\rm (b)}(\rho) = 
\frac{1}{12} \Bigl( \frac{3}{2e} \Bigr)^{3\delta/2} 
\frac{1}{\Gamma(1+{\scriptstyle \frac{3}{2}}\delta)}\,
(\log 2 + \gamma_E) 
\nonumber\\
&&
~~~~~~
- {\rm Re}
\int_0^\infty \!\!\!ds \left\{
\frac{\Gamma(-x)}{\Gamma(1+\frac{x}{2} \delta)} \, \rho^{-is}\,
e^{-\pi i x/2} \biggl( \frac{x}{2e}\biggr)^{x \delta /2} +
\frac{i}{2x(3\!-\! x)} \biggl( \frac{3}{2e}\biggr)^{3\delta/2} 
\frac{1}{\Gamma(1+{\scriptstyle \frac{3}{2}}\delta)}
\right\}_{x = 3-i\,s }
.
\nonumber\\
\eea
The asymptotic forms of $d^{\rm (b)}(\rho)$ are given by
\bea
d^{\rm (b)}(\rho) \sim
\frac{1}{12} \Bigl( \frac{3}{2e} \Bigr)^{3\delta/2} 
\frac{1}{\Gamma(1+{\textstyle \frac{3}{2}}\delta)}\,
\left[
2\,\log|\log\rho| + \log \mbox{$ \frac{9}{2}$} + \gamma_E 
\right] ,
~~~~~~~
\rho \to 0 
~~\mbox{or}~~
\rho \to \infty ,
\eea
and in the intermediate region both asymptotic forms are smoothly interpolated.
%\vspace{3mm}\\
\item
${\cal D}(r,\infty,\xi)$ 
in case (b) with $a_1=0$ and $2/3<\xi<1$:
\bea
&&
{\cal D}^{\rm (b)}(r,\infty,\xi)\Bigr|_{a_1=0}=
- \frac{4C_F}{\beta_0 } \,
\bigl( \Lambda_{\overline{\rm MS}}^{\mbox{\scriptsize 2-loop}} \bigr)^{3\xi}\, r^{3\xi-1}
\,
\nonumber\\ &&
~~~~~~~~~~~~~~~~~~~~~~~~~
\times
{\rm Re}
\int_0^\infty \!\!\!ds \,
\frac{\Gamma(-x)}{\Gamma(1+\frac{x}{2} \delta)} \, 
\Bigl( r \,\Lambda_{\overline{\rm MS}}^{\mbox{\scriptsize 2-loop}}
\Bigr)^{-is}\,
e^{-\pi i x/2} \biggl( \frac{x}{2e}\biggr)^{x \delta /2} 
\Biggr|_{x = 3\,\xi-i\,s }
. 
\eea
Its asymptotic forms are given by
\bea
{\cal D}^{\rm (b)}(r,\infty,\xi)\Bigr|_{a_1=0} \sim
\frac{
\bigl( \Lambda_{\overline{\rm MS}}^{\mbox{\scriptsize 2-loop}} \bigr)^{3\xi}\, 
r^{3\xi-1}
}{\log \Bigl( 
r \,\Lambda_{\overline{\rm MS}}^{\mbox{\scriptsize 2-loop}}
\Bigr)} \times
\frac{4C_F}{\beta_0 } \,
\frac{\Gamma(-3\xi)}{\Gamma(1+\frac{3}{2}\xi\delta)} \, 
\sin\Bigl( \frac{3}{2}\pi\xi \Bigr) \,
\left( \frac{3\xi}{2e} \right)^{3\xi\delta/2} ,
\nonumber\\
r \to 0~~~{\rm or}~~~ r\to\infty .
\eea
\end{itemize}
The expressions when $a_1\ne 0$ or in case (c)
are more complicated and lengthy.
%In any case, they are qualitatively similar
%(apart from the overall normalization)
%to those already presented in the case of
%large-$\beta_0$ approximation.

%
\begin{figure}
\begin{center}
\psfrag{case(b)}{Case (b)}
\psfrag{xi=1}{\hspace{0mm}$\xi=1$}
\psfrag{nl=0}{$n_l=0$}
\psfrag{r=1/Lambda}{\hspace{-3mm}
$r =1/\Lambda_{\overline{\rm MS}}^{\mbox{\scriptsize 2-loop}}$} 
\psfrag{Lambda*r}{\hspace{3mm}
$r \,\Lambda_{\overline{\rm MS}}^{\mbox{\scriptsize 2-loop}}$} 
\psfrag{Vc(r)+Cr}{\hspace{0mm}$V_C(r)+{\cal C}\, r$} 
\psfrag{lefttitle}
{\hspace{-2mm}$V_N(r)$~~and~~$V_C(r)+{\cal C}\,r$} 
\psfrag{N=10}{\hspace{-5mm}$N=10$}
\psfrag{N=30}{\hspace{-5mm}$N=30$}
\psfrag{N=100}{\hspace{-5mm}$N=100$}
\includegraphics[width=10cm]{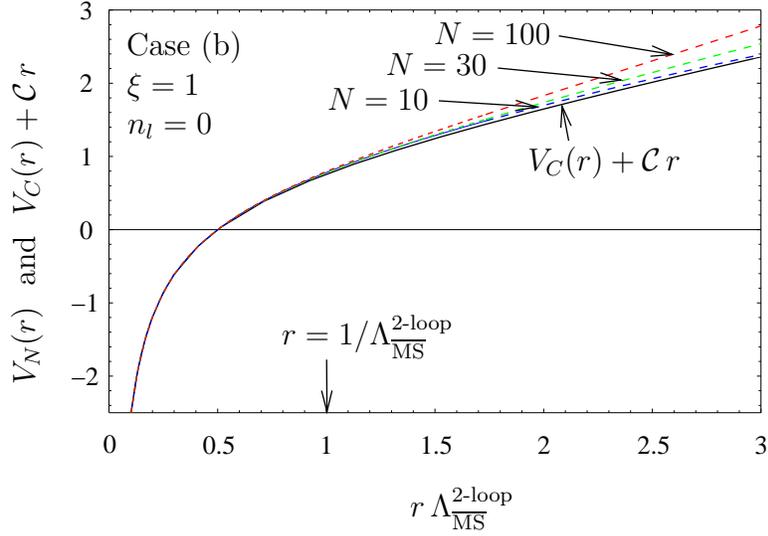}
\end{center}
\vspace*{-.5cm}
\caption{\small
[Case (b): NLL] 
$V_N(r)$ for $N=10$, 30, 100 and $\xi=1$
(dashed lines).
For comparison, the ``Coulomb''+linear potential
$V_C(r)+{\cal C}\,r$ is also plotted (solid black).
Constants have been added to $V_N(r)$ and
 $V_C(r)+{\cal C}\,r$
to make them coincide 
at $r \, \Lambda_{\overline{\rm MS}}^{\mbox{\scriptsize 2-loop}}
=0.5$.
We set $n_l=0$.
\label{VN-NLL}}
\end{figure}
\begin{figure}
\begin{center}
\psfrag{Case (b), nl=0}{\hspace{-5mm}Case (b), $n_l=0$}
\psfrag{xi=0.9}{\hspace{1mm}$\xi=0.9$}
\psfrag{]}{\hspace{4mm}$\left.\rule[0mm]{0mm}{6mm} \right\}$}
\psfrag{N=30, xi=1.1}{\hspace{-10mm}$N=30,~\xi=1.1$}
\psfrag{rho=Lambda*r}{\hspace{5mm}$\rho = \widetilde{\Lambda}\, r$} 
\psfrag{vc+C*rho}{\hspace{-10mm}$V_C(r)+{\cal C}\, r$} 
\psfrag{lefttitle}
{\hspace{-3mm}$V_N(r)$~~and~~$V_C(r)+{\cal C}\,r$} 
\psfrag{N=10}{\hspace{-2mm}$N=10$}
\psfrag{N=300}{\hspace{0mm}$N=300$}
\psfrag{r=1/Lambda}{\hspace{-6mm}
$r =1/\Lambda_{\overline{\rm MS}}^{\mbox{\scriptsize 2-loop}}$} 
\psfrag{Lambda*r}{\hspace{6mm}
$r \,\Lambda_{\overline{\rm MS}}^{\mbox{\scriptsize 2-loop}}$} 
\includegraphics[width=11cm]{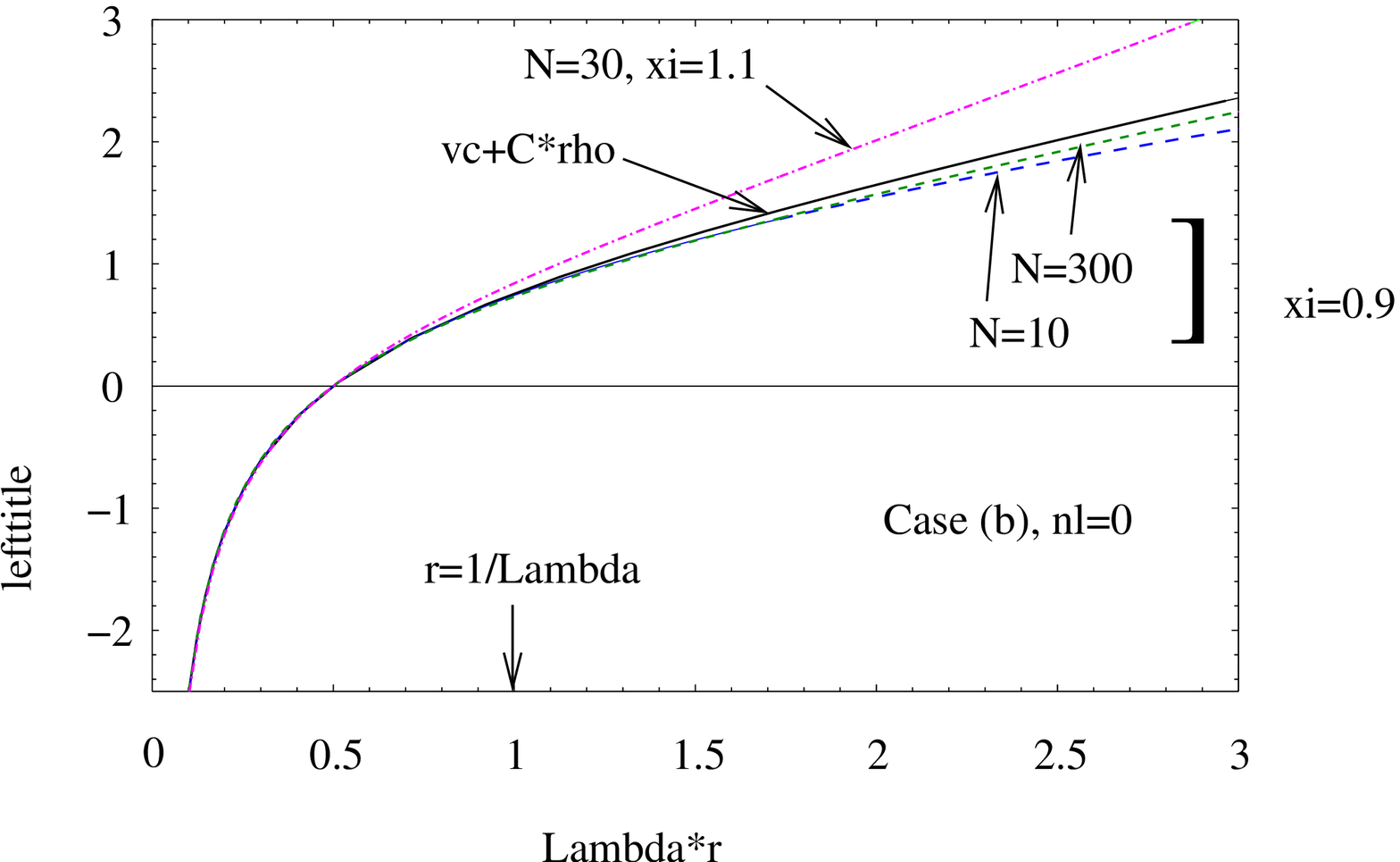}
\end{center}
\vspace*{-.5cm}
\caption{\small
[Case (b): NLL] 
$V_N(r)$ for different values of $\xi$ and $N$.
(Dashed lines for $\xi=0.9$ and dot-dahed line for
$\xi=1.1$.)
For comparison, the ``Coulomb''+linear potential
$V_C(r)+{\cal C}\,r$ is also shown (solid line).
Other conventions are same as in Fig.~\ref{VN-NLL}.
\label{VN-NLL-vary-xi}}
\end{figure}
\begin{figure}
\begin{center}
\psfrag{case(c)}{Case (c)}
\psfrag{xi=1}{\hspace{0mm}$\xi=1$}
\psfrag{nl=0}{$n_l=0$}
\psfrag{r=1/Lambda}{\hspace{-3mm}
$r =1/\Lambda_{\overline{\rm MS}}^{\mbox{\scriptsize 3-loop}}$} 
\psfrag{Lambda*r}{\hspace{3mm}
$r \,\Lambda_{\overline{\rm MS}}^{\mbox{\scriptsize 3-loop}}$} 
\psfrag{Vc(r)+Cr}{\hspace{0mm}$V_C(r)+{\cal C}\, r$} 
\psfrag{lefttitle}
{\hspace{-2mm}$V_N(r)$~~and~~$V_C(r)+{\cal C}\,r$} 
\psfrag{N=10}{\hspace{-7mm}$N=10$}
\psfrag{N=30}{\hspace{-7mm}$N=30$}
\psfrag{N=100}{\hspace{-7mm}$N=100$}
\includegraphics[width=10cm]{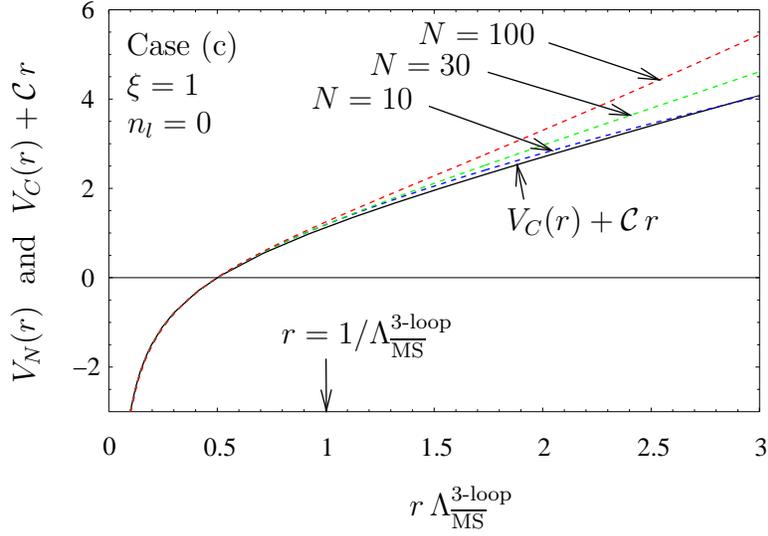}
\end{center}
\vspace*{-.5cm}
\caption{\small
[Case (c): NNLL] 
$V_N(r)$ for $N=10$, 30, 100 and $\xi=1$
(dashed lines).
For comparison, the ``Coulomb''+linear potential
$V_C(r)+{\cal C}\,r$ is also plotted (solid black).
Other conventions are same as in Fig.~\ref{VN-NLL}.
\label{VN-NNLL}}
\end{figure}
\begin{figure}
\begin{center}
\psfrag{Case (c), nl=0}{\hspace{-5mm}Case (c), $n_l=0$}
\psfrag{xi=0.9}{\hspace{1mm}$\xi=0.9$}
\psfrag{]}{\hspace{4mm}$\left.\rule[0mm]{0mm}{6mm} \right\}$}
\psfrag{N=30, xi=1.1}{\hspace{-10mm}$N=30,~\xi=1.1$}
\psfrag{rho=Lambda*r}{\hspace{5mm}$\rho = \widetilde{\Lambda}\, r$} 
\psfrag{vc+C*rho}{\hspace{-10mm}$V_C(r)+{\cal C}\, r$} 
\psfrag{lefttitle}
{\hspace{-3mm}$V_N(r)$~~and~~$V_C(r)+{\cal C}\,r$} 
\psfrag{N=10}{\hspace{-2mm}$N=10$}
\psfrag{N=300}{\hspace{0mm}$N=300$}
\psfrag{r=1/Lambda}{\hspace{-6mm}
$r =1/\Lambda_{\overline{\rm MS}}^{\mbox{\scriptsize 3-loop}}$} 
\psfrag{Lambda*r}{\hspace{6mm}
$r \,\Lambda_{\overline{\rm MS}}^{\mbox{\scriptsize 3-loop}}$} 
\includegraphics[width=11cm]{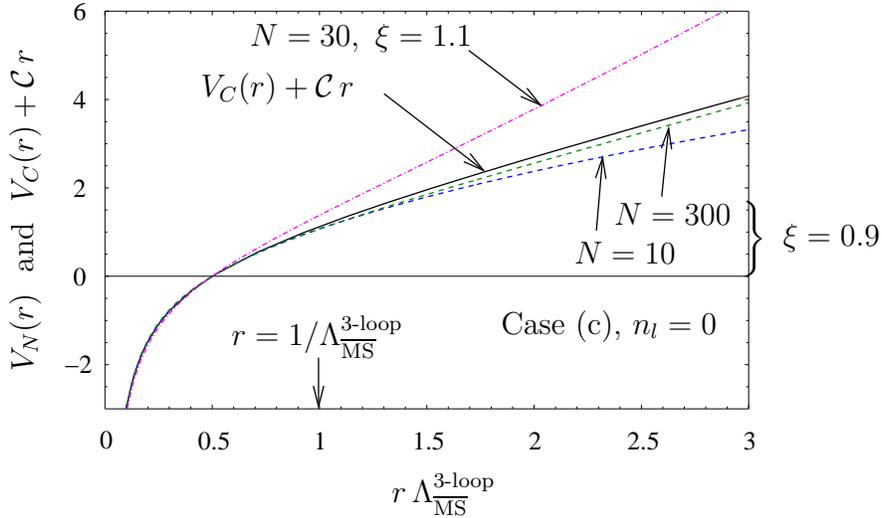}
\end{center}
\vspace*{-.5cm}
\caption{\small
[Case (c): NNLL] 
$V_N(r)$ for different values of $\xi$ and $N$.
(Dashed lines for $\xi=0.9$ and dot-dahed line for
$\xi=1.1$.)
For comparison, the ``Coulomb''+linear potential
$V_C(r)+{\cal C}\,r$ is also shown (solid line).
Other conventions are same as in Fig.~\ref{VN-NLL}.
\label{VN-NNLL-vary-xi}}
\end{figure}
In Figs.~\ref{VN-NLL}--\ref{VN-NNLL-vary-xi}, 
we show $V_N(r)$ for different values
of $N$ and $\xi$ in cases (b) and (c).
(We set $n_l=0$ in these figures.)
They are compared with the 
``Coulomb''+linear potential
$V_C(r)+{\cal C}\,r$.
The corresponding figures in case (a) can be obtained 
by simple rescaling of
Figs.~\ref{VN-CplusL} and \ref{VN-vary-xi-N}.
Apart from the overall normalization, we see similar general features.
Most importantly, $V_N(r)$ approximates well 
$V_C(r)+{\cal C}\,r$ at $r \simlt \Lambda_{\overline{\rm MS}}^{-1}$
for a reasonably wide range of $\xi$ and $N$.
For fixed $\xi$, $V_N(r)$ becomes steeper at
$r \simgt \Lambda_{\overline{\rm MS}}^{-1}$ as $N$ increases
[cf.\ eqs.~(\ref{DrhoN}), (\ref{logNinRG})].
For fixed $N$, $V_N(r)$ is steeper for larger $\xi$ at
$r \simgt \Lambda_{\overline{\rm MS}}^{-1}$;
this is because, if $\alpha_S(\mu)$ is kept fixed and the truncation
order is increased, all the higher-order terms additionally
included contribute with positive sign.
An only qualitative difference between case (a) and cases (b),(c)
is that, for the same value of $\xi$ and $N$, 
$V_N(r)$ is slightly steeper (in comparison to 
$V_C(r)+{\cal C}\,r$) at
$r \simgt \Lambda_{\overline{\rm MS}}^{-1}$ in cases (b),(c)
than in case (a).
We postpone
comparisons between cases (a),(b),(c) or comparisons with lattice
computations of $V_{\rm QCD}(r)$ until Sec.~\ref{s5}.

The effects of US logs in case (c$'$) are very small (if we ignore
shifts by $r$-independent constant).
For instance, if we superimpose plots of
$V_N(r)$ and $V_C(r)+{\cal C}\,r$ of case (c$'$)
on Fig.~\ref{VN-NNLL}, 
as we vary 
$\mu_f
\Lambda_{\overline{\rm MS}}^{\mbox{\scriptsize 3-loop}}$ between 
$2$ -- $5$,
they are hardly 
distinguishable from the corresponding lines of case (c).
(Only $V_N(r)$ for $N=100$ is visibly raised at 
$r\,\Lambda_{\overline{\rm MS}}^{\mbox{\scriptsize 3-loop}} \simgt 2$.)
The smallness of contributions from US logs stems from the
small coefficient $C_A^3/(6\beta_0)$ and suppression 
by $\log[\alpha_S(q)/\alpha_S(\mu_f)]$
in eq.~(\ref{NNLLUSlog}).

Conclusions are 
essentially the same as those in the large-$\beta_0$ approximation,
because qualitative behaviors of $V_N(r)$ are similar:
The ``Coulomb''+linear potential,
$V_C(r)+{\cal C}\, r$, can be regarded as a genuine part
of our prediction,
while we may associate ${\cal D}(r,N,\xi)$ with an
${\cal O}(\LQ^3 r^2)$ uncertainty (and beyond) due to IR
renormalons.
Taking the variations of $V_N(r)$, corresponding to the
different values of $\xi$ and $N$
shown in Figs.~\ref{VN-NLL}-\ref{VN-NNLL-vary-xi}, 
as a measure of uncertainties  of the predictions
for $V_N(r)$, the uncertainties are fairly small in the distance
region $r < \Lambda_{\overline{\rm MS}}^{-1}$.

Let us compare our results in Secs.~\ref{s3.2} and \ref{s3.3}
with the results of the existing literature.
The scale-fixing prescription according to the
principle of minimal sensitivity was advocated 
and studied originally in \cite{Stevenson:1981vj}.
In \cite{Beneke:1992ea}, a scale-fixing prescription close
to eq.~(\ref{xi}) was advocated, based on an analysis of large-order
behavior of perturbative series \`a la renormalons;
the prescription was used to suppress an ambiguity induced by UV renormalon,
which is located closer to the origin than IR renormalons in the Borel plane.
We studied in \cite{Sumino:2003yp} 
the large-order behaviors of $V_N(r)$ using the scale-fixing
condition eq.~(\ref{xi}) but restricting to the case $\xi=1$, in the large-$\beta_0$
approximation and using the estimates by RG.
Ref.~\cite{VanAcoleyen:2003gc} extended these analyses:
Within the large-$\beta_0$ approximation, and using the scale-fixing
condition eq.~(\ref{xi}), a general formula for the
large-order behavior
of a wide class of perturbative series was obtained and
the relation to the Borel summation was elucidated;
furthermore, the relation
to the principle of minimal sensitivity was
studied.
Our present analysis is a direct extension of \cite{Sumino:2003yp};
our results in the large-$\beta_0$
approximation in Sec.~\ref{s3.2} are consistent with the general formula
of \cite{VanAcoleyen:2003gc} when the formula is applied to $V_{\beta_0}(r)$.
(Since the assumed singularity structure in the Borel plane is slightly different
from that of $V_{\beta_0}(r)$, slight modification of the formula is necessary).
Unique aspects of  \cite{Sumino:2003yp} and the present work, besides
being a dedicated examination of the QCD potential, are
(a) the specific way of decomposition (close to Laurent expansion in $r$),
and (b) inclusion of 2-loop and 3-loop running
of $\alpha_S(q)$.
Furthermore, the separation into scale-independent (prescription-independent)
part and scale-dependent (prescription-dependent) part is unique to
the present work.

\subsection{\boldmath $[\alpha_V^{\rm PT}(q)]_N$ as $N \to \infty$}
\label{s3.4}

In order to understand the properties of $V_N(r)$ given in the
previous two subsections,
we examine behaviors of the truncated $V$-scheme
coupling at $N \to \infty$, defined by 
\bea
[\alpha_V^{\rm PT}(q)]_\infty \equiv \lim_{N \to \infty}
[\alpha_V^{\rm PT}(q)]_N .
\eea
The relation (\ref{xi}) between $\alpha_S(\mu)$ and $N$ is
understood in taking the limit.

\begin{figure}
\begin{center}
\psfrag{a(q)}{$\beta_0\,\alpha_{V,\beta_0}^{\rm PT}(q)$}
\psfrag{[a(q)]_N}{$\beta_0\,[\alpha_{V,\beta_0}^{\rm PT}(q)]_N$}
\psfrag{lefttitle}
{\hspace{-8mm}
$\beta_0\,[\alpha_{V,\beta_0}^{\rm PT}(q)]_N
$~~~and~~~$\beta_0\,\alpha_{V,\beta_0}^{\rm PT}(q)$}
\psfrag{q/Lambda}{\hspace{0mm}
$q/\widetilde{\Lambda}$}
\psfrag{N=infty, xi=0.9}{\scriptsize $N=\infty,~\xi=0.9$}
\psfrag{N=10, xi=1}{\scriptsize $N=10,~\xi=1$}
\psfrag{N=infty, xi=1}{\hspace{-2mm}\scriptsize $N=\infty,~\xi=1$}
\psfrag{N=infty, xi=1.1}{\hspace{-6mm}\scriptsize $N=\infty,~\xi=1.1$}
\includegraphics[width=11cm]{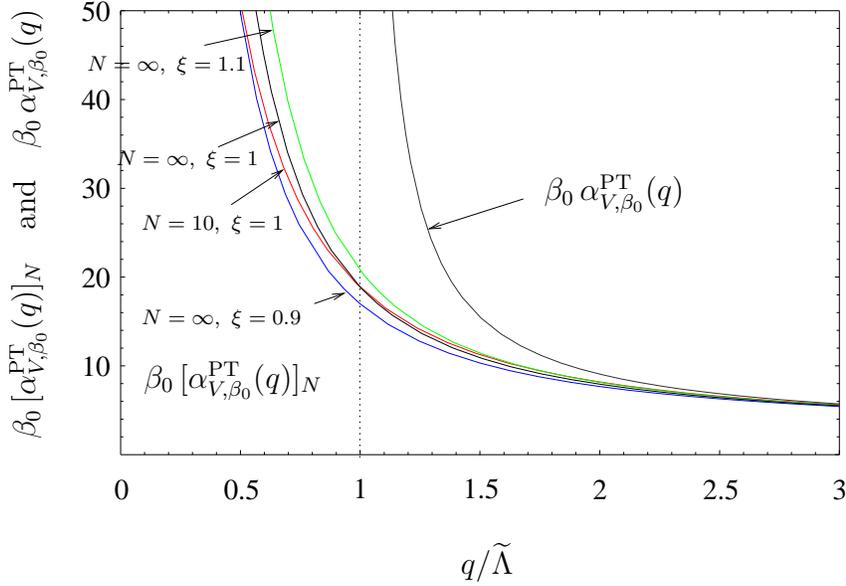}
\end{center}
\vspace*{-.5cm}
\caption{\small
$\beta_0\,[\alpha_{V,\beta_0}^{\rm PT}(q)]_N$ and 
$\beta_0\,\alpha_{V,\beta_0}^{\rm PT}(q)$
vs.\ $q/\widetilde{\Lambda}$ for different values of $\xi$ and $N$.
\label{alfinf}}
\end{figure}

In the large-$\beta_0$ approximation, one easily finds
\bea
&&
[\alpha_{V,\beta_0}^{\rm PT}(q)]_\infty
=  \lim_{N \to \infty}
\left[ \alpha_S(\mu)\,
\frac{1}{1-L}
\right]_N
=\lim_{N \to \infty} \alpha_S(\mu) \,
\frac{1-L^N}{1-L}
\nonumber
\\ &&
~~~~~~~~~~~~~~
= \frac{2\pi}{\beta_0 \log \bigl({q}/{\widetilde{\Lambda}}\bigr)}
\Biggl\{ 1 - \biggl( \frac{\widetilde{\Lambda}}{q} \biggr)^{3\xi} \Biggr\} ,
\label{alfinfty_beta0}
\eea
where
$L = \frac{\beta_0\alpha_S(\mu)}{2\pi} 
\log \Bigl(\frac{\mu e^{5/6}}{q}\Bigr)
= 1 + \frac{3\xi}{N}\log \Bigl(\frac{\widetilde{\Lambda}}{q}\Bigr)$.
There is no singularity at $q = \widetilde{\Lambda}$, and
$[\alpha_{V,\beta_0}^{\rm PT}(q)]_\infty$ is regular at $0 < |q| < \infty$
in the complex $q$ plane.
For $\xi \sim 1$, the first term in the curly bracket is dominant at
UV ($q \to \infty$), whereas the second term is dominant at
IR ($q \to 0$).
Hence, $[\alpha_{V,\beta_0}^{\rm PT}(q)]_\infty$ can be regarded
as a modified coupling, regularized 
in the IR region, $|q| \simlt  \widetilde{\Lambda}$;
by including a power correction, the Landau pole of
the original coupling, $\alpha_{V,\beta_0}^{\rm PT}(q) = 
2\pi/[\beta_0 \log \bigl({q}/{\widetilde{\Lambda}}\bigr)]$, 
has been removed.
(See Fig.~\ref{alfinf}.)

One can test sensitivity of the prediction to the IR behavior
of the regularized coupling by varying $\xi$.
The results of Sec.~\ref{s3.2} show that $B(N)$ and $D(\rho,N)$, 
associated with IR renormalons, are sensitive
to the IR behavior of $[\alpha_{V,\beta_0}^{\rm PT}(q)]_\infty$,
in accord with our expectation.
On the other hand, $\xi$-independence of $v_C(\rho) + C\rho$ 
shows that $v_C(\rho) + C\rho$ is determined only by the original coupling
$\alpha_{V,\beta_0}^{\rm PT}(q)$, and that  it is insensitive
to the IR behavior of the regularized coupling.

If we estimate the higher-order terms using RG,
$[\alpha_{V}^{\rm PT}(q)]_\infty$
in case (a) is obtained simply by replacing
$\widetilde{\Lambda}$ by
$\Lambda_{\overline{\rm MS}}^{\mbox{\scriptsize 1-loop}}$
in eq.~(\ref{alfinfty_beta0}).
$[\alpha_{V}^{\rm PT}(q)]_\infty$
in case (b) can be analyzed using 
a one-parameter integral representation.
To this end, let us first analyze
the two-loop running coupling constant $\alpha_{S}(q)$
(defined by eq.~(\ref{RGeq}) with $\beta_n=0$ for $n\ge 2$)
and $[\alpha_S(q)]_\infty$.
Motivated by eq.~(\ref{alfinfty_beta0}), we separate 
$[\alpha_S(q)]_\infty$ into $\alpha_S(q)$
and 
$\Delta \alpha_S(q) \equiv [\alpha_S(q)]_\infty - \alpha_S(q) $.
They can be expressed in one-parameter integral forms, respectively,
as
\bea
&&
\alpha_{S}(q) = \int_0^\infty dx\, f_1(x;q) ,
~~~~~~
\Delta \alpha_S(q) = - \int_{3\xi}^\infty dx\,  f_1(x;q) ,
\label{one-param-int}
\eea
with
\bea
%\\ &&
f_1(x;q) = \frac{2\pi}{\beta_0}\,
\left( {q}/{\Lambda_{\overline{\rm MS}}^{\mbox{\scriptsize 2-loop}} }
\right)^{-x}
\left( \frac{x}{2e} \right)^{x\,\delta  /2}
\frac{1}{\Gamma ( 1 + x\,\delta  /2)} .
\eea
(See App.~\ref{appB} for derivation.)
These expressions are valid for a complex argument $q$ if
$|q| > q_* \equiv \delta^{-\delta/2} \,
\Lambda_{\overline{\rm MS}}^{\mbox{\scriptsize 2-loop}}$;
they can be analytically continued to other regions
by deforming the integral contour of $x$.
$\alpha_S(q)$ and $\Delta \alpha_S(q)$, respectively,
are singular at the Landau singularity $q = q_*$
(branch point), whereas their sum
$[\alpha_S(q)]_\infty$ is regular at $0 < |q| < \infty$.
One may find asymptotic behaviors of $\alpha_S(q)$
and $\Delta \alpha_S(q)$ from the above expressions.
The well-known asymptotic behavior of $\alpha_S(q)$ as
$q \to \infty$ is reproduced by rescaling
$x \log ({q}/{\Lambda_{\overline{\rm MS}}^{\mbox{\scriptsize 2-loop}} })
\to x$ and expanding the integrand in
$1/\log ({q}/{\Lambda_{\overline{\rm MS}}^{\mbox{\scriptsize 2-loop}} })$.
The asymptotic behavior of $\alpha_S(q)$ as
$q \to 0$, when $q$ is varied along the path $C_1$ of Fig.~\ref{path},
can be obtained as follows.
First rotate the integral contour clockwise around the origin,
$x = e^{-i\pi} y$ ($0<y<\infty$),
then rescale 
$y \log ({\Lambda_{\overline{\rm MS}}^{\mbox{\scriptsize 2-loop}} }/q)
\to y$ and expand the integrand in
$1/\log ({\Lambda_{\overline{\rm MS}}^{\mbox{\scriptsize 2-loop}} }/q)$.
The asymptotic behavior of $[\alpha_S(q)]_\infty$ as
$q \to \infty$
can be obtained by expanding the integrand of
eq.~(\ref{one-param-int}) about $x = 3\xi$ except for the
factor 
$( {q}/{\Lambda_{\overline{\rm MS}}^{\mbox{\scriptsize 2-loop}} }
)^{-x}$.
The asymptotic behavior of $[\alpha_S(q)]_\infty$ as
$q \to 0$, when $q$ is varied along the path $C_1$,
can be obtained similarly, by first
rotating the integral contour clockwise around the origin.
The results read
\bea
&&
\alpha_S(q) \sim \frac{2\pi}{\beta_0 \, \log 
({q}/{\Lambda_{\overline{\rm MS}}^{\mbox{\scriptsize 2-loop}} }) },
~~~~~~~~~~~~~~~~~~~~~~~~~~~~~~~~~~~~~~~~~~~~~~~~
\mbox{$q \to 0$~~or~~$q \to \infty$},
\label{asym-alf2L}
\\
&&
\Delta \alpha_S(q) \sim - \frac{2\pi}{\beta_0 \, \log 
({q}/{\Lambda_{\overline{\rm MS}}^{\mbox{\scriptsize 2-loop}} }) 
}
 \left(
\frac{\Lambda_{\overline{\rm MS}}^{\mbox{\scriptsize 2-loop}} }
{q}
\right)^{3\xi} 
\!\! \times
\frac{( {3\xi}/({2e}) )^{3\,\xi\,\delta  /2}}{\Gamma ( 1 + 3\,\xi\,\delta  /2)} 
,
~~~~~
\mbox{$q \to 0$~~or~~$q \to \infty$} .
\label{asym-deltaalf2L}
\eea
Thus, apart from the overall normalization, the 
leading asymptotic
behaviors are identical with the one-loop running case,
eq.~(\ref{alfinfty_beta0}).\footnote{
There are qualitative differences in the subleading 
asymptotic behaviors.
}

$\alpha_{V}^{\rm PT}(q)$
in case (b) is given by 
$\alpha_S(q) + \bigl(\frac{a_1}{4\pi}\bigr)\,\alpha_S(q)^2$
in terms of the two-loop running coupling constant.
We also separate $[\alpha_S(q)^2]_\infty$ into
$\alpha_S(q)^2$ and $\Delta \alpha_S(q)^2
 \equiv [\alpha_S(q)^2]_\infty - \alpha_S(q)^2 $, which
can be analyzed similarly using the one-parameter
integral expressions
\bea
\alpha_{S}(q)^2 = \int_0^\infty dx\, f_2(x;q) ,
~~~~~~
\Delta \alpha_S(q)^2 = - \int_{3\xi}^\infty dx\,  f_2(x;q) ,
\eea
with
\bea
%\\ &&
f_2(x;q) = \frac{8\pi^2}{\beta_1}
\left( {q}/{\Lambda_{\overline{\rm MS}}^{\mbox{\scriptsize 2-loop}} }
\right)^{-x}\,
\delta^{-x\,\delta/2}\,
\frac{\gamma(1+x\,\delta/2,x\,\delta/2)}{\Gamma ( 1 + x\,\delta  /2)} .
\eea
The asymptotic forms of $\Delta \alpha_S(q)^2$ are given by
\bea
\Delta \alpha_S(q)^2 \sim - \frac{2\pi}{\beta_0 \, \log 
({q}/{\Lambda_{\overline{\rm MS}}^{\mbox{\scriptsize 2-loop}} }) 
}
\, \left(
\frac{\Lambda_{\overline{\rm MS}}^{\mbox{\scriptsize 2-loop}} }
{q}
\right)^{3\xi} 
\times
\frac{4\pi\beta_0}{\beta_1}\,
\delta^{-3\,\xi\,\delta/2}\,
\frac{\gamma(1+3\,\xi\,\delta/2,3\,\xi\,\delta/2)}
{\Gamma ( 1 + 3\,\xi\,\delta  /2)} 
, ~~~
&&
\nonumber\\
\mbox{$q \to 0$~~or~~$q \to \infty$},
&&
\label{asym-deltaalf2Lsq}
\eea
while the asymptotic behaviors of $\alpha_S(q)^2$ are
given simply by the square of eq.~(\ref{asym-alf2L}).
Thus,
$\bigl(\frac{a_1}{4\pi}\bigr)\,\alpha_S(q)^2$ term does not
affect the asymptotic behaviors of
$\alpha_{V}^{\rm PT}(q)$ and 
$\Delta \alpha_{V}^{\rm PT}(q)
 \equiv [\alpha_{V}^{\rm PT}(q)]_\infty - \alpha_{V}^{\rm PT}(q)$,
apart from the overall normalization.

The same is true in case (c).
The leading asymptotic behaviors of 
$\alpha_{V}^{\rm PT}(q)$ and 
$\Delta \alpha_{V}^{\rm PT}(q)$ as 
$q\to 0$ 
(when $q$ is varied along the path $C_1$)
and $q \to \infty$ are same as those in case (a) (one-loop running),
besides the overall normalization.
These IR and UV behaviors determine the behaviors of
$V_N(r)$ at $r \to \infty$ and $r \to 0$ for large $N$.
For this reason,
the truncated potentials $V_N(r)$ have
qualitatively similar features in all the cases
examined in Secs.~\ref{s3.2} and \ref{s3.3}.
We also note that, since $[\alpha_{V}^{\rm PT}(q)]_\infty$ has
no singularity along the positive real axis, and since 
$\Delta \alpha_{V}^{\rm PT}(q)$ is more dominant than
$\alpha_{V}^{\rm PT}(q)$ in IR,
the leading IR behavior of $[\alpha_{V}^{\rm PT}(q)]_\infty$,
when $q$ is sent to $+0$ along the positive real axis, is same as
that of $\Delta \alpha_{V}^{\rm PT}(q)$ as $q \to 0$
(when $q$ is varied along the path $C_1$).

\subsection{\boldmath How to decompose $V_N(r)$}
\label{s3.5}

We explain how we decompose $V_N(r)$
into 4 parts, as given in Secs.~\ref{s3.2} and \ref{s3.3}.
Integrating over the angular variables in eq.~(\ref{defVN}),
one obtains
\bea
V_N(r) = - \frac{2C_F}{\pi} \, \int_0^\infty dq \,
\frac{\sin qr}{qr} \,
[ \alpha_{V}^{\rm PT}(q) ]_N
= - \frac{2C_F}{\pi} \, {\rm Im} \, \int_0^\infty dq \,
\frac{e^{iqr}}{qr} \,
[ \alpha_{V}^{\rm PT}(q) ]_N .
\label{defVN-1param}
\eea
We separate the integral into two parts, 
according to the different asymptotic behaviors of 
$[ \alpha_{V}^{\rm PT}(q) ]_\infty$, i.e.\
$\alpha_{V}^{\rm PT}(q)$ and 
$\Delta \alpha_{V}^{\rm PT}(q)
= [\alpha_{V}^{\rm PT}(q)]_\infty - \alpha_{V}^{\rm PT}(q)$:
\bea
&&
V_N(r) = U_1(r) + U_2(r,N,\xi) ,
\\
&&
U_1(r) = 
- \frac{2C_F}{\pi} \, {\rm Im} \, \int_{C_1} dq \,
\frac{e^{iqr}}{qr} \,
\alpha_{V}^{\rm PT}(q)  ,
\\ &&
U_2(r,N,\xi) = 
- \frac{2C_F}{\pi} \, {\rm Im} \, \int_{C_1} dq \,
\frac{e^{iqr}}{qr} \,
\Bigl\{
[\alpha_{V}^{\rm PT}(q)]_N - \alpha_{V}^{\rm PT}(q)
\Bigr\}  .
\label{U2}
\eea
We deformed the integral
contour in order to avoid the
Landau singularity on the positive real axis; see Fig.~\ref{path}(i).
Contributions from the Landau singularity 
cancel between $U_1$ and $U_2$, 
since the original integral (\ref{defVN-1param}) does not 
contain the singularity.

Since
$
[\alpha_{V}^{\rm PT}(q)]_N - \alpha_{V}^{\rm PT}(q)
 \sim  q^{-3\xi}/\log q$
as $N \to \infty$,
the integral in eq.~(\ref{U2})
becomes IR divergent in this limit for $\xi \sim 1$.
On the other hand, the negative power of $q$ induces the positive
power behavior of $r$ 
in $U_2$ in the large $N$ limit.
Assuming $\xi > 2/3$, let us define
\bea
U_2(r,N,\xi) = \frac{\cal A}{r} + {\cal B}(N,\xi) + {\cal C}\, r 
+ {\cal D}(r,N,\xi) 
+(\mbox{terms that vanish as $N\to\infty$}),
\eea
where ${\cal D}(r,N,\xi)$ is subleading as compared to 
$ {\cal C}\, r$
at short-distances.
${\cal A}$ and ${\cal C}$ can be extracted as follows.
\bea
&&
{\cal A} = \lim_{r \to 0} r \, U_2
= \lim_{r \to 0}
- \frac{2C_F}{\pi} \, {\rm Im} \, \int_{C_1} dq \,
\frac{e^{iqr}}{q} \,
\Bigl\{
[\alpha_{V}^{\rm PT}(q)]_N - \alpha_{V}^{\rm PT}(q)
\Bigr\}  
\nonumber
\\ &&
~~~
= \frac{C_F}{\pi i} \, \int_{C_2} dq \,
\frac{1}{q} \,
\Bigl\{
[\alpha_{V}^{\rm PT}(q)]_N - \alpha_{V}^{\rm PT}(q)
\Bigr\}  
= - \frac{C_F}{\pi i} \, \int_{C_2} dq \,
\frac{\alpha_{V}^{\rm PT}(q)}{q} \,
= -\frac{4\pi C_F}{\beta_0} ,
\label{calA}
\\ &&
{\cal C} =
\lim_{r \to 0} \frac{1}{2}
\frac{\partial^2}{\partial r^2} ( r \, U_2 )
= \lim_{r \to 0}
 \frac{2C_F}{\pi} \, {\rm Im} \, \int_{C_1} dq \,
{e^{iqr}}\,{q} \,
\Bigl\{
[\alpha_{V}^{\rm PT}(q)]_N - \alpha_{V}^{\rm PT}(q)
\Bigr\}  
\nonumber
\\ &&
~~~
=  \frac{C_F}{2\pi i} \, \int_{C_2} dq \,
q\,{\alpha_{V}^{\rm PT}(q)} .
\label{calC}
\eea
To show the last equality of eq.~(\ref{calA}), 
we may use the RG equation (\ref{RGeq}), 
or, we can evaluate the integral
explicitly using $\alpha_{V}^{\rm PT}(q)$ at LL, NLL and NNLL.
$[\alpha_{V}^{\rm PT}(q)]_N$ does not contribute because it has
no singularity inside the contour $C_2$,
hence, both ${\cal A}$ and ${\cal C}$ are independent of $\xi$ and $N$.
Similarly, ${\cal B}$ and ${\cal D}$ are given by
\bea
&&
{\cal B}(N,\xi) =
\lim_{r \to 0} 
\frac{\partial}{\partial r} ( r \, U_2 )
= \lim_{r \to 0}
- \frac{2C_F}{\pi} \, {\rm Im} \,\, i  \int_{C_1} dq \,
{e^{iqr}}\,
\Bigl\{
[\alpha_{V}^{\rm PT}(q)]_N - \alpha_{V}^{\rm PT}(q)
\Bigr\}  ,
\label{calB}
\\ &&
{\cal D}(r,N,\xi) = U_2(r,N,\xi) - \left[\frac{\cal A}{r} 
+ {\cal B}(N,\xi) + {\cal C} \, r \right] 
\nonumber\\
&&
~~~~~~~~~
=- \frac{2C_F}{\pi} \, {\rm Im} \, \int_{C_1} dq \,
\frac{e^{iqr}-[1+iqr+\frac{1}{2}(iqr)^2]}{qr} \,
\Bigl\{
[\alpha_{V}^{\rm PT}(q)]_N - \alpha_{V}^{\rm PT}(q)
\Bigr\}  .
\label{calD}
\eea
In the limit $N \to \infty$,
${\cal B}$ is IR divergent, and if $\xi \ge 1$, 
${\cal D}$ is also IR divergent.
We would want to factor out divergent part as $N \to \infty$
in these cases.\footnote{
The integrals (\ref{calA})--(\ref{calD})
are convergent in the UV region, assuming the double
limits eq.~(\ref{limits}).
}

If $\xi <1$, ${\cal D}$ is finite as $N \to \infty$.
Then, we may insert the expression for $\Delta \alpha_{V}^{\rm PT}(q)$
obtained in the previous subsection.
In case (a) or in the large-$\beta_0$ approximation,
it is convenient to deform the integral contour into the upper
half plane by setting $q = ix/r$ ($0<x<\infty$).
Then, one readily obtains eq.~(\ref{Dxi}).
To find the asymptotic forms, eq.~(\ref{Dxi-asym}) and
subleading terms, one
expands the integrand (inside ${\rm Im}[...]$) by $\log x$.
In cases (b) and (c), we may
insert the integral expressions for 
$\Delta \alpha_{V}^{\rm PT}(q)$ to eq.~(\ref{calD}) and
integrate over $q$.
Thus, we obtain one-parameter integral expressions for ${\cal D}$,
except for the coefficient of $a_2$ in case (c).\footnote{
We were able to reduce the coefficient of $a_2$ in case (c)
only to a 2-dimensional integral form.
}
We may find asymptotic forms of ${\cal D}$, for instance,
if we insert the asymptotic expansion of 
$\Delta \alpha_{V}^{\rm PT}(q)$ 
[eqs.~(\ref{asym-deltaalf2L}), (\ref{asym-deltaalf2Lsq}) 
and subleading terms] and proceed as in case (a).

When $\xi=1$, we may factor out the divergence as
\bea
&&
{\cal D}(r,N,1) 
\approx
- \frac{2C_F}{\pi} \, {\rm Im} \, \int_{C_1} dq \,
\frac{e^{iqr}-[1+iqr+\frac{1}{2}(iqr)^2+\theta (q_0-|q|)\frac{1}{6}(iqr)^3]}{qr} \,
\Delta \alpha_{V}^{\rm PT}(q) \Bigr|_{\xi=1}
\nonumber\\
&&
~~~~~~~~~~~~~~~
- \frac{2C_F}{\pi} \, {\rm Im} \, \int_0^{q_0} dq \,
\frac{(iqr)^3}{6qr}
\Bigl\{
[\alpha_{V}^{\rm PT}(q)]_N - \alpha_{V}^{\rm PT}(q)
\Bigr\}_{\xi=1},
\eea
where $q_0$ is an IR cutoff to remove the IR divergence as
$q \to 0$.
In all the cases (a)--(c), 
one may extract the divergent part from the
second term, which may be taken as
\bea
\int_0^1 d\tilde{q} \, \frac{\tilde{q}^2}{\log \tilde{q}} \,
\biggl[ \Bigl( 1 - \frac{3}{N} \log\tilde{q} \Bigr)^N - 1 \biggr]
= \frac{1}{2}\, (
\log N + \log 2 + \gamma_E ) + {\cal O}\biggl(\frac{1}{\sqrt{N}} \biggr) ,
\eea
apart from an overall normalization (proportional to $r^2$).
Here, we have rescaled $q$ to a dimensionless variable $\tilde{q}$.
If $\xi > 1$, one may factor out the divergences in a 
similar manner; one should subtract powers of $r$
as many times as needed to remove all the IR divergences.
It is even simpler to factor out divergences from
${\cal B}(N,\xi)$; for instance,
see eq.~(21) of \cite{Sumino:2003yp} for $\xi=1$ and in the
large-$\beta_0$ approximation.

We define the ``Coulomb'' potential (with logarithmic
corrections at short-distances) as
\bea
V_C(r) = U_1(r) + \frac{\cal A}{r} .
\eea
It is determined by
$\alpha_{V}^{\rm PT}(q)$, therefore, it is
independent of $\xi$ and $N$.
Since the leading behavior of $V_N(r)$ as
$r\to 0$ is ~${\rm const.}/(r \log r)$ as determined by 
RG equation, the ${\cal A}/r$
term of $U_2$ must be cancelled by the $1/r$
term contained in $U_1$.
To compute the asymptotic forms of $U_1(r)$ as $r \to 0$ and
$r \to \infty$, one may insert the integral expressions
for $\alpha_{V}^{\rm PT}(q)$, given in the previous subsection, and integrate over $q$;
then rescale
$x \log (1/{r\Lambda_{\overline{\rm MS}}^{\mbox{\scriptsize 2-loop}} })
\to x$ and expand the integrand in
$1/\log (1/{r\Lambda_{\overline{\rm MS}}^{\mbox{\scriptsize 2-loop}} })$.
In this method, however, one should carefully choose the integral contour
for $x$ to avoid singularities.
Another method is as follows.
Consider first the case (a).
We deform the integral
contour into the upper half plane on the complex $q$-plane
and integrate by parts:
\bea
&&
U_1(r) = \frac{4C_F}{\beta_0 r} \,
{\rm Im} \int_0^\infty \frac{dx}{x} \,
\frac{e^{-x}}{\log (x/\rho) + i \pi /2}
\nonumber\\
&&
~~~~~
= -\frac{4C_F}{\beta_0 r} \,
{\rm Im} \int_0^\infty\! dx \,
e^{-x} \, \log \biggl( \log {\rho}-\log {x} -\frac{i\pi}{2} \biggr),
\eea
where $\rho = r\Lambda_{\overline{\rm MS}}^{\mbox{\scriptsize 1-loop}}$.
By expanding the integrand in $\log x$, we obtain the asymptotic
forms eq.~(\ref{asym-vC}).
In cases (b) and (c), we may proceed in parallel with the
above steps, after
appropriate change of variables in the integration.
Then eqs.~(\ref{asym-VC-1}) and (\ref{asym-VC-2}) are obtained.
\medbreak

Let us summarize our algorithm for decomposing $V_N(r)$ into
$V_C(r)+{\cal B} + {\cal C}r + {\cal D}(r)$.
First we separate $[\alpha_{V}^{\rm PT}(q)]_N$ into 2 parts
according to the different asymptotic behaviors of $[\alpha_{V}^{\rm PT}(q)]_\infty$
as
$q \to 0$ and $q \to \infty$.
The separation is particularly simple in
the one-loop running case, eq.~(\ref{alfinfty_beta0}).
Next we deform the integral contour into upper-half plane
to avoid the Landau singularity.
Thus, $V_N(r)$ is separated into  $U_1(r)$ and $U_2(r)$.
$U_2(r)$ has a power series expansion in $r$ from ${\cal O}(r^{-1})$ to 
${\cal O}(r)$.
Beyond that order, power series expansion in $r$
breaks down due to 
non-analyticity in $r$.\footnote{
For simplicity, we restrict ourselves
to the case $2/3\leq \xi \le 1$ in this and the next
paragraph.
}
Hence, $U_2(r)$ is naturally decomposed into 
${\cal A}/r + {\cal B} + {\cal C}r + {\cal D}(r)$.
Then, ${\cal A}/r$ 
is combined with $U_1(r)$
to form $V_C(r)$, which behaves as a Coulombic potential
in the entire range of $r$ apart from logarithmic corrections.
Thus, $V_N(r)$ is decomposed into
$\{ V_C(r), {\cal B} , {\cal C}r , {\cal D}(r)\}$, which correspond
to $\{r^{-1},r^0,r^1,r^{3\xi-1}\}$ terms;
the $r^{-1}$ and $r^{3\xi-1}$ terms include logarithmic corrections.
The ``Coulomb''+linear potential, $V_C(r) + {\cal C}r$, 
is determined only by $\alpha_{V}^{\rm PT}(q)$,
hence, it is independent of $\xi$ and $N$.
%We may regard $V_C(r) + {\cal C}r$ as a genuine
%part of the perturbative prediction.
%It is also consistent with the argument given at the end of Sec.~.

If we choose a different prescription to avoid the Landau singularity
in defining $U_1(r)$ and $U_2(r)$,
each of them will change (but the sum $V_N(r)$ will not).
Consequently,  
$V_C(r)$ and $ {\cal D}(r)$ will also change.
Since, however, the contribution of Landau singularity does 
not have a simple power-like form in $r$, in other prescriptions
$V_C(r) = U_1(r) + {\cal A}/{r}$ is not Coulomb-like at
large-distances (rather oscillatory).
We consider our decomposition 
natural,
in the sense that it is closest to a decomposition into
terms with simple powers in $r$ (cf.\
argument at the end of Sec.~\ref{s2.4}), as well as since
$V_C(r)+{\cal C}r$ is a good approximation of $V_N(r)$
at $r \simlt \Lambda_{\overline{\rm MS}}^{-1}$
for a reasonably wide range of $\xi$ and $N$
(cf.\ Figs.~\ref{VN-vary-xi-N},\ref{VN-NLL}--\ref{VN-NNLL-vary-xi}).

\section{\boldmath Renormalization Schemes in
OPE of $V_{\rm QCD}(r)$}
\label{s4}
\clfn

We analyze $V_{\rm QCD}(r)$ in OPE
when $r \ll \Lambda_{\rm QCD}^{-1}$.
As we have seen in Sec.~\ref{s2.4}, the leading short-distance contribution
is given by the singlet potential $V_S(r)$.
In this section, we provide renormalization prescriptions
for $V_S(r)$ explicitly.
We also show that the `Coulomb''+linear potential,
extracted in the previous section, can be qualified as
a singlet potential (Wilson coefficient) 
defined in a specific renormalization scheme.
These renormalized singlet potentials
are free from IR renormalons and IR divergences; hence, they can be
computed systematically and (in principle)
we can improve the predictions to arbitrary precision.
Correspondingly, the non-perturbative contributions are
unambiguously defined.

\subsection{Factorization scheme vs.\
``Coulomb''+linear potential}
\label{s4.1}

Following the argument of Sec.~\ref{s2.4},
let us define a renormalized singlet potential,
in a scheme where the IR divergences and IR renormalons are subtracted, as
\bea
V_S^{({\rm R})}(r;\mu_f\!) = 
 - \frac{2C_F}{\pi} 
\int_{\mu_f}^\infty \! dq \, \frac{\sin (qr)}{qr} \, 
\alpha_{V_S}^{({\rm R})}(q;\mu_f) 
\label{defVSR}
\eea
with
\bea
\alpha_{V_S}^{({\rm R})}(q;\mu_f)=
\alpha_V^{\rm PT}\! (q) + \delta\alpha_V\! (q;\mu_f) .
\label{VSR}
\eea
$\delta\alpha_V(q;\mu_f)$ is the counter term which subtracts the
IR divergences of $\alpha_V^{\rm PT}\! (q)$, 
given as multiple poles in $\epsilon$.
(We assume that $\alpha_V^{\rm PT}\! (q)$ is computed in
dimensional regularization.)

Let us consider two schemes in particular for defining $\delta\alpha_V(q;\mu_f)$.
First one is to subtract the IR divergences of
$\alpha_V^{\rm PT}\! (q)$ in the $\overline{\rm MS}$ scheme.
Explicitly, at NNNLO, we set
\bea
\delta\alpha_V(q;\mu_f) = \alpha_S(\mu)\,
\left(\frac{\alpha_S(\mu)}{4\pi}\right)^3
\times 72\pi^2 \left[
- \frac{1}{\epsilon}-8\log \left(
\frac{\mu}{q} \right)+4 \gamma_E - 4\log(4\pi) + 2 \log \left(
\frac{\mu_f}{q} \right) \right] ,
\label{delalfV-MSbar}
\eea
where we retained (only) the physical US logarithm
according to the argument given below eq.~(\ref{a3}).
Here, $\mu_f$ represents the scale at which loop momenta are
effectively cut off.
The logarithms induced by running of $\alpha_S$
can be resummed  up to NNNLL by setting $\mu \to q$
in $\alpha_{V_S}^{({\rm R})}(q;\mu_f)$.
As we saw in Sec.~\ref{s3.3}, resummation of
US logarithms does not give sizable effects,
so we will not try to resum US logs but rather include 
US logs only up to
NNNLO, as given in eq.~(\ref{delalfV-MSbar}).\footnote{
Resummation of US logs up to NNLL
is achieved if we omit $\log(\mu_f/q)$ in
eq.~(\ref{delalfV-MSbar}) and make the replacement
eq.~(\ref{NNLLUSlog}).
Resummation of US logs up to NNNLL
has not been computed yet.
}

Second scheme is 
to regularize the IR divergences 
by expanding $\alpha_V^{\rm PT}\! (q)$ as a double
series in $\alpha_S$ and $\log \alpha_S$.
Then, no artificial subtraction from the IR region of
loop momenta is made, and
$\alpha_{V_S}^{({\rm R})}(q;\mu_f)$ becomes independent of
$\mu_f$.
At NNNLO, $\delta\alpha_V(q;\mu_f)$ is obtained
from the Fourier transform of
eq.~(\ref{deltaEUS-pert}):
\bea
&&
\delta\alpha_V(q;\mu_f) =
\alpha_S(\mu) \left( \frac{\alpha_S(\mu)}{4\pi}\right)^3 
\nonumber \\ &&
~~~~~~~~~~~~~~~~
\times
{72 \pi^2}\left[ \,
-\frac{1}{\epsilon} - 4 \left\{ 2\log \Bigl(\frac{\mu}{q}\Bigr) +\log(4\pi) \right\}
+6\gamma_E -\frac{5}{3} + 2\log\Bigl(3\alpha_S(\mu)\Bigr)
\right] .
\label{delalfV-2nd-scheme}
\eea
Indeed it is independent of $\mu_f$.
This prescription introduces
a physical scale $\Delta V(r) =V_O(r)-V_S(r) \approx C_A \alpha_S/r$ as an
IR regulator
in loop integrals, hence, contributions from $q < \Delta V(r)$
are suppressed \cite{Appelquist:es}.
Below,  we will resum powers of
$\log(\mu/q)$ associated with the running of $\alpha_S$ but not
powers of US $\log \alpha_S$, for the same reason as in the first scheme.
In Sec.~\ref{s5.1}, we will compare the two schemes 
eqs.~(\ref{delalfV-MSbar}) and (\ref{delalfV-2nd-scheme})
numerically.

Dependence of $V_S^{({\rm R})}\!(r;\mu_f\!)$
on $\mu_f$ is introduced through
subtraction of the IR divergences (in the first scheme)
and of the IR renormalons.
The subtraction of the IR divergences 
in the first scheme induces logarithmic
dependences on $\mu_f$
[eq.~(\ref{delalfV-MSbar})], 
while the subtraction of the IR
renormalons induces power-like dependences on $\mu_f$.
The former resides in the counter term $\delta\alpha_V(q;\mu_f)$,
and the latter arises from the lower cutoff of the integral
in eq.~(\ref{defVSR}).
Roughly,
\bea
\frac{\partial }{\partial \mu_f}V_S^{\rm (R)}(r;\mu_f)
 \sim {\cal O}( \mu_f^2 r^2 ) ,
\label{muf-dep-VS}
\eea
neglecting the $r$-independent part.
Note that $\mu_f r \ll 1$ due to the hierarchy (\ref{hierarchy}).

Corresponding to the above definitions of $V_S^{({\rm R})}\!(r;\mu_f\!)$, 
we define the ``Coulomb''+linear potential 
from $\alpha_{V_S}^{({\rm R})}(q;\mu_f)$ by
\bea
&&
V_{\rm C+L}(r) = V_C^{({\rm R})}(r) + {\cal C}^{({\rm R})}\,r ,
%\rule[-5mm]{0mm}{6mm}
\label{VCplusL}
\eea
where
\bea
&&
V_C^{({\rm R})}(r) = 
- \frac{C_F}{\pi  i\, r} \, \int_{C_2} dq \,
\frac{\alpha_{V_S}^{({\rm R})}(q;\mu_f)}{q} \,
- \frac{2C_F}{\pi} \, {\rm Im}
\int_{C_1}\! dq \, \frac{e^{iqr}}{qr} \, \alpha_{V_S}^{({\rm R})}(q;\mu_f)
,
\label{VcR}
\rule[-6mm]{0mm}{6mm}
\\ &&
{\cal C}^{({\rm R})} = \frac{C_F}{2\pi i} \int_{C_2}\! dq \, q \, 
 \alpha_{V_S}^{({\rm R})}(q;\mu_f) .
\label{calCR}
\eea
Up to NNLL, it is natural to take
$\delta\alpha_V(q;\mu_f)=0$, hence, in this case,
$V_{\rm C+L}(r)$ coincides with the ``Coulomb''+linear
potential obtained in the large-order analysis,
eqs.~(\ref{genform1}) and (\ref{genform3}) [cf.\ eq.~(\ref{calA})];
in particular, $V_{\rm C+L}(r)$ is independent of $\mu_f$
up to this order.

Below we will show that 
\bea
V_S^{({\rm R})}(r;\mu_f) - V_{\rm C+L}(r) = 
{\rm const.} + {\cal O}(\mu_f^3r^2) .
\label{result}
\eea
Eqs.\,(\ref{deltaEUS}) and (\ref{result}) imply that, 
within the framework of OPE, 
short-distance contributions ($q>\mu_f$) determine the
``Coulomb''+linear part of the QCD potential, 
hence it is predictable
in perturbative QCD.
The residual term (apart from an $r$-independent constant)
is of order $\mu_f^3r^2$, which mixes with
the non-perturbative contribution $\delta E_{\rm US}(r)$, and 
is subleading at $r \ll \mu_f^{-1}$.
These features are consistent with our expectation
discussed at the end of Sec.~\ref{s2.4}.

\begin{figure}
\begin{center}
\psfrag{C3}{$C_3$} 
\psfrag{q}{$q$} 
\psfrag{muf}{$\mu_f$} 
\psfrag{q*}{$q_*$} 
\psfrag{0}{\hspace{-1mm}\raise-1mm\hbox{$0$}}
\includegraphics[width=7cm]{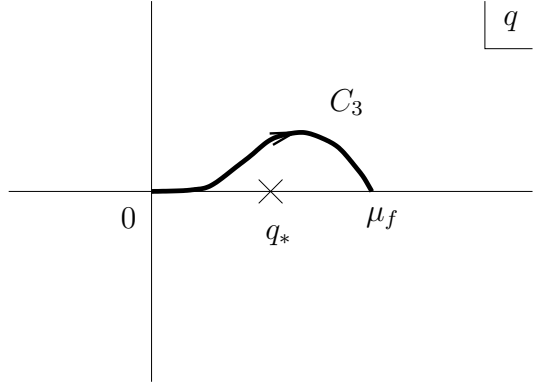}
\caption{
Integral path $C_3$ in the complex $q$-plane.
$q_*$ denotes the Landau singularity of $\alpha_S(q)$.
For 1-loop running, $q_*$ is a pole; beyond 1-loop running,
$q_*$ is a branch point.
In the latter case, branch cut is on the real axis starting from $q_*$
to $-\infty$.
\label{pathC3}}
\end{center}
\end{figure}

Eq.~(\ref{result}) can be shown as follows.
According to eqs.~(\ref{defVSR}), (\ref{VCplusL})--(\ref{calCR}),
\bea
&&
V_S^{({\rm R})}(r;\mu_f)  - V_{\rm C+L}(r) 
\nonumber\\
&&
= 
\frac{C_F}{\pi  i\, r} \, \int_{C_2} dq \,
\frac{\alpha_{V_S}^{({\rm R})}(q;\mu_f)}{q} 
 + \frac{2C_F}{\pi} \, {\rm Im}
\int_{C_3}\! dq \, \frac{e^{iqr}}{qr} \, \alpha_{V_S}^{({\rm R})}(q;\mu_f)
-{\cal C}^{({\rm R})}\, r ,
\label{proof}
\eea
where the integral path $C_3$ is shown in Fig.~\ref{pathC3}.
%%%%%%%%%%%%%%%%%%%%%%%%%%%%%%%%%%%%%%%%%%%%%%%%%%%%%%%%%%%%%%
%%%%%%%%%%%%%%%%%%%%%%%%%%%%%%%%%%%%%%%%%%%%%%%%%%%%%%%%%%%%%%
Since $\mu_f r \ll 1$, we may expand the Fourier factor as
$e^{iqr} = 1 + iqr - \frac{1}{2}(qr)^2 + \dots$
in the integral along $C_3$.
Then the leading term of the expansion
cancels against the first term of
eq.~(\ref{proof}), while the third term of the
expansion [$-\frac{1}{2}(qr)^2 $]
cancels against $-{\cal C}^{({\rm R})}\, r$
of eq.~(\ref{proof}).
Therefore, only remaining terms on the right-hand-side of
eq.~(\ref{proof}) are\, 
${\rm const.} + {\cal O}(\mu_f^3r^2) $.

One may think that in defining $V_S^{({\rm R})}\!(r;\mu_f\!)$
subtracting the integral 
eq.~(\ref{int-renormalon}) is not sufficient for 
subtracting all the IR renormalons.
The relation eq.~(\ref{result}) is unchanged, even
if one subtracts the IR renormalon contributions
using whatever other sophisticated method for estimating them.
This is because the IR renormalons in $V_S^{({\rm R})}(r;\mu_f)$ 
take the form\,
${\rm const.} + {\cal O}(\Lambda_{\rm QCD}^3r^2)$.

The perturbative expansion of $V_S^{({\rm R})}(r;\mu_f)$ 
may still be an asymptotic series (due to e.g.\  UV
renormalons\footnote{
Nevertheless, we note that up to now UV renormalons have not been
identified in $V_{\rm QCD}(r)$.
}).
Since the IR renormalons have been subtracted and the factorization
scale is set as $\mu_f \gg \Lambda_{\rm QCD}$, we may expect that 
$V_S^{({\rm R})}(r;\mu_f)$ is Borel summable.\footnote{
This is up to the uncertainties caused by the instanton-induced singularities 
in the Borel plane, which we neglect in our analysis.
}
(At least, the Borel integral is convergent 
in the large-$\beta_0$ approximation.)
Then, we may define $V_S^{({\rm R})}(r;\mu_f)$ from the perturbative series
either by Borel summation or according to the prescription of
\cite{Beneke:1992ea,VanAcoleyen:2003gc}.
Thus, $V_S^{({\rm R})}(r;\mu_f)$ can be computed systematically
(based on perturbative QCD).

\subsection{\boldmath $V_{\rm C+L}(r)$ as a 
$\mu_f$--independent renormalized singlet potential}
\label{s4.2}
\clfn

Up to NNLL, the ``Coulomb''+linear potential 
$V_{\rm C+L}(r) = V_C(r) + {\cal C}r$
was extracted from $V_N(r)$ as a 
prescription-independent part, corresponding to
a renormalon-free part, in Sec.~\ref{s3}.
We have also seen that $V_{\rm C+L}(r)$ coincides,
up to ${\cal O}(r^2)$,
with $V_S^{({\rm R})}(r;\mu_f)$, which is the Wilson coefficient 
free from IR renormalons.
Therefore, unlike original perturbative expansion of 
$V_{\rm QCD}(r)$, we expect that $V_{\rm C+L}(r)$
is free from intrinsic uncertainties.
In fact, $V_{\rm C+L}(r)$ can be computed systematically
and its accuracy can be improved (in principle) to arbitrary
precision as follows.
%Since the singlet potential
%$V_S^{({\rm R})}(r;\mu_f)$ is free from IR renormalons, we expect
%that the ``Coulomb'' +linear potential
%$V_{\rm C+L}(r)$ is also free from IR renormalons.
%In fact, this can be seen as follows.
Since $\alpha_{V_S}^{({\rm R})}(q;\mu_f)$ does not
contain IR renormalons,\footnote{
In a more sophisticated estimate of renormalons 
than eq.~(\ref{int-renormalon}), one
may find that
$\alpha_{V_S}^{({\rm R})}(q;\mu_f)$ still contains IR renormalons.
If this turns out to be the case, according to our philosophy,
it is appropriate to subtract the renormalons
by modifying the counter term $\delta\alpha_V(q;\mu_f)$.
} 
we may compute $\alpha_{V_S}^{({\rm R})}(q;\mu_f)$ 
and improve accuracy of its value along the 
contours $C_1$ and $C_2$ by including corrections
at LL, NLL, NNLL, and so on.
(We further improve the series
by Borel summation or by the prescription of
\cite{Beneke:1992ea,VanAcoleyen:2003gc} 
if necessary.)
In the infinite-order limit,
$\alpha_{V_S}^{({\rm R})}(q;\mu_f)$ is expected to be finite
everywhere along these contours.\footnote{
There is no known source of divergence except the
instanton-induced singularities, which we neglect in this paper.
}
Then, via eqs.~(\ref{VcR}) and (\ref{calCR}), $V_{\rm C+L}(r)$ can be
computed.
(It is beyond our scope to prove convergence of $V_{\rm C+L}(r)$.
Here, we have eliminated all of the known sources of divergences.)

In the definition of $V_{\rm C+L}(r)$, renormalization scheme
dependence enters through the definition of 
$\alpha_{V_S}^{({\rm R})}(q;\mu_f)$ or of
$\delta\alpha_V(q;\mu_f)$.
Below, we will focus on the second scheme,
discussed in the previous subsection:
To regularize the IR divergences 
by expanding $\alpha_V^{\rm PT}\! (q)$ as a double
series in $\alpha_S$ and $\log \alpha_S$.
Then, 
$V_{\rm C+L}(r)$ becomes independent of
$\mu_f$.
In view of the construction of OPE of the QCD potential,
the ``Coulomb''+linear potential $V_{\rm C+L}(r)$
defined in this scheme can be qualified, without any problem, 
as a renormalized singlet potential (Wilson coefficient) 
defined in a specific scheme.
We remind the reader that the bare
singlet potential $V_S(r)$ coincides with the
perturbative expansion of $V_{\rm QCD}(r)$, and that
the renormalized singlet potential is defined by subtracting
IR renormalons and IR divergences from it.
According to the large-order analysis and the
definition (\ref{VCplusL})--(\ref{calCR}),
$V_{\rm C+L}(r)$ matches this requirement.
Furthermore, eq.~(\ref{result}) ensures
consistency of identifying $V_{\rm C+L}(r)$ as a
renormalized singlet potential.
$V_{\rm C+L}(r)$ is well-defined, systematically computable, and
free from ambiguities induced
by IR renormalons or IR divergences.
The only notable difference from the
ordinary OPE is that, in this scheme, 
$V_{\rm C+L}(r)$ is independent of the factorization scale $\mu_f$.
%Qualitatively, we have 
%subtracted only the IR renormalons from the double
%series expansion of $V_{\rm QCD}(r)$, without
%introducing an IR cutoff $\mu_f$.

Eq.~(\ref{result}) shows that $V_S^{\rm (R)}(r;\mu_f)$ 
approaches $V_{\rm C+L}(r)$ as we reduce  $\mu_f (\gg \LQ)$.
This matches our naive expectation:
$V_{\rm C+L}(r)$ is obtained 
from the bare $V_S(r)$ by subtracting 
the part corresponding to IR renormalons, which reside in
the region $q \sim \LQ$;
on the other hand, $V_S^{\rm (R)}(r;\mu_f)$ is obtained
by cutting off a larger domain $q<\mu_f$.

\subsection{\boldmath $\delta E_{\rm US}(r)$:
C+L scheme and factorization scheme}
\label{s4.3}

We may replace $V_S(r)$ in eq.~(\ref{deltaEUS}) by 
$V_{\rm C+L}(r)$ and $V_S^{({\rm R})}\!(r;\mu_f\!)$, respectively,
and define the non-perturbative contributions
$\delta E_{\rm US}(r)$ corresponding to both schemes.
Alternatively, \hspace{-1.5mm}\footnote{
As for $\delta E_{\rm US}(r)$ in the
factorization scheme, contributions from gluons 
close to the UV cutoff $q \sim \mu_f$ can be computed reliably
in expansion in $\alpha_S$ using eq.~(\ref{deltaEUS}) (although the
entire $\delta E_{\rm US}(r)$ cannot be computed reliably).
Then, one can show explicitly that the 
$\mu_f$-dependence of $V_S^{({\rm R})}\!(r;\mu_f\!)$
[cf.\ eq.~(\ref{muf-dep-VS})] are cancelled by the 
$\mu_f$-dependence of $\delta E_{\rm US}(r)$ \cite{Brambilla:1999xf},
showing consistency of defining $\delta E_{\rm US}(r)$
in two ways.
}
via eq.~(\ref{OPE}), we may identify them, respectively, with
$V_{\rm QCD}(r) - V_{\rm C+L}(r)$
and 
$V_{\rm QCD}(r) - V_S^{\rm (R)}(r;\mu_f)$.
Let us call the former as $\delta E_{\rm US}(r)$ in
the C+L scheme and the latter as $\delta E_{\rm US}(r)$ in
the factorization scheme.
As mentioned above, there are no intrinsic uncertainties
in $V_{\rm C+L}(r)$ and $V_S^{({\rm R})}\!(r;\mu_f\!)$, so that $\delta E_{\rm US}(r)$
in both schemes are unambiguously defined.
IR renormalons have been subtracted from the bare $V_S(r)$ and
absorbed into $\delta E_{\rm US}(r)$.
$\delta E_{\rm US}(r)$ in
the C+L scheme is independent of $\mu_f$, while $\delta E_{\rm US}(r)$ in
the factorization scheme depends on $\mu_f$.

According to the argument in Sec.~\ref{s2.4}, 
$r$-dependence of $\delta E_{\rm US}(r)$ in the
factorization scheme can be predicted.
(We always set $\mu_f \gg \LQ$ in the factorization scheme.)
It is either of order $\mu_f^4 r^3$ (very small $r$) or
of order $\mu_f^3 r^2$ (small $r$), depending on the
relation between $\Delta V(r)$ and $\mu_f$.
We expect the latter $r$-dependence in the 
distance region of our interest.
On the other hand, the $r$-dependence
of $\delta E_{\rm US}(r)$ in the
C+L scheme can be estimated in parallel with that of
$\delta E_{\rm US}(r)$ in the
factorization scheme with $\mu_f \sim \LQ$.
Therefore,
$r$-dependence in the
C+L scheme can be predicted for small $r$
(corresponding to $\Delta V(r) \gg \LQ$) to be order
$\LQ^4 r^3$.
If $r$ is not sufficiently small ($\Delta V(r) \sim \LQ$), 
precise $r$-dependence is not known.
Since, however, $\delta E_{\rm US}(r)$ in the C+L scheme
contains no other scale than $\LQ$, it should be
at most order $\LQ$ at $r \simlt \LQ^{-1}$.
Namely, it is
much smaller than $\delta E_{\rm US}(r)$ in the
factorization scheme (order $\mu_f^3 r^2$),
provided $\mu_f$ is
sufficiently large.
It means that $V_{\rm C+L}(r)$ is much closer to $V_{\rm QCD}(r)$
than $V_S^{({\rm R})}\!(r;\mu_f\!)$
in the distance region of our interest.
This gives us a good motivation to analyze $V_{\rm C+L}(r)$
in OPE, in addition to the more conventional factorization 
($\mu_f$-dependent) scheme.

All the above arguments are based on order-of-magnitude
estimates. 
We would like to make the statements clearer by 
making a quantitative analysis.

\section{Determinations of
\boldmath $\delta E_{\rm US}(r)$ and
$r_0 \, \Lambda_{\overline{\rm MS}}$}
\label{s5}
\clfn

In this section, we compare the singlet potentials, which
we defined in
different schemes in the previous section, and recent lattice data
for the static QCD potential.
According to previous analyses, 
(a) accuracy of the prediction
for $V_S^{\rm (R)}(r;\mu_f)$ or $V_{\rm C+L}(r)$
can be improved systematically;
(b)
the difference $\delta E_{\rm US}(r)=V_{\rm QCD}(r)-V_S^{\rm (R)}(r;\mu_f)$ 
[$V_{\rm QCD}(r)-V_{\rm C+L}(r)$]
is expected to be 
${\cal O}(\mu_f^3 r^2)$ 
[${\cal O}(\LQ^4r^3)$ at small distances, whereas precise form is
unknown at larger distances] and is non-perturbative.
We verify these properties numerically
by comparison to lattice computations of $V_{\rm QCD}(r)$.
Then we determine the size of the non-perturbative contribution
$\delta E_{\rm US}(r)$.
As a byproduct, we determine the relation between
$\Lambda_{\overline{\rm MS}}$ and lattice scale 
(Sommer scale) at the same time.

Throughout this section, we use 
lattice data in the quenched approximation, since in
this case 
lattice data are most accurate in the short-distance region.
In computations of $V_S^{\rm (R)}(r;\mu_f)$ and $V_{\rm C+L}(r)$:
we set $n_l=0$ accordingly; 
except where stated otherwise,
we take the input parameter 
as $\alpha_S(Q)=0.2$, which corresponds to\footnote{
As well known, when the strong coupling constant at some large
scale, e.g.\ $\alpha_S(m_b)$, is fixed, the values of
$\Lambda_{\overline{\rm MS}}^{\mbox{\scriptsize 1-loop}}$, 
$\Lambda_{\overline{\rm MS}}^{\mbox{\scriptsize 2-loop}}$, and
$\Lambda_{\overline{\rm MS}}^{\mbox{\scriptsize 3-loop}}$
differ substantially.
As a result, if we take a common value of $\Lambda_{\overline{\rm MS}}$ as the
input parameter, $V_{\rm C+L}(r)$ up to different orders differ
significantly at small distances,
where the predictions are supposed to be more accurate.
}
$\Lambda_{\overline{\rm MS}}^{\mbox{\scriptsize 1-loop}}/Q =0.057$,
$\Lambda_{\overline{\rm MS}}^{\mbox{\scriptsize 2-loop}}/Q =0.13$, 
$\Lambda_{\overline{\rm MS}}^{\mbox{\scriptsize 3-loop}}/Q =0.12$;
at NNNLL,
except where stated otherwise,
we use the estimate of $\bar{a}_3$ by Pineda in eq.~(\ref{a3}),
$\bar{a}_3=292\times4^3=18688$ \cite{Pineda:2002se}.
An arbitrary
$r$-independent constant has been added to each potential and 
each lattice data set to facilitate comparisons in the figures.
Methods for numerically evaluating $V_S^{\rm (R)}(r;\mu_f)$ and
$V_{\rm C+L}(r)$ are shown in App.~\ref{appD}.

We relate the
scale for each lattice data set to 
$\Lambda_{\overline{\rm MS}}$ in the following manner.
For each lattice data set we calculate 
(or use the given value of) the Sommer scale $r_0$ 
defined by \cite{Sommer:1993ce}
\bea  
r^2 \,{dV_{\rm QCD}\over dr} \bigg|_{r=r_0} = 1.65 . 
\label{def-Sommer}
\eea
(For reference to the real world, it is customary to
interpret $r_0 = 0.5~{\rm fm} \approx 2.5~{\rm GeV}^{-1}$.)
Then the lattice data are expressed in units of $r_0$.
In Sec~\ref{s5.1}, we convert the units into 
$\Lambda_{\overline{\rm MS}}^{\mbox{\scriptsize 3-loop}}$
using the central value of the relation
\bea
r_0 \,\Lambda_{\overline{\rm MS}}^{\mbox{\scriptsize 3-loop}}=0.602 \pm 0.048 ,
\label{r0Lambda}
\eea
as obtained by \cite{Capitani:1998mq}.
In contrast, in Sec.~\ref{s5.2}, we will not use the relation between
$r_0$ and $\Lambda_{\overline{\rm MS}}^{\mbox{\scriptsize 3-loop}}$
as an input
but rather determine this relation from a fit to the
data for $\delta E_{\rm US}(r)$.
We will explain the mechanism why this is possible.

\subsection{Consistency checks}
\label{s5.1}

Here, we verify various properties of the singlet potentials
$V_S^{\rm (R)}(r;\mu_f)$ and $V_{\rm C+L}(r)$ and of the
corresponding non-perturbative contributions.

\begin{figure}
\begin{center}
\hspace{-35mm}
\includegraphics[width=17cm]{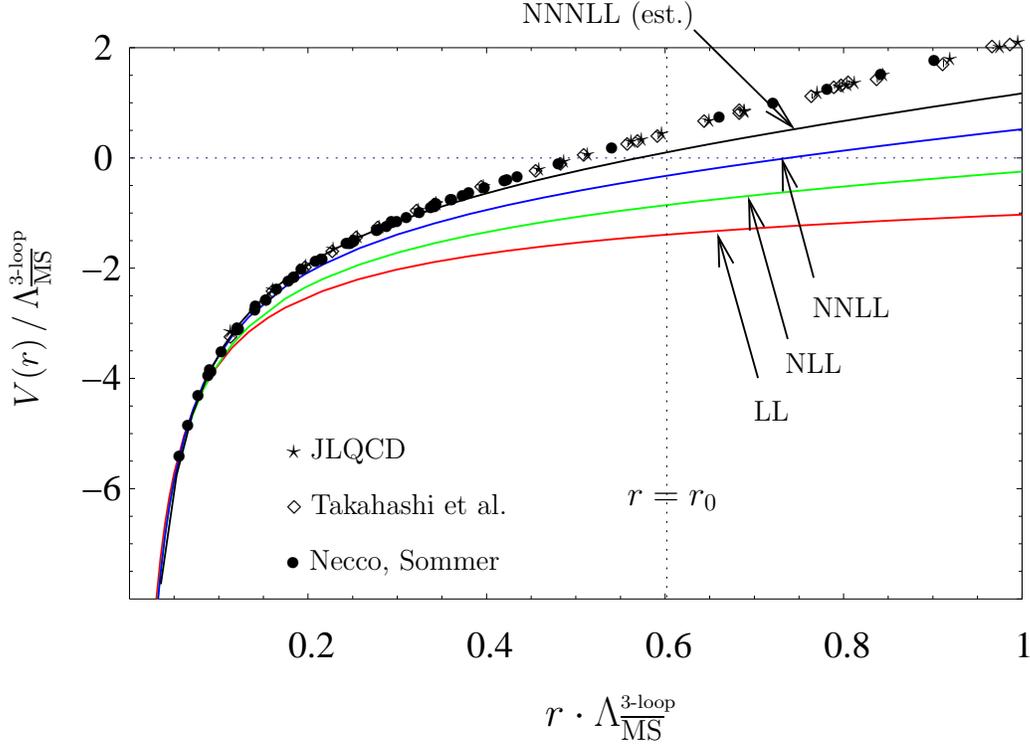}
\end{center}
\vspace*{-.5cm}
\caption{\small
Comparison of $V_{\rm C+L}(r)$ in the $\mu_f$-independent
scheme (solid lines) and the lattice data in
quenched approximation
[Takahashi et al.\ \cite{Takahashi:2002bw} ($\diamond$), 
Necco/Sommer \cite{Necco:2001xg} ($\bullet$), and JLQCD \cite{Aoki:2002uc} ($\star$)].
Input parameters for $V_{\rm C+L}(r)$ are $\alpha_S(Q)=0.2$
and $n_l=0$;
at NNNLL, Pineda's estimate for $\bar{a}_3$ is used.
\label{comp-lat}}
\end{figure}

First we compare the 
``Coulomb''+linear potential
$V_{\rm C+L}(r)$ up to different orders,
in the $\mu_f$-independent
scheme defined in Secs.~\ref{s4.1} and \ref{s4.2}.
Up to NNLL, they coincide with $V_C(r) + {\cal C} \, r$
of Sec.~\ref{s3.3}.
We also compare them with
lattice calculations of the QCD potential.
See Fig.~\ref{comp-lat}.\footnote{
If we use Chishtie-Elias's estimate of $\bar{a}_3$, the NNNLL
line in Fig.~\ref{comp-lat} hardly changes.
If we use the estimate of $\bar{a}_3$
by large-$\beta_0$ approximation,
the NNNLL line is located between the
present NNNLL line and NNLL line.
}
We see that $V_{\rm C+L}(r)$ up to different orders agree
well with one another
at small distances, whereas at large distances $V_{\rm C+L}(r)$
becomes steeper as we include higher-order terms via RG;
cf.\ Tab.~\ref{tab-sigma}.
\begin{table}
\begin{center}
\begin{tabular}{l|cccc}
\hline
& LL & NLL & NNLL & NNNLL\\
\hline
${\cal C}^{\rm (R)}/\Bigl(\Lambda_{\overline{\rm MS}}^{\mbox{\scriptsize 3-loop}}\Bigr)^2$
& 0.1836 & 0.6950 & 1.261 & 1.758 \\
\hline
\end{tabular}
\end{center}
\caption{\small
Coefficient of the linear potential [eq.~(\ref{calCR})] in units of
$\bigl(\Lambda_{\overline{\rm MS}}^{\mbox{\scriptsize 3-loop}}\bigr)^2$.
Conventions are same as in Fig.~\ref{comp-lat}.
\label{tab-sigma}}
\end{table}
This feature is in accordance with the qualitative understanding 
within perturbative QCD, in which the potential
becomes steeper due to the running of the strong coupling
constant,
since $\alpha_V^{\rm PT}(q)$ increases more rapidly
at IR as we include higher-order terms.
The lattice data and $V_{\rm C+L}(r)$ also agree well
at small distances,
while they deviate at larger distances.
More terms we include in $V_{\rm C+L}(r)$, up to larger
distances the potential agrees with the lattice data.\footnote{
It is worth noting that the NNLL line in Fig.~\ref{comp-lat}
is numerically very close to the NNLO prediction obtained with principle of minimal
sensitivity in \cite{Recksiegel:2002um}.
}
Theoretically, we expect $V_{\rm C+L}(r)$ to converge
as we increase the order.
The current status seems to be consistent with this expectation,
since the lines in Fig.~\ref{comp-lat} apparently converge
to the lattice data.
One may adjust the input value $\alpha_S(Q)$ such that
convergence becomes fast at some particular $r$;
present choice $\alpha_S(Q)=0.2$ leads to
a fast convergence at 
$r\Lambda_{\overline{\rm MS}}^{\mbox{\scriptsize 3-loop}}
<0.1$.
We may increase the value of input $\alpha_S(Q)$ such 
that convergence becomes fast 
at larger $r$.
Then, $V_{\rm C+L}(r)$ up to different orders come closer 
to one another at $r \,
\Lambda_{\overline{\rm MS}}^{\mbox{\scriptsize 3-loop}} > 0.1$.
[The relation between $V_{\rm C+L}(r)$ up to NNLL
and the lattice data remains unchanged, since we use the
3-loop RG relation to fix the lattice scale, i.e.\
this relation is invariant under 3-loop RG evolution.]

\begin{figure}
\begin{center}
\hspace{-35mm}
\includegraphics[width=19cm]{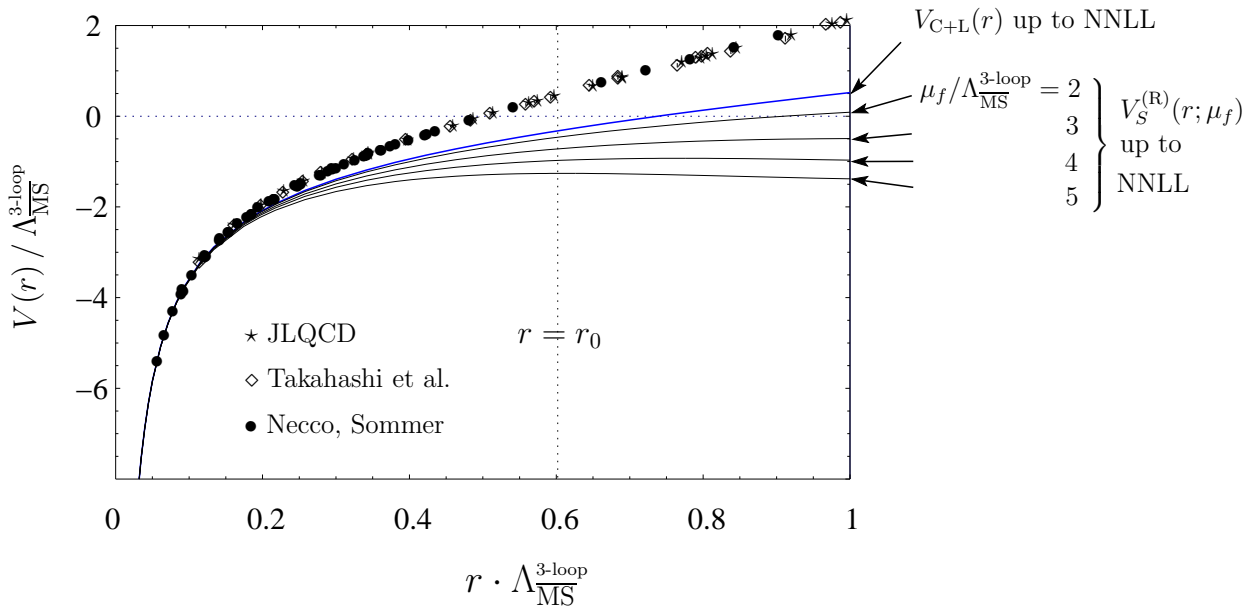}
\end{center}
\vspace*{-.5cm}
\caption{\small
Comparison of lattice data and
$V_S^{\rm (R)}(r;\mu_f)$
up to NNLL for 
$\mu_f/
\Lambda_{\overline{\rm MS}}^{\mbox{\scriptsize 3-loop}}
=2$ to 5 (downwards).
As a reference, we also plot $V_{\rm C+L}(r)$ up to NNLL.
The lattice data and parameters
for $V_S^{\rm (R)}(r;\mu_f)$ and $V_{\rm C+L}(r)$
are same as in Fig.~\ref{comp-lat}.
\label{comp-lat-2}}
\end{figure}

Next we compare the renormalized singlet potential
$V_S^{\rm (R)}(r;\mu_f)$ for different
values of the factorization scale $\mu_f$.
In Fig.~\ref{comp-lat-2}, we plot $V_S^{\rm (R)}(r;\mu_f)$
up to NNLL for 
$\mu_f/
\Lambda_{\overline{\rm MS}}^{\mbox{\scriptsize 3-loop}}
=2$,3,4,5.
Note that, up to NNLL, there is no distinction between the
first scheme and the second scheme of Sec.~\ref{s4.1}.
We see that $V_{\rm C+L}(r)$ is located closer
to the lattice data than $V_S^{\rm (R)}(r;\mu_f)$ 
for all $\mu_f$.
As we vary $\mu_f$, the variation of $V_S^{\rm (R)}(r;\mu_f)$ is
larger at larger distances.
As we lower $\mu_f$, $V_S^{\rm (R)}(r;\mu_f)$
approaches $V_{\rm C+L}(r)$.
These features are in agreement with the argument given 
in Sec.~\ref{s4.3}.
Note that we cannot
lower $\mu_f$ below the Landau singularity
$q_*^{\mbox{\scriptsize 3-loop}} \approx 1.53\,
\Lambda_{\overline{\rm MS}}^{\mbox{\scriptsize 3-loop}}$.
$V_S^{\rm (R)}(r;\mu_f)$ up to different orders behave
similarly:
As $\mu_f$ is lowered, $V_S^{\rm (R)}(r;\mu_f)$ is
raised at larger distances, approaching $V_{\rm C+L}(r)$
up to the corresponding order.
For fixed $\mu_f$, $V_S^{\rm (R)}(r;\mu_f)$ agrees with
lattice data up to larger distances as we increase the order.

\begin{figure}
\begin{center}
%\hspace{-35mm}
\begin{tabular}{cc}
\psfrag{LL}{\footnotesize  LL}
\psfrag{NLL}{\hspace{2mm}\footnotesize  NLL}
\psfrag{NNLL}{\footnotesize NNLL}
\psfrag{NNNLL}{\hspace{0mm}\footnotesize NNNLL}
\psfrag{Large-beta0}{\hspace{-3mm}\footnotesize large-$\beta_0$ appr.}
\psfrag{r0}{\hspace{-6mm} $r=r_0$}
\psfrag{delEUS}{\hspace{-20mm} $\delta E_{\rm US}(r)/
\Lambda_{\overline{\rm MS}}^{\mbox{\scriptsize 3-loop}}$~~ (C+L scheme)}
\psfrag{r}
{\hspace{-3mm}
$r \cdot \Lambda_{\overline{\rm MS}}^{\mbox{\scriptsize 3-loop}}$}
\includegraphics[width=7.8cm]{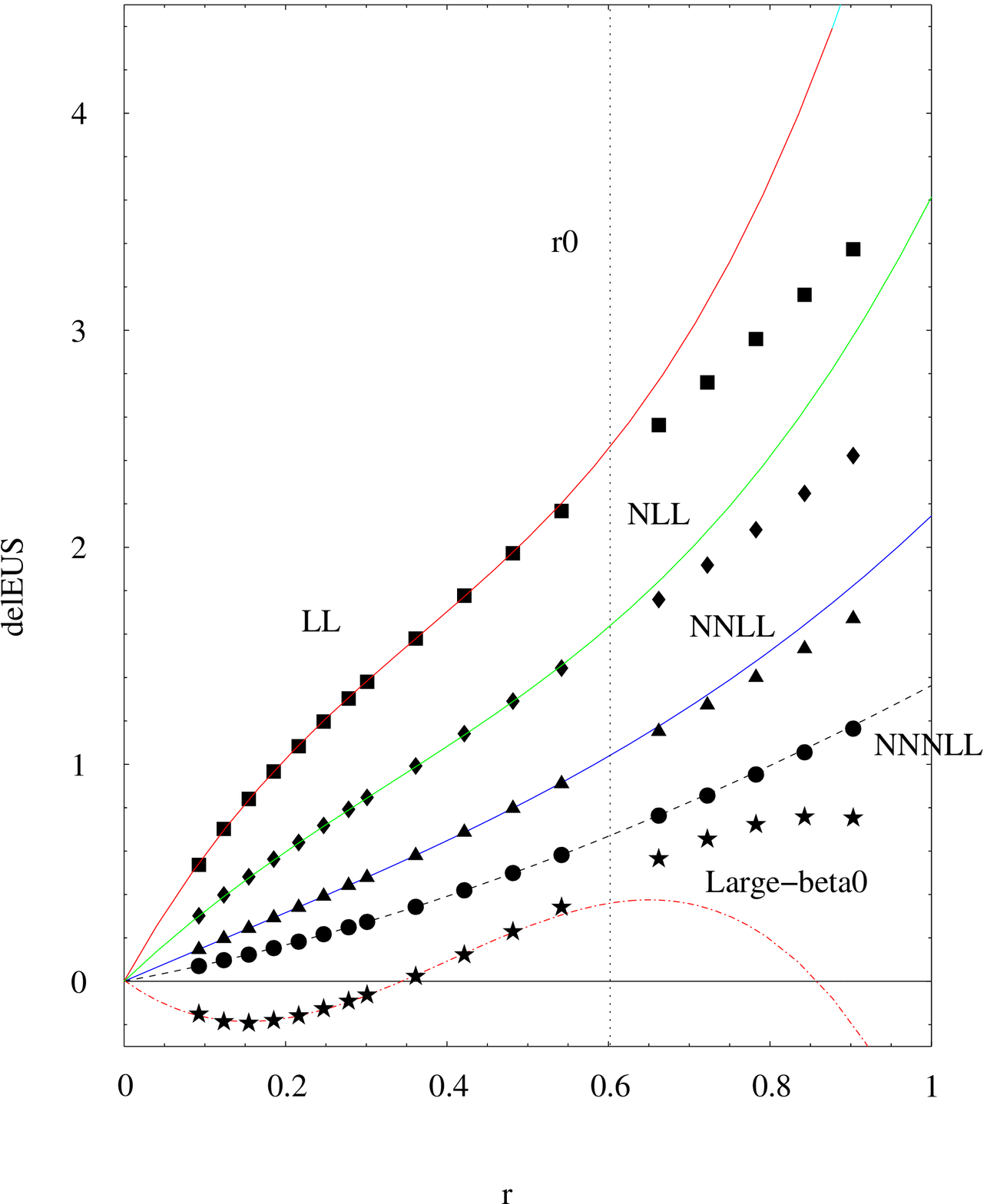}
&
\hspace{8mm}
\psfrag{NNLL, muf=3}{\hspace{-12mm}\footnotesize NNLL, $\mu_f=3\,
\Lambda_{\overline{\rm MS}}^{\mbox{\tiny 3-loop}}$}
\psfrag{NNLL, muf=5}{\hspace{-12mm}\footnotesize NNLL, $\mu_f=5\,
\Lambda_{\overline{\rm MS}}^{\mbox{\tiny 3-loop}}$}
\psfrag{NNNLL, muf=3}{\hspace{-1mm}\footnotesize NNNLL, $\mu_f=3\,
\Lambda_{\overline{\rm MS}}^{\mbox{\tiny 3-loop}}$}
\psfrag{NNNLL, muf=5}{\hspace{-1mm}\footnotesize NNNLL, $\mu_f=5\,
\Lambda_{\overline{\rm MS}}^{\mbox{\tiny 3-loop}}$}
\psfrag{r0}{\hspace{-6mm} $r=r_0$}
\psfrag{delEUS}{\hspace{-30mm} $\delta E_{\rm US}(r)/
\Lambda_{\overline{\rm MS}}^{\mbox{\scriptsize 3-loop}}$~~(factorization scheme)}
\psfrag{r}
{\hspace{-2mm}
$r \cdot \Lambda_{\overline{\rm MS}}^{\mbox{\scriptsize 3-loop}}$}
\raise0.5mm\hbox{
\includegraphics[width=7.7cm]{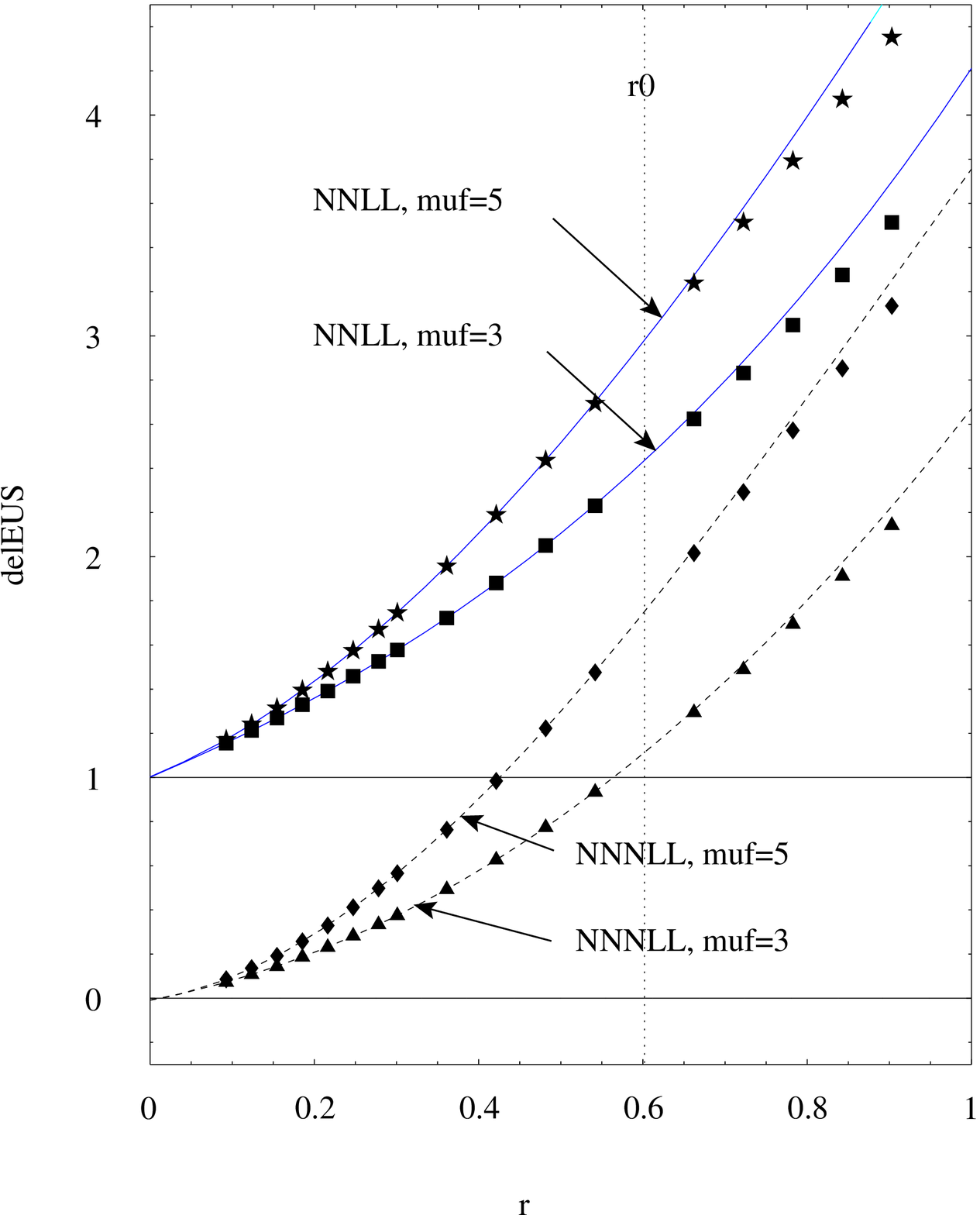}
}
\end{tabular}
\end{center}
\vspace*{0mm}
\hspace{45mm}(a)\hspace{88mm}(b)
\vspace*{0cm}
\caption{\small
$\delta E_{\rm US}(r)/
\Lambda_{\overline{\rm MS}}^{\mbox{\scriptsize 3-loop}}$ vs.\
$r \, \Lambda_{\overline{\rm MS}}^{\mbox{\scriptsize 3-loop}}$.
The lattice data from Tab.~2 of \cite{Necco:2001xg} are used.
Parameters for $V_{\rm C+L}(r)$ and $V_S^{\rm (R)}(r;\mu_f)$
are same as in Fig.~\ref{comp-lat}.
Lines represent fits to the data points in the range
$r \, \Lambda_{\overline{\rm MS}}^{\mbox{\scriptsize 3-loop}}<0.5$
by third-order polynomials,
given explicitly in Tab.~\ref{tab-fitfn}.
(a) C+L scheme, up to LL, NLL, NNLL and NNNLL;
C+L scheme in the large-$\beta_0$ approximation
is also shown.
(b) first scheme within the factorization scheme
(subtraction of IR divergences by $\overline{\rm MS}$ scheme);
for display purposes, 
$\delta E_{\rm US}(r)/
\Lambda_{\overline{\rm MS}}^{\mbox{\scriptsize 3-loop}}$ 
at NNLL are shifted
by +1 vertically.
\label{delEUS}}
\end{figure}

We turn to the measurement of 
$\delta E_{\rm US}(r)$.
%and 
%$\delta E_{\rm US}(r;\mu_f) = V_{\rm QCD}(r) - V_S^{\rm (R)}(r;\mu_f)$
%[factorization scheme].
In computing $\delta E_{\rm US}(r)$, we 
use the lattice data from Table 2 of
\cite{Necco:2001xg} for a non-perturbative computation of $V_{\rm QCD}(r)$,
since this data set seems to be most accurate at short distances.
In Fig.~\ref{delEUS}(a), we plot $\delta E_{\rm US}(r)$ in
the C+L scheme
[$V_{\rm QCD}(r)-V_{\rm C+L}(r)$]
in units of
$\Lambda_{\overline{\rm MS}}^{\mbox{\scriptsize 3-loop}}$.
The errors of the data points, due to the errors of the lattice
data for $V_{\rm QCD}(r)$, are comparable to or smaller than the sizes of
the symbols used for the plot.
Also shown in the same figure are fits to the data points
of the form $A_1 \, \rho + A_2\, \rho^2 + A_3\, \rho^3$, where
$\rho = r \, \Lambda_{\overline{\rm MS}}^{\mbox{\scriptsize 3-loop}}$.
Only the data points in the range
$r \, \Lambda_{\overline{\rm MS}}^{\mbox{\scriptsize 3-loop}}<0.5$
were used for the fits.
We have added $r$-independent constants such that
all the fits go through the origin.
As we increase the order, $\delta E_{\rm US}(r)$ becomes
smaller.
We see that the cubic fit becomes a better approximation
in a wider range in
$r \, \Lambda_{\overline{\rm MS}}^{\mbox{\scriptsize 3-loop}}<1$
as we increase the order.
[Here, we determine $\delta E_{\rm US}(r)$
in expansion in $r$, hence,
we fit the data in the small
$r$ region
($r \, \Lambda_{\overline{\rm MS}}^{\mbox{\scriptsize 3-loop}}<0.5$);
it helps to enhance sensitivity to the coefficients of
$r^n$ for small $n$.
Since 
$\delta E_{\rm US}(r) \sim {\cal O}(\LQ^4r^3)$ at
sufficiently small $r$, i.e.\ $\delta E_{\rm US}(r) \to 0$ 
as $r \to 0$, it would make sense to perform a polynomial
fit;
however, in the next subsection, we reconsider
this naive picture and give a more complete
analysis.]

We can learn more detailed features from
the explicit polynomials obtained from the fits,
shown in Tab.~\ref{tab-fitfn}.
%Generally, quality of the fits by cubic polynomials becomes better as
%we include higher-order corrections;
%at each order,
%the coefficient of the linear term is more stable than 
%that of quadratic term, which is more stable than the
%coefficient of the cubic term.
%These features can be verified by changing the number of
%data points used for the fits or by changing the order of
%the polynomial fits.
%For instance, in this way, at NNNLL in the C+L scheme
%in Tab.~\ref{tab-fitfn}, 
%we estimate the error of the coefficient of the linear term to be
%about 5\%, 
%the error of the coefficient of the quadratic term to be
%about 10\%;
%at NNLL in the C+L scheme,  
%we estimate the error of the coefficient of the linear term to be
%about 10\%, 
%the error of the coefficient of the quadratic term to be
%about 300\%.
\begin{table}
\begin{tabular}{l|lll}
\hline
& \hspace{10mm}C+L scheme & factorization scheme & factorization scheme \\
&&\hspace{5mm}($\mu_f=3\Lambda_{\overline{\rm MS}}^{\mbox{\tiny 3-loop}}$)
&\hspace{5mm}($\mu_f=5\Lambda_{\overline{\rm MS}}^{\mbox{\tiny 3-loop}}$)
\\
\hline
LL & ~~\,$6.7 \, \rho - 9.6 \, \rho^2 + 8.7 \, \rho^3$ &  & \\
NLL & ~~\,$3.5 \, \rho - 3.6 \, \rho^2+ 3.6 \, \rho^3$ & & \\
NNLL & ~~\,$1.6 \, \rho - 0.3 \, \rho^2 + 0.8 \, \rho^3$ & 
$1.6\,\rho + 0.9\,\rho^2+0.8\,\rho^3$
& $1.5\,\rho +3.5\,\rho^2-0.8\,\rho^3$ \\
NNNLL, 1st scheme &  & $0.7\, \rho + 1.8\,\rho^2 + 0.1\,\rho^3$
 & $0.6\,\rho+5.0\,\rho^2-1.8\,\rho^3$ \\
NNNLL, 2nd scheme & ~~\,$0.7 \, \rho + 0.8 \, \rho^2 - 0.1 \, \rho^3$ 
& $0.7\, \rho + 2.0\,\rho^2 - 0.0\,\rho^3$
 & $0.6\,\rho+5.1\,\rho^2-1.9\,\rho^3$ \\
Large-$\beta_0$ appr. & $-2.7\,\rho+10.9\,\rho^2-9.1\,\rho^3$ & & \\
\hline
\end{tabular}
\caption{\small
Fits of $\delta E_{\rm US}(r)/
\Lambda_{\overline{\rm MS}}^{\mbox{\tiny 3-loop}}$ 
by cubic polynomials in the region $\rho < 0.5$, where
$\rho=r  \Lambda_{\overline{\rm MS}}^{\mbox{\tiny 3-loop}}$.
(See Fig.~\ref{delEUS} for plots.)
The lattice data \cite{Necco:2001xg} are used.
Parameters for $V_{\rm C+L}(r)$ and $V_S^{\rm (R)}(r;\mu_f)$
are same as in Fig.~\ref{comp-lat}.
\label{tab-fitfn}}
\end{table}
The coefficient of the linear potential decreases as we increase
the orders, from LL to NNNLL.
Up to NNNLL, the coefficient in
units of $(\Lambda_{\overline{\rm MS}}^{\mbox{\tiny 3-loop}})^2$ is
about 0.7.
We may compare this value with the string tension
(coefficient of linear potential) extracted from the
large-distance behavior of the lattice data
$\sigma \approx 3.8\,
(\Lambda_{\overline{\rm MS}}^{\mbox{\tiny 3-loop}})^2$ \cite{Necco:2001xg}.
Thus, the linear potential in $\delta E_{\rm US}(r)$ is quite small
comparatively at our current best
knowledge.
We are interested in $\delta E_{\rm US}(r)$ in
the infinite-order limit.
Up to our current best knowledge, however,
$\delta E_{\rm US}(r)$ in the C+L scheme
has not stabilized yet,
although we see a tendency that it approaches a quadratic form.
We conclude that the present status is consistent with
$\delta E_{\rm US}(r)$ in the infinite-order limit 
being order $\LQ^3 r^2$ at 
$r \,\Lambda_{\overline{\rm MS}}^{\mbox{\scriptsize 3-loop}}\simlt 1$
(vanishing linear potential).
From Figs.~\ref{comp-lat} and \ref{delEUS}(a), it is not
clear whether the limit would also be consistent with 
order $\LQ^4 r^3$ or with zero.
We will clarify this point quantitatively in the
next subsection.

In Fig.~\ref{delEUS}(b) are plotted $\delta E_{\rm US}(r)$
in the first scheme (IR divergences of $\alpha_V^{\rm PT}(q)$ 
are subtracted in 
$\overline{\rm MS}$ scheme) within the factorization scheme
[$V_{\rm QCD}(r)-V_S^{\rm (R)}(r;\mu_f)$].
In this figure and in Tab.~\ref{tab-fitfn}, we see that
$\delta E_{\rm US}(r;\mu_f)$ approximate the
quadratic form ${\cal O}(\mu_f^3 r^2)$ expected from OPE.
Approximation to the quadratic form is more evident 
than in the C+L scheme, since the coefficents of the
quadratic terms are larger.
Differences between the first and second scheme 
in the factorization scheme are
tiny:
If we plot, in the same figure, $\delta E_{\rm US}(r)$
in the second scheme (IR divergences of $\alpha_V^{\rm PT}(q)$ 
are regularized by
double expansion in $\alpha_S$ and $\log \alpha_S$),
they are hardly distinguishable from the lines of first
scheme.
This is also confirmed by the explicit forms of the
polynomial fits in Tab.~\ref{tab-fitfn}.

We may compare the fits of $\delta E_{\rm US}(r)$
in the factorization
scheme and those in the C+L scheme in Tab.~\ref{tab-fitfn}.
The coefficients of linear terms are almost
same in both schemes, up to NNLL and up to
NNNLL, respectively.
This confirms consistency with eq.~(\ref{result}).
The difference of the coefficients of quadratic terms
between both schemes is expected to be proportional
to $\mu_f^3$ in the limit $\mu_f \gg \LQ$,
according to eq.~(\ref{result}).
This relation is roughly satisfied in Tab.~\ref{tab-fitfn} as well.\footnote{
This is not a test of eq.~(\ref{result}) or eq.~(\ref{proof});
this is a test of the quality of the cubic fits.
A comparison with the direct computation of eq.~(\ref{proof})
in Tab.~\ref{tab-fitfn2} indicates that the coefficients of
the linear terms
are determined with good accuracy, whereas the coefficients
of the quadratic terms are determined with about 20\% accuracy.
}

\begin{figure}
\begin{center}
\hspace{-35mm}
\includegraphics[width=15cm]{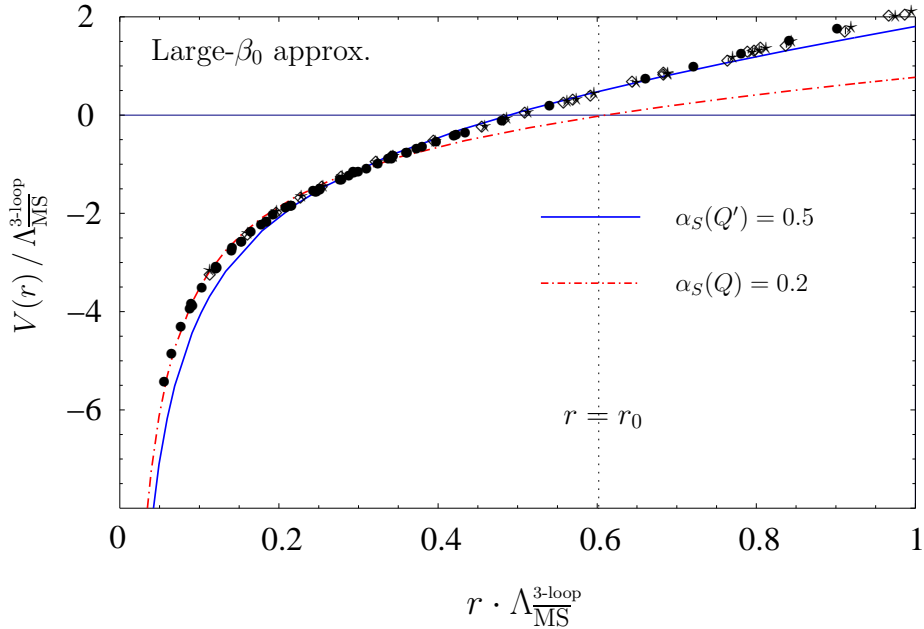}
\end{center}
\vspace*{-.5cm}
\caption{\small
Comparison of $V_{\rm C+L}(r)$ in
the large-$\beta_0$ approximation and lattice data.
Inputs for $V_{\rm C+L}(r)$ are $\alpha_S(Q)=0.2$ and $\alpha_S(Q')=0.5$.
Other conventions are same as in Fig.~\ref{comp-lat}.
\label{comp-lat-large-beta0}}
\end{figure}

Next we compare $V_{\rm C+L}(r)$ in the
large-$\beta_0$ approximation,\footnote{
Here, we mean that the entire
$V_{\rm C+L}(r)$ is evaluated in the
large-$\beta_0$ approximation, i.e.\
we set $a_n=(5\beta_0/3)^n$ in eq.~(\ref{alfVRG})
and all $\beta_n=0$ except $\beta_0$ in eq.~(\ref{RGeq}).
This should not be confused with $V_{\rm C+L}(r)$ up to 
NNNLL where (only) $\bar{a}_3$ is evaluated in the 
large-$\beta_0$ approximation.
} 
analyzed in Sec.~\ref{s3.2}, with
the lattice data.
We take two different values of the input parameter:
$\alpha_S(Q)=0.2$ and $\alpha_S(Q')=0.5$, which
correspond to 
$Q/\Lambda_{\overline{\rm MS}}^{\mbox{\scriptsize 1-loop}} =17.4$
and
$Q'/\Lambda_{\overline{\rm MS}}^{\mbox{\scriptsize 1-loop}}=3.1$,
respectively.
See Fig.~\ref{comp-lat-large-beta0}.
Since the large-$\beta_0$ approximation incorporates
only 1-loop running of $\alpha_S$, the prediction
for $V_{\rm C+L}(r)$ is
fairly scale dependent.\footnote{
c.f.\ The scale dependence of
$V_{\rm C+L}(r)$ up to NNNLL is by far smaller.
}
Nevertheless, considering the crudeness of the approximation,
agreement of the prediction with the lattice data 
is remarkably good.
$V_{\rm C+L}(r)$
with $\alpha_S(Q)=0.2$ in Fig.~\ref{comp-lat-large-beta0}
is much closer to the lattice data
than $V_{\rm C+L}(r)$ at LL in Fig.~\ref{comp-lat}.
On the other hand, a more detailed examination reveals
limitations of the large-$\beta_0$ approximation.
In Fig.~\ref{delEUS}, we plot
$\delta E_{\rm US}(r)$ computed from $V_{\rm C+L}(r)$
in the large-$\beta_0$ approximation with $\alpha_S(Q)=0.2$;
a cubic fit using the data points at
$r  \Lambda_{\overline{\rm MS}}^{\mbox{\tiny 3-loop}}<0.5$
is shown in the same figure and in Tab.~\ref{tab-fitfn}.
Although the magnitude of $\delta E_{\rm US}(r)$ is
small, $\delta E_{\rm US}(r)$ is oscillatory.
The sizes of the coefficients of the polynomial fit are 
unnaturally large, and the fit
does not reproduce $\delta E_{\rm US}(r)$ beyond
$r  \Lambda_{\overline{\rm MS}}^{\mbox{\tiny 3-loop}}=0.5$.
Since there is no reason to believe that the 
large-$\beta_0$ approximation is very close to the 
infinite-order limit, 
we consider the different behaviors between $\delta E_{\rm US}(r)$
in this case and that of RG analysis (LL -- NNNLL)
to be an indication that 
the consistency checks performed in this subsection
have sensitivity to the details.

\subsection{Determinations of $ r_0\Lambda_{\overline{\rm MS}}$
and \boldmath $\delta E_{\rm US}(r)$
in C+L scheme
}
\label{s5.2}
\clfn

From the above analysis, we conclude that all the theoretical
expectations derived from OPE are either positively confirmed or
consistent with lattice data
within the present level of uncertainties.
In this subsection, with the 
aid of theoretical predictions of OPE, we estimate the
infinite-order limit of
$\delta E_{\rm US}(r)$ in the C+L scheme.\footnote{
We note that the analysis in this subsection requires a rather
high accuracy in the numerical evaluations of $V_{\rm C+L}(r)$.
}

Up to this point, we used the central value (0.602) of the
relation between
$r_0$ and $\Lambda_{\overline{\rm MS}}^{\mbox{\scriptsize 3-loop}}$
in eq.~(\ref{r0Lambda}).
Now we examine how our determination of $\delta E_{\rm US}(r)$
will be affected if we vary this relation.
%We can show that indeed our fit has a good
%sensitivity to this relation.
To simplify the argument, for the moment
let us suppose that the lattice data set 
(when expressed in units of $r_0$) has no errors
and $V_{\rm C+L}(r)$ in the infinite-order limit is known.
Let $V_{\rm latt}(r;x)$
represent the lattice data set converted to units of 
$\Lambda_{\overline{\rm MS}}^{\mbox{\scriptsize 3-loop}}$
using a given value of 
$x \equiv r_0\Lambda_{\overline{\rm MS}}^{\mbox{\scriptsize 3-loop}}$.
Then, if $x$ equals
its precise value $x_{\rm true}$, according to OPE,
$\delta E_{\rm US}(r) = 
V_{\rm latt}(r;x_{\rm true})
-V_{\rm C+L}(r)$
goes to zero as $r \to 0$ (ignoring an $r$-independent constant), 
hence, it would be approximated
reasonably well by a polynomial of $r$.
When
$x$ 
differs from $x_{\rm true}$, we thus expect
\bea
&&
 V_{\rm diff}(r;x) \equiv 
V_{\rm latt}(r;x)
-V_{\rm C+L}(r)
\nonumber
\\ &&
~~~~~~ ~~~~~~~\,
=
V_{\rm latt}(r;x_{\rm true})
-V_{\rm C+L}(r) + 
V_{\rm latt}(r;x)
-V_{\rm latt}(r;x_{\rm true})
\nonumber
\\ &&
~~~~~~ ~~~~~~~\,
\approx
P(r) + \Delta (r;x,x_{\rm true}) ,
\label{delEUS-vary-r0Lambda}
\eea
where $P(r)$ is a polynomial of $r$, and
\bea
\Delta (r;x,x')
\equiv
V_{\rm latt}(r;x)
-V_{\rm latt}(r;x') .
\eea
If we know $V_{\rm latt}(r;x)$ for some value of $x$, we can
find $V_{\rm latt}(r;x)$ for other values of $x$ via
\bea
V_{\rm latt}(r;x') = V_{\rm latt}\left(r\frac{x}{x'};x\right)\,
\frac{x}{x'} .
\eea
On the other hand, we know that $V_{\rm latt}(r;x)$
tends to~ ${\rm const.}\times (r \log |r\Lambda_{\overline{\rm MS}}| )^{-1}$
at short distances, while it tends to a linear potential at
large distances.
Then, one readily finds that 
$
\Delta (r;x,x_{\rm true})
$
has approximately a ``Coulomb''+linear form
at $r\, \Lambda_{\overline{\rm MS}}^{\mbox{\scriptsize 3-loop}} \simlt 1$ if
$x \neq x_{\rm true}$.

In fact, by varying 
$x=r_0\Lambda_{\overline{\rm MS}}^{\mbox{\scriptsize 3-loop}}$
within the range given by eq.~(\ref{r0Lambda}),
we find a very good fit of $V_{\rm diff}(r;x)$:
\bea
&&
V_{\rm diff}(r;x)
\approx
\Delta (r;x,0.596)
+ ( 0.8 \, \rho + 0.7 \, \rho^2 )\,
\Lambda_{\overline{\rm MS}}^{\mbox{\scriptsize 3-loop}}
\biggr|_{\rho = 
r\Lambda_{\overline{\rm MS}}^{\mbox{\tiny 3-loop}}
} .
~~~~~\mbox{[NNNLL, $\bar{a}_3$(Pineda)]}
\nonumber\\
\label{fitNNNLLPineda}
\eea
Inclusion of $\Delta(r;x,x')$ into the fitting function 
stabilizes the fit considerably when
$r_0\Lambda_{\overline{\rm MS}}^{\mbox{\tiny 3-loop}}$
is far from its optimal value (0.596):
The coefficients of the linear and quadratic terms are
much less affected
even if we include cubic term in the fitting function or even if
we use data points up to larger $r$ for the fit,
whereas they are very unstable without $\Delta(r;x,x')$
in the fitting function.
A further examination shows that it is the inclusion of a
Coulombic term (it does not matter very much whether
log corrections are included or not) that
stabilizes the fit when 
$r_0\Lambda_{\overline{\rm MS}}^{\mbox{\tiny 3-loop}}$
is far from its optimal value.
In principle we should have included $\Delta(r;x,x')$
in the fitting function in the 
previous subsection.
A posteriori, it is because the central value of eq.~(\ref{r0Lambda})
happened to be close to the optimal value in eq.~(\ref{fitNNNLLPineda})
that polynomial fits were relatively stable (in particular as we increase the
order) and 
all the features seemed quite consistent with the theoretical expectations.

We note that the error of the input
$r_0\Lambda_{\overline{\rm MS}}^{\mbox{\scriptsize 3-loop}}$
in eq.~(\ref{r0Lambda})
is sizable with regard to the accuracy of
$V_{\rm C+L}(r)$ 
in our analysis.
Indeed,  in \cite{Pineda:2002se}, the error of the input 
$r_0\Lambda_{\overline{\rm MS}}^{\mbox{\scriptsize 3-loop}}$
was the largest source of errors in the determination of
$\delta E_{\rm US}(r)$.
Conversely, this means that our analysis has a sensitivity to determine
the relation between $r_0$ and
$\Lambda_{\overline{\rm MS}}^{\mbox{\scriptsize 3-loop}}$
by itself (and possibly reduce the error of
$\delta E_{\rm US}(r)$ at the same time).
Hence, in our determination of
$\delta E_{\rm US}(r)$, we will determine the value of
$x=r_0\Lambda_{\overline{\rm MS}}^{\mbox{\scriptsize 3-loop}}$
simultaneously, by performing a fit to the data for
$V_{\rm diff}(r;x) = V_{\rm latt}(r;x)-V_{\rm C+L}(r)$.

We approximate
$\delta E_{\rm US}(r)$ by quadratic function
$(A_0+A_1\,\rho + A_2\, \rho^2 )\Lambda_{\overline{\rm MS}}^{\mbox{\tiny 3-loop}}$
at $\rho =r\,\Lambda_{\overline{\rm MS}}^{\mbox{\scriptsize 3-loop}}
\simlt 0.5$.
As for errors, we take into account 
three types of sources:
(i) errors of the lattice data,
(ii) error of $\bar{a}_3$, and
(iii) error due to higher-order corrections.
Then we compute the probability density distribution for the
parameters 
$(A_0, A_1, A_2, x)$
in the following way.
Define
\bea
&&
\chi_V^{\, 2}
= \sum_{i=1}^{11} \left[
\frac{V_{\rm latt}(r_i;x)
-\{ V_{\rm C+L}(r_i;s) + t \, \delta V_{\rm C+L}(r_i) \}
-(A_0+A_1\,\rho_i + A_2\, \rho_i^2 )\Lambda_{\overline{\rm MS}}^{\mbox{\tiny 3-loop}}}
{\delta^V_i(x)}
\right]^2 ,
\rule[-7mm]{0mm}{6mm}
\label{chisq}
\\ &&
P^V(A_0, A_1, A_2, x) = {\cal N}_V^{\, -1}
\int ds \,dt\,\, e^{- \chi_V^2 /2} \, P_s(s) \, P_t (t) ,
\eea
where
$x=r_0\,\Lambda_{\overline{\rm MS}}^{\mbox{\scriptsize 3-loop}}$ and
$\rho_i = r_i \, \Lambda_{\overline{\rm MS}}^{\mbox{\scriptsize 3-loop}}$.
The normalization constant ${\cal N}_V$ is chosen such that the integral
of $P^V(A_0, A_1, A_2, x)$ over the entire range is unity.
$s$ and $t$ parametrize errors of the theoretical
prediction for $V_{\rm C+L}(r)$.
Details are as follows.
\begin{enumerate}
\renewcommand{\labelenumi}{(\roman{enumi})}

\item
We use the first 11 lattice data points given in Tab.~2 of \cite{Necco:2001xg}.
$r_i = r_i(x)$
denotes the distance $r$ of the $i$-th
lattice data point (given originally in units
of $r_0$ and $r_c$) after conversion to units of
$\Lambda_{\overline{\rm MS}}^{\mbox{\scriptsize 3-loop}}$, using\footnote{
We fix $r_c/r_0=0.5133$ \cite{Necco:2001xg};
its error is small, which we neglect in our analysis.
}
$x=r_0\,\Lambda_{\overline{\rm MS}}^{\mbox{\scriptsize 3-loop}}$;
$\delta^V_i(x)$ denotes the error of the $i$-th
lattice data point (given originally in units
of $r_0$) after conversion to units of
$\Lambda_{\overline{\rm MS}}^{\mbox{\scriptsize 3-loop}}$.
The first 11 data points correspond to
$\rho_i < 0.5$
when $x=0.602$.

\item
$V_{\rm C+L}(r;s)$ denotes $V_{\rm C+L}(r)$ up to NNNLL
evaluated with
$\bar{a}_3 = s \times \bar{a}_3({\rm Pineda})$.
We scan $s$ between 0 and 2 with equal weight,
i.e.\ $P_s(s)=1/2$ if $0 \leq s \leq 2$ and $P_s(s)=0$ otherwise.
The interval $0\leq s \leq2$ covers within its range 
the estimates of $\bar{a}_3$
by Pineda \cite{Pineda:2002se}, by Chishtie-Elias \cite{Chishtie:2001mf} and by 
large-$\beta_0$ approximation.

\item
$t \, \delta V_{\rm C+L}(r)$ represents an estimate of the difference between
$V_{\rm C+L}(r)$ up to NNNLL and $V_{\rm C+L}(r)$ in the infinte-order
limit.
$\delta V_{\rm C+L}(r)$ is estimated by the difference between $V_{\rm C+L}(r)$ up to NNLL
and that up to NNNLL.
We scan $t$ between $-1$ and 1 with equal weight,
i.e.\ $P_t(t)=1/2$ if $|t|\leq 1$ and $P_t(t)=0$ otherwise.

\end{enumerate}

Alternatively we may use the QCD force $F_{\rm QCD}(r) = dV_{\rm QCD}/dr$ 
in the determination of $\delta E_{\rm US}(r)$ and $x$.
Since we are not interested in the $r$-independent part of the
potential, we can extract information on the relevant parameters
using the force as well.
Similarly to before, we define
\bea
&&
\chi_F^{\, 2}
= \sum_{i=1}^{10} \left[
\frac{F_{\rm latt}(r_i;x)
-\{ V_{\rm C+L}'(r_i;s) + t \, \delta V_{\rm C+L}'(r_i) \}
-(A_1 + 2 \, A_2\, \rho_i )\Lambda_{\overline{\rm MS}}^{\mbox{\tiny 3-loop}}}
{\delta^F_i(x)}
\right]^2 ,
\rule[-7mm]{0mm}{6mm}
\label{chisqF}
\\ &&
P^F( A_1, A_2, x) = {\cal N}_F^{\, -1}
\int ds \,dt\,\, e^{- \chi_F^2 /2} \, P_s(s) \, P_t (t) .
\eea
We use the first 10 points of the
lattice data for $\{ F_{\rm latt}(r_i), \, \delta^F_i(r_i) \}$,
corresponding to 
$\rho_i(0.602) < 0.5$, 
given in the same
table (Tab.~2) in \cite{Necco:2001xg}.
Other details are same as in the case using the potential.

There is a considerable difference between the use of the potential and 
the force in the determination of $\delta E_{\rm US}(r)$ and $x$.
The difference stems from different correlations of the errors 
($\{ \delta^V_i\}$, $\{ \delta^F_i \}$) of
the respective lattice data sets and from our treatment of
these errors.
It is known that there exists a high correlation
among the errors of the lattice data at different $r_i$,
for the QCD potential or for the force.
It is also known that the error correlation of the force is smaller
than that of the potential \cite{Sommer-priv-comm}.
On the other hand, up to now, the covariance matrix of the errors 
for neither of these quantities is
available.
We therefore decided to use the lattice data with Gaussian errors,
neglecting the correlations; see eqs.~(\ref{chisq}) and (\ref{chisqF}).
This treatment should, in general, result in overestimates of 
errors in the determination
of $(A_1,A_2,x)$.

\begin{figure}
\begin{center}
\begin{tabular}{cc}
{\large 
Using the QCD Potential
}
\vspace*{-10mm}\\
\psfrag{A1}{$A_1$} 
\psfrag{A2}{$A_2$} 
\psfrag{68pcCL}{\footnotesize 68\%} 
\psfrag{95pcCL}{\footnotesize 95\%} 
\includegraphics[width=8.8cm]{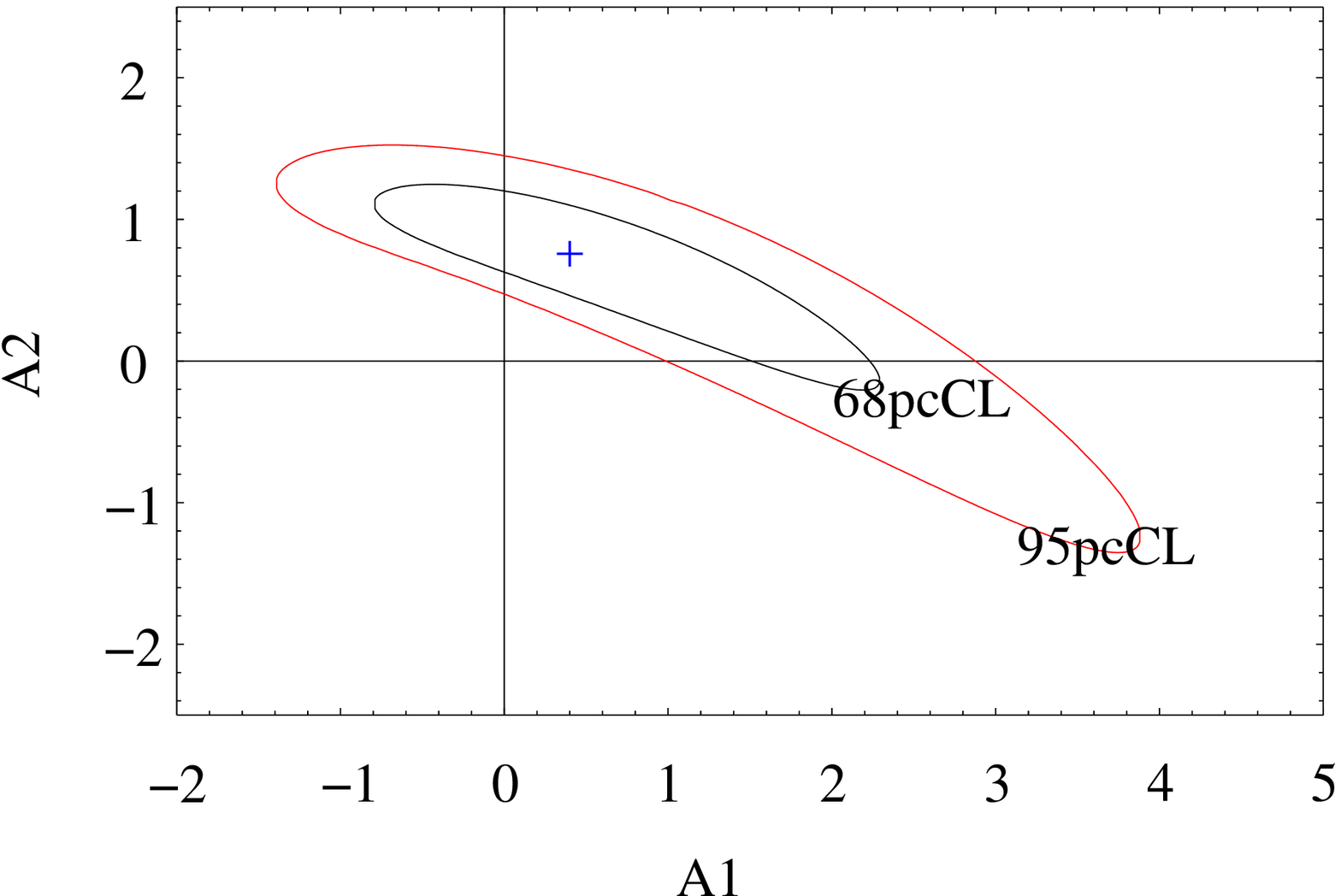}
& 
\hspace{0mm}
\psfrag{68pc}{\footnotesize 68\%} 
\psfrag{95pc}{\footnotesize 95\%} 
\psfrag{r0}{\hspace{-6mm} $r=r_0$}
\psfrag{delEUS}{\hspace{-24mm} $\delta E_{\rm US}(r)/
\Lambda_{\overline{\rm MS}}^{\mbox{\scriptsize 3-loop}}$~~(C+L scheme)}
\psfrag{r}
{\hspace{-11mm}
$\rho=
r \cdot \Lambda_{\overline{\rm MS}}^{\mbox{\scriptsize 3-loop}}$}
\raise0.5mm\hbox{
\includegraphics[width=6.7cm]{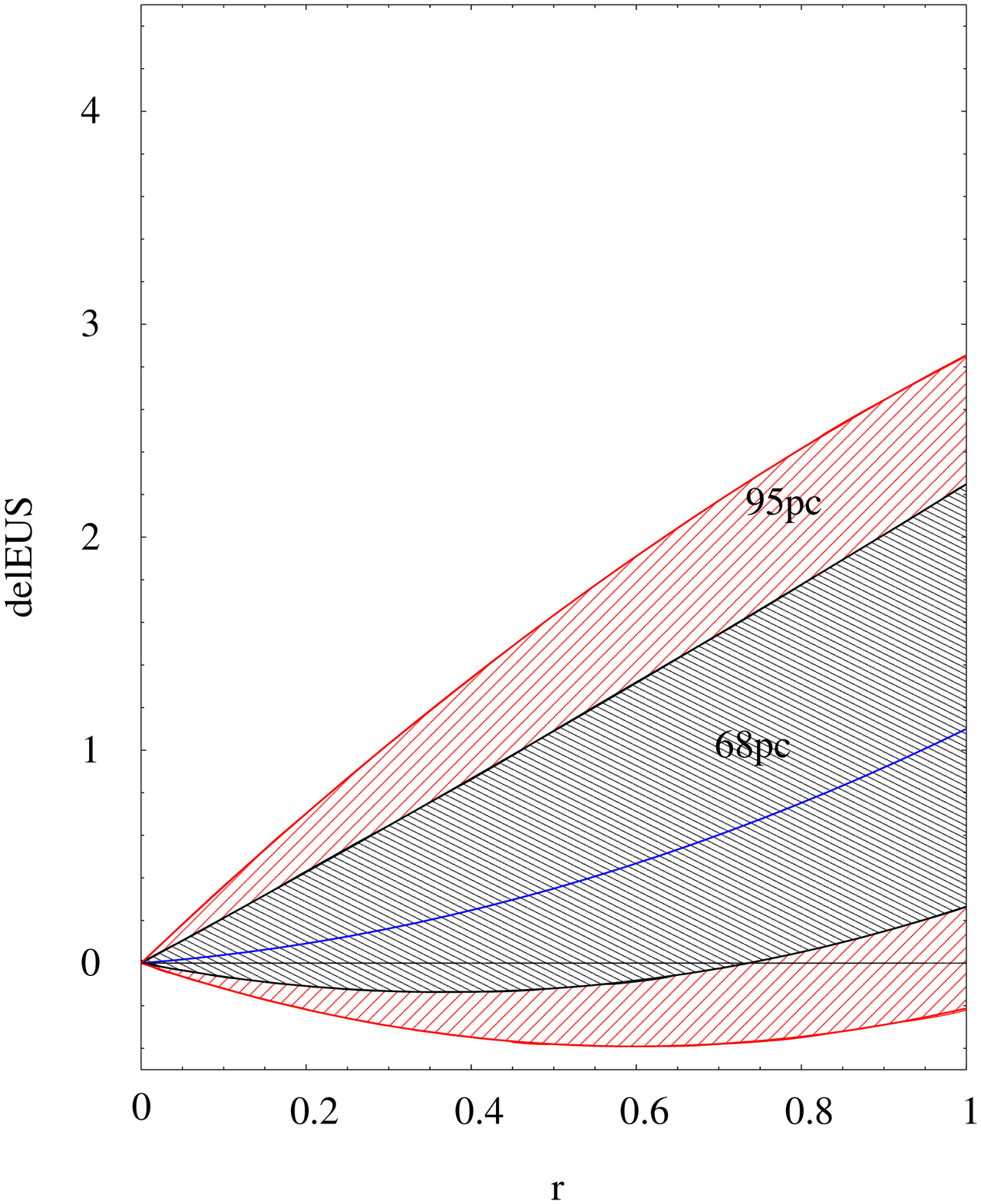} 
}
\vspace{3mm}
\\
\hspace{5mm}(i) & \hspace{12mm}(ii)
\end{tabular}
\end{center}
\vspace*{-.5cm}
\caption{\small 
(i) Contour plot of the probability
density distribution $P^V_{A_1A_2}(A_1,A_2)$, corresponding
to 68\% and 95\% CL regions.
Cross represents $(A_1,A_2)$ with the highest probability density,
$(A_1,A_2)=(0.40,0.76)$.
(ii) Bounds on $\delta E_{\rm US}(r)$ in the C+L scheme corresponding to the
regions of (i).
Quadratic fit with the highest
probability density is also plotted (blue line).
\label{bound-A1A2}}
%\end{figure}
%
%\begin{figure}
\begin{center}
\vspace*{10mm}
\begin{tabular}{cc}
{\large 
Using the QCD Force
}
\vspace*{-10mm}\\
\psfrag{A1}{$A_1$} 
\psfrag{A2}{$A_2$} 
\psfrag{68pcCL}{\footnotesize 68\%} 
\psfrag{95pcCL}{\footnotesize 95\%} 
\includegraphics[width=8.8cm]{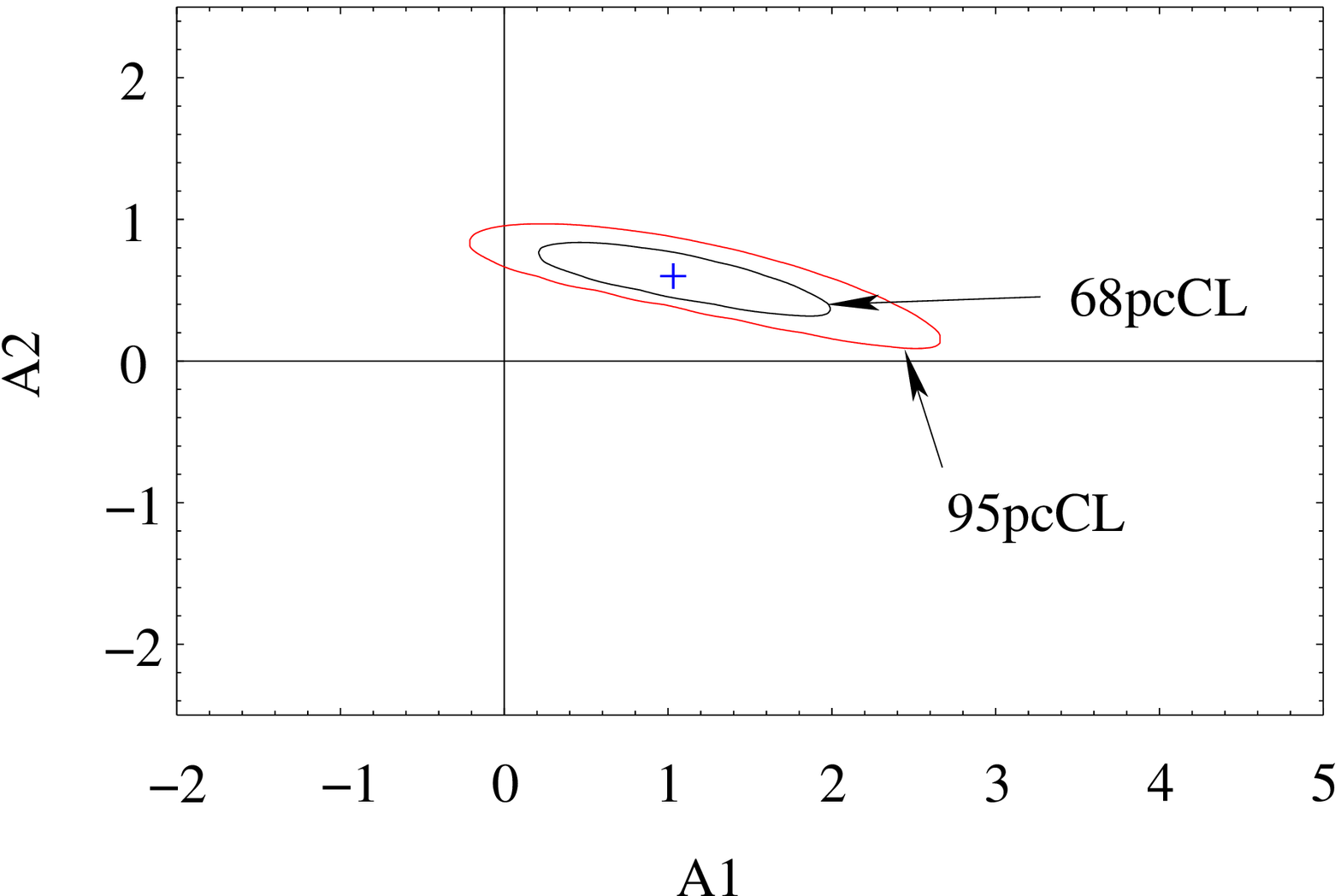}
& 
\hspace{0mm}
\psfrag{68pc}{\footnotesize 68\%} 
\psfrag{95pc}{\footnotesize 95\%} 
\psfrag{r0}{\hspace{-6mm} $r=r_0$}
\psfrag{delEUS}{\hspace{-24mm} $\delta E_{\rm US}(r)/
\Lambda_{\overline{\rm MS}}^{\mbox{\scriptsize 3-loop}}$~~(C+L scheme)}
\psfrag{r}
{\hspace{-11mm}
$\rho=
r \cdot \Lambda_{\overline{\rm MS}}^{\mbox{\scriptsize 3-loop}}$}
\raise0.5mm\hbox{
\includegraphics[width=6.7cm]{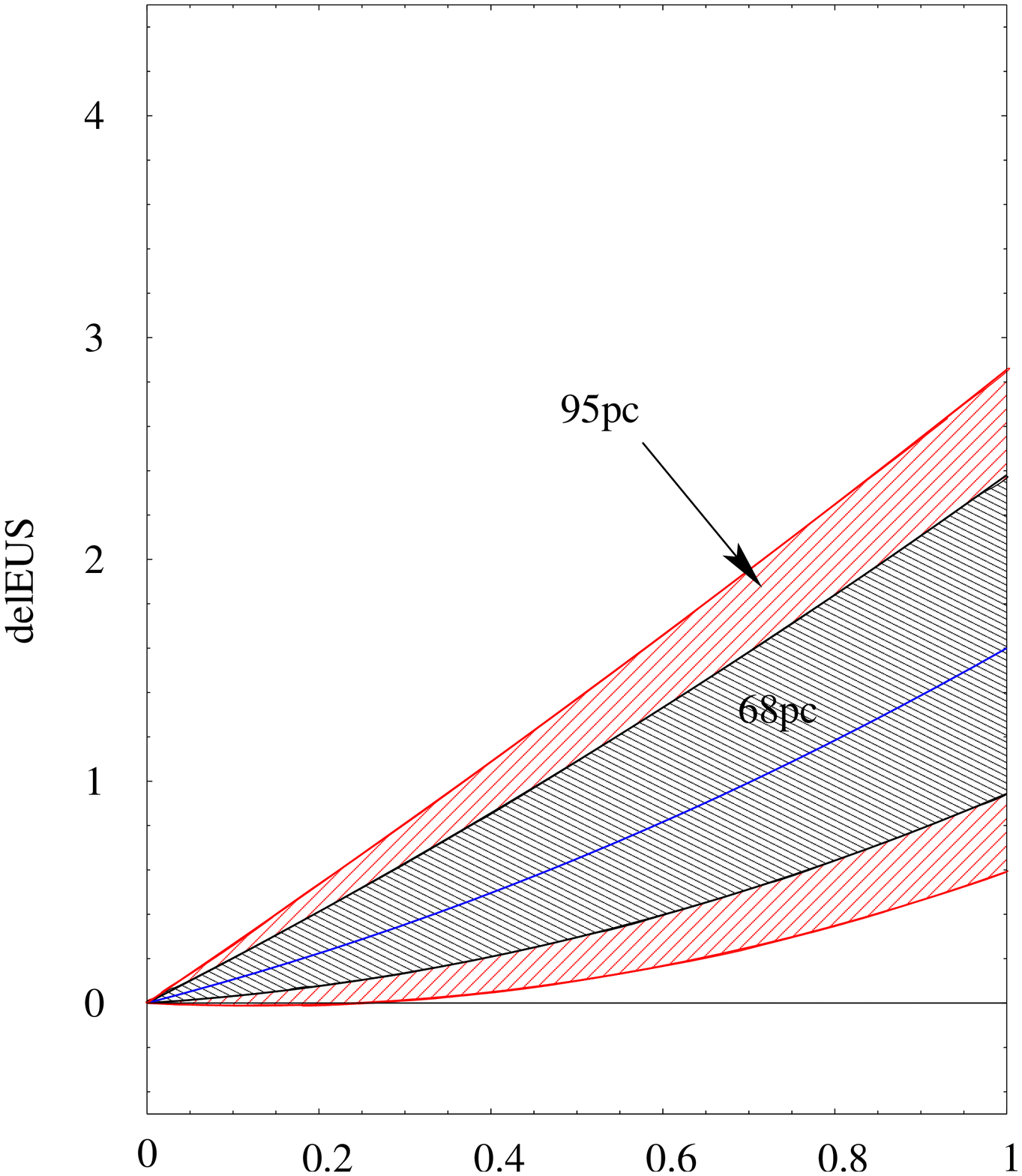} 
}
\vspace{3mm}
\\
\hspace{5mm}(i) & \hspace{12mm}(ii)
\end{tabular}
\end{center}
\vspace*{-.5cm}
\caption{\small 
(i) Contour plot of the probability
density distribution $P^F_{A_1A_2}(A_1,A_2)$, corresponding
to 68\% and 95\% CL regions.
Cross represents $(A_1,A_2)$ with the highest probability density,
$(A_1,A_2)=(1.03,0.60)$.
(ii) Bounds on $\delta E_{\rm US}(r)$ in the C+L scheme
corresponding to the
regions of (i).
Quadratic fit with the highest
probability density is also plotted (blue line).
\label{bound-A1A2-force}}
\end{figure}
\begin{figure}
\begin{center}
\psfrag{x}{\hspace{-9mm}\large
$x=r_0 \, \Lambda_{\overline{\rm MS}}^{\mbox{\scriptsize 3-loop}}$}
\psfrag{pdf}{\hspace{-11mm}Probability density distribution}
\psfrag{using potential}{\hspace{7mm}$P^{V}_x(x)$}
\psfrag{using force}{\hspace{-2mm}$P^{F}_x(x)$}
\includegraphics[width=10cm]{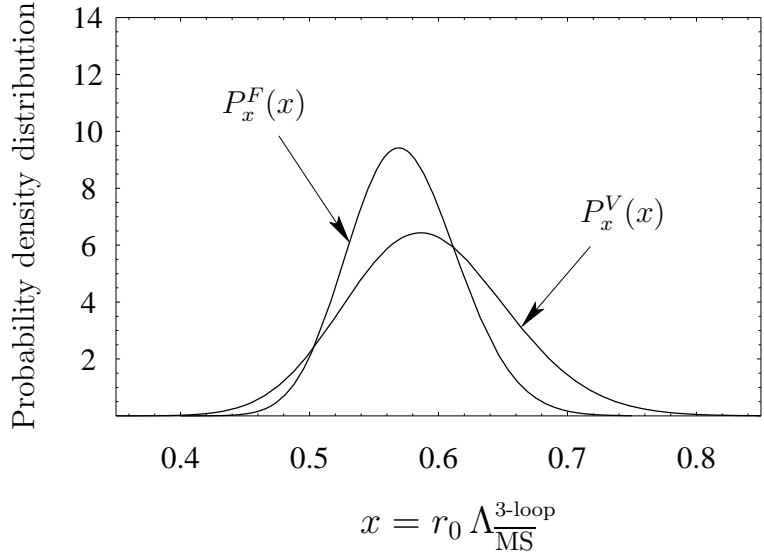} 
\end{center}
\caption{\small 
Probability density distributions 
$P^{V,F}_x(x)$ vs.\ $x$.
\label{PDF-Px}
}
\end{figure}
Bounds on $\delta E_{\rm US}(r)$ can be obtained
from the probability
density distributions for $(A_1,A_2)$, defined by
\bea
&&
P^V_{A_1A_2}(A_1,A_2) = \int dx \, dA_0 \,\, P^V(A_0, A_1, A_2, x) ,
\\ &&
P^F_{A_1A_2}(A_1,A_2) = \int dx \,\, P^F( A_1, A_2, x) .
\eea
Figs.~\ref{bound-A1A2},\ref{bound-A1A2-force}(i) show contour plots of 
these probability density distributions corresponding to the
68\% and 95\% confidence level (CL) regions.
The corresponding bounds on $\delta E_{\rm US}(r)$
in the C+L scheme
are given by [see Figs.~\ref{bound-A1A2},\ref{bound-A1A2-force}(ii)]
\bea
&&
\mbox{\it Using the potential:}
\nonumber \\ 
&&
~~~~~~~~~~~~
\left\{
\begin{array}{ll}
-0.7 \, \rho + 1.0 \, \rho^2 < 
\delta E_{\rm US}(r) /
\Lambda_{\overline{\rm MS}}^{\mbox{\scriptsize 3-loop}}
< 2.1\, \rho + 0.2\, \rho^2
~~~~~ &
(\mbox{68\% CL})
\rule[-4mm]{0mm}{6mm}
\\ 
-1.3 \, \rho + 1.1 \, \rho^2 < 
\delta E_{\rm US}(r) /
\Lambda_{\overline{\rm MS}}^{\mbox{\scriptsize 3-loop}}
< 3.7\, \rho -0.8\, \rho^2
~~~~~ &
(\mbox{95\% CL}) 
\end{array}
\right.
\label{res-delE}
\rule[-10mm]{0mm}{6mm}
\\
&&
\mbox{\it Using the force:}
\nonumber \\ 
&&
~~~~~~~~~~~~
\left\{
\begin{array}{ll}
~~\, 0.2 \, \rho + 0.7 \, \rho^2 < 
\delta E_{\rm US}(r) /
\Lambda_{\overline{\rm MS}}^{\mbox{\scriptsize 3-loop}}
< 2.0\, \rho +0.4\, \rho^2
~~~~~ &
(\mbox{68\% CL})
\rule[-4mm]{0mm}{6mm}
\\ 
-0.2 \, \rho + 0.8 \, \rho^2 < 
\delta E_{\rm US}(r) /
\Lambda_{\overline{\rm MS}}^{\mbox{\scriptsize 3-loop}}
< 2.6\, \rho +0.2\, \rho^2
~~~~~ &
(\mbox{95\% CL}) 
\end{array}
\right.
\label{res-delE-force}
\eea
These bounds are mutually consistent, and
the latter bounds are tighter.
Since the origin $(A_1,A_2)=(0,0)$ lies outside the 95\% CL
regions in Figs.~\ref{bound-A1A2},\ref{bound-A1A2-force}(i),
we conclude that $\delta E_{\rm US}(r)$ being ${\cal O}(\LQ^4r^3)$ or
$\delta E_{\rm US}(r)=0$ is disfavored.
We see that positive $A_2$ is favored, in agreement with the
fits in Tab.~\ref{tab-fitfn}.

The probability density distributions for
$x=r_0\,\Lambda_{\overline{\rm MS}}^{\mbox{\scriptsize 3-loop}}$
are defined by
\bea
&&
P^V_x(x)=\int dA_0\,dA_1\,dA_2 \,\, P^V(A_0, A_1, A_2, x) ,
\\ &&
P^F_x(x)=\int dA_1\,dA_2 \,\, P^F( A_1, A_2, x) .
\eea
They are shown in Fig.~\ref{PDF-Px}.
Each distribution is close to Gaussian, so
we simply quote the mean and the standard
deviation for $x$:
\bea
&&
\mbox{\it Using the potential:}
~~~~~~~~~
x=0.592 \pm 0.062 \, ,
~~~~~~~~~~~~~~
\label{res-x}
\\ &&
\mbox{\it Using the force:}
~~~~~~~~~~~~~~
x=0.574\pm 0.042 \, .
\label{res-x-force}
\eea
Note that we did not use the relation (\ref{r0Lambda}) at all
to obtain these results.
Our results are mutually consistent, as well as 
in excellent agreement with eq.~(\ref{r0Lambda}).\footnote{
We also note the value obtained by \cite{Necco:2001xg},
$0.586\pm 0.048$.
This value is closer to our values.
}
The error in eq.~(\ref{res-x-force}) is of similar size
to (slightly smaller than) that in eq.~(\ref{r0Lambda}).

Some comments are in order.

%\begin{itemize}
%\item

As explained above,
the sensitivity to 
$x=r_0\,\Lambda_{\overline{\rm MS}}^{\mbox{\scriptsize 3-loop}}$
originates from the mixing of a Coulombic term in
$V_{\rm diff}(r;x)$ when $x$ is different from its true value.
The only assumption we made in the determination of $x$ is that 
$\delta E_{\rm US}(r)$ can be approximated by a quadratic polynomial
of $r$ at $r\,\Lambda_{\overline{\rm MS}}^{\mbox{\scriptsize 3-loop}}
\simlt 0.5$.
Since $\delta E_{\rm US}(r)$ goes to zero at sufficiently small
$r$ according to OPE, a polynomial fit should be reasonable.
Effects of higher powers of $r$ is expected to be suppressed at small $r$.
In fact, we made a consistency check 
by including $A_3 \rho^3$ term 
into the fits of $\delta E_{\rm US}(r)$
and/or by varying the number of data points used for the fits.
As we include more data points,
errors determined from $P^V_x(x)$ and $P^F_x(x)$ decrease, respectively,
while the quality of the quadratic fits tends to 
get worse (quality is better with the cubic fits).
In any case, the obtained bounds on $x$ are consistent with
eqs.~(\ref{res-x}) and (\ref{res-x-force}).
We found that our present choice, quadratic fit with 
the first 11 (10) data points, 
is close to optimal in performing the fits.

When we scan $s$ and $t$ between the intervals $(0,2)$ and
$(-1,1)$, respectively,
the minimum values of $\chi_V^2$ and $\chi_F^2$ vary
between (0.5, 0.6) and (0.1, 0.2), respectively.
Since the number of degrees of freedom is
$N_{\mbox{\scriptsize dof}}=11-4=10-3=7$,
both $(\chi_V^2)_{\rm min}/N_{\mbox{\scriptsize dof}}$ and
$(\chi_F^2)_{\rm min}/N_{\mbox{\scriptsize dof}}$ are below 10\%.
This is consistent with existence of
high correlations among the lattice errors 
at different $r_i$, which we mentioned already.
We note that 
our errors in 
eqs.~(\ref{res-delE}),(\ref{res-delE-force}),(\ref{res-x}),(\ref{res-x-force})
may be overestimated for this reason.

The difference between the bounds obtained by using the potential and
force can be attributed to the
difference of the correlations of the lattice errors.
Since the correlation is larger for the potential, the errors
of the lattice data are effectively more enhanced (overestimated) 
in our treatment, hence the bounds are wider when we use
the potential.
The lattice errors are the dominant source of errors in
the determination of $(A_1,A_2)$ and $x$, both when
we use the potential and force.
In this sense, the covariance matrices of the lattice data 
are highly demanded.

Since the errors for $V_{\rm C+L}(r)$ 
(parametrized by $s$ and $t$) are much larger than
the errors of the lattice data,
one may wonder why the latter can be dominant
source of errors 
with regard to the former.
[Note that in Fig.~\ref{delEUS} the errors of the lattice data are
smaller or comparable to the size of the symbols used for the plot;
the variation of $V_{\rm C+L}(r;s)$ with $s$ is comparable in size
to $\delta V_{\rm C+L}(r)$.]
This can be understood as follows:\\
(1)
Practically, the measurement of $x$ is sensitive only to the ``Coulomb''
part of $V_{\rm diff}(r;x)$, hence only
the ``Coulomb'' part of the errors matters.
Let us denote by $V^{(n)}_{\rm C+L}(r)$ the difference between
$V_{\rm C+L}(r)$ up to N$^n$LL and $V_{\rm C+L}(r)$ up to N$^{n-1}$LL.
Then we perform fits of 
$V^{(n)}_{\rm C+L}(r)/\Lambda_{\overline{\rm MS}}^{\mbox{\scriptsize 3-loop}}$
in the form $c_{-1}\,\rho^{-1}+c_1\,\rho+c_2\,\rho^2$, using the
11 data points evaluated at $r=r_i$.
The results for $n=1,2,3$ are shown
in Tab.~\ref{Coulomb-in-V(n)C+L}.
\begin{table}
\begin{center}
\begin{tabular}{ll}
\hline
$V^{(1)}_{\rm C+L}(r)/\Lambda_{\overline{\rm MS}}^{\mbox{\scriptsize 3-loop}}$
&$ -0.0153 \, \rho^{-1} + 1.1\,\rho -0.5\,\rho^2$\\
$V^{(2)}_{\rm C+L}(r)/\Lambda_{\overline{\rm MS}}^{\mbox{\scriptsize 3-loop}}$
&$ -0.0087 \, \rho^{-1} + 0.8\,\rho -0.2\,\rho^2$\\
$V^{(3)}_{\rm C+L}(r;s=1)/\Lambda_{\overline{\rm MS}}^{\mbox{\scriptsize 3-loop}}$
&$ -0.0028 \, \rho^{-1} + 0.5\,\rho -0.05\,\rho^2$\\
\hline
$[V^{(3)}_{\rm C+L}(r;s=2)-V^{(3)}_{\rm C+L}(r;s=1)]/
\Lambda_{\overline{\rm MS}}^{\mbox{\scriptsize 3-loop}}$
&$ -0.0038 \, \rho^{-1} + 0.6\,\rho -0.08\,\rho^2$\\
$[V^{(3)}_{\rm C+L}(r;s=0)-V^{(3)}_{\rm C+L}(r;s=1)]/
\Lambda_{\overline{\rm MS}}^{\mbox{\scriptsize 3-loop}}$~~~
&$ +0.0038 \, \rho^{-1} - 0.6\,\rho +0.08\,\rho^2$\\
\hline
\end{tabular}
\end{center}
\caption{
Fits of $V^{(n)}_{\rm C+L}(r_i)/\Lambda_{\overline{\rm MS}}^{\mbox{\scriptsize 3-loop}}$
in the form $c_{-1}\,\rho^{-1}+c_1\,\rho+c_2\,\rho^2$, using the
11 data points evaluated at $r=r_i(0.602)$.
\label{Coulomb-in-V(n)C+L}
}
\end{table}
Magnitudes of all the coefficients $c_k^{(n)}$ decrease as the order increases,
if we take $s=1$ as a reference for $V^{(3)}_{\rm C+L}(r)$.
This is natural, since $V^{(n)}_{\rm C+L}(r) \to 0$ as $n \to \infty$, therefore
all $c_k^{(n)} \to 0$.
Thus, it is quite reasonable to estimate the ``Coulomb'' part of
the errors of $V_{\rm C+L}(r)$ by
the ``Coulomb'' part included in
$t \, V^{(3)}_{\rm C+L}(r;s=1)$ or by that included in
$V^{(3)}_{\rm C+L}(r;s) - V^{(3)}_{\rm C+L}(r;s=1)$.
In fact, these error estimates are encoded in our analysis.
Noting that we neglect the correlation of the errors of the lattice data,
the errors of the lattice data are indeed larger than the 
``Coulomb'' part of
the errors of $V_{\rm C+L}(r)$.
[Of course, the argument given here is only for demonstration to understand
the errors better.
When we derived our results 
eqs.~(\ref{res-delE}),(\ref{res-delE-force}),(\ref{res-x}),(\ref{res-x-force}),
neither did we do a fit of the form
$c_{-1}\,\rho^{-1}+c_1\,\rho+c_2\,\rho^2$
nor extract a ``Coulomb'' part.]
\\
(2)
There is a similar mechanism in the determination of $(A_1,A_2)$.
Since $\Delta(r;x,x')$ has a ``Coulomb''+linear form,
if a small admixture of ``Coulomb'' part is allowed in 
$V_{\rm diff}(r)$ due to
the errors of the lattice data, the linear term attached to
the ``Coulomb'' part mixes in as well.
Hence, if the lattice errors are larger, allowing a ``Coulomb'' part
to mix in,
the bounds on $(A_1,A_2)$ spread mainly in the $A_1$ direction.
This explains the difference of Figs.~\ref{bound-A1A2}(i) and
\ref{bound-A1A2-force}(i).
Again, it is the size of the ``Coulomb'' part of the errors
that matters.

Our results can be
compared with the determination of $\delta E_{\rm US}(r)$
by Pineda \cite{Pineda:2002se}.
It is the only study that determined
the non-perturbative contribution using OPE, preceding
our current work.
There are some important differences between
Pineda's analysis and ours.
\begin{itemize}
\item 
Pineda used 
$x=r_0\,\Lambda_{\overline{\rm MS}}^{\mbox{\scriptsize 3-loop}}$
as an input parameter, given by eq.~(\ref{r0Lambda}).
Its error turns out to be the dominant source of errors in
the determination of $\delta E_{\rm US}(r)$ 
(defined in RS scheme \cite{Pineda:2001zq,Pineda:2002se}).
On the other hand, in our analysis, we determine $x$
from a fit to the data.
\item
We estimate the error of the singlet potential (in the C+L scheme)
by varying $s$ and $t$ between $0\leq s\leq 2$ and $|t|\leq 1$.
On the other hand, Pineda estimates the error of the singlet potential
(in RS scheme) by varying $\bar{a}_3$ between the range
corresponding to $1/2 < s <3/2$, while there is no
estimate corresponding to variation of $t$, i.e.\ $t$ is fixed to zero.
Thus, our error estimate of the singlet potential
is more conservative.
\item
Pineda does not incorporate errors of the lattice data at all.
On the other hand, in our analysis,  they are included neglecting
the correlation.
Since our analysis is sensitive to a ``Coulomb'' part, 
the lattice errors are the major source of errors.
\item
The singlet
potential in RS scheme contains ${\cal O}(\LQ^3r^2)$
renormalon.
It means that
the singlet potential has an intrinsic uncertainty
of this order.
This essentially
prevents a determination of $\delta E_{\rm US}(r)$
with better than ${\cal O}(\LQ^3r^2)$ accuracy, because
$\delta E_{\rm US}(r) = V_{\rm QCD}(r) - V_S(r)$ cannot
be defined with better accuracy.
On the other hand, our potential is free from the
${\cal O}(\LQ^3r^2)$ renormalon (and also from the rest of IR renormalons).
So, at least conceptually, there is a difference in the achievable accuracies
between the two analyses.

\end{itemize}
Due to these differences, comparisons between
our bounds on $\delta E_{\rm US}(r)$ and those of Pineda
are not straightforward.
An explicit bound obtained by Pineda, assuming a form
$\delta E_{\rm US}(r) = {\rm const.} \times r^2$, reads
\bea
|\delta E_{\rm US}(r)| < 2.8~ 
\bigl( \Lambda_{\overline{\rm MS}}^{\mbox{\scriptsize 3-loop}} \bigr)^3\, r^2
~~~(\mbox{RS scheme, Pineda \cite{Pineda:2002se}}) .
\eea
One way of comparison may be to perform a fit by setting
$A_1=0$ in our analysis.
From the probability density distribution
$P^F_{A_1A_2}(0,A_2)$, we obtain
\bea
0.7~ 
\bigl( \Lambda_{\overline{\rm MS}}^{\mbox{\scriptsize 3-loop}} \bigr)^3\, r^2
< \delta E_{\rm US}(r) <
0.9~
\bigl( \Lambda_{\overline{\rm MS}}^{\mbox{\scriptsize 3-loop}} \bigr)^3\, r^2
~~~~~\left(
\begin{tabular}{c}
C+L scheme, $A_1=0$\\
68\% CL using force
\end{tabular}
\right)
 .
\eea

%\end{itemize}

\subsection{Determination of \boldmath $\delta E_{\rm US}(r)$
in factorization scheme}
\label{s5.3}

We can also determine the size of $\delta E_{\rm US}(r;\mu_f)$ in the factorization
scheme.
In fact, in the second scheme (within the factorization scheme), the purely
non-perturbative contribution is common to that in the C+L scheme.
This is because the difference of $\delta E_{\rm US}(r)$ is merely
the difference of the Wilson coefficients $V_{\rm C+L}(r)$
and $V_S^{\rm (R)}(r;\mu_f)$;
it is given by (minus)
the right-hand-side of eq.~(\ref{proof}), which is
systematically computable by means of perturbative expansion and 
log resummation via RG.
So, our task reduces to estimating
the infinit-order limit of eq.~(\ref{proof}).
\begin{table}
\begin{center}
\begin{tabular}{l|c}
\hline
&{\hspace{0mm}$[V_{\rm C+L}(r)-V_S^{\rm (R)}(r;\mu_f)]/
\Lambda_{\overline{\rm MS}}^{\mbox{\tiny 3-loop}}$} \\
\hline
LL & $0.53 \,\rho^2 + 0.00\,\rho^3 + \cdots$ \\
NLL & $0.85 \,\rho^2 + 0.07\,\rho^3 + \cdots$ \\
NNLL & $1.04\,\rho^2 + 0.27\,\rho^3 + \cdots$  \\
NNNLL & $1.02\,\rho^2 + 0.58\,\rho^3 + \cdots$ \\
\hline
\end{tabular}
\end{center}
\caption{\small
Expansion of $[V_{\rm C+L}(r)-V_S^{\rm (R)}(r;\mu_f)]/
\Lambda_{\overline{\rm MS}}^{\mbox{\scriptsize 3-loop}}$ 
in $\rho=r  \Lambda_{\overline{\rm MS}}^{\mbox{\tiny 3-loop}}$,
computed using eq.~(\ref{proof}).
We neglect $\rho$-independent constants.
$V_S^{\rm (R)}(r;\mu_f)$ is computed in the second scheme
and with $\mu_f=3\Lambda_{\overline{\rm MS}}^{\mbox{\tiny 3-loop}}$.
Other parameters for $V_{\rm C+L}(r)$ and $V_S^{\rm (R)}(r;\mu_f)$
are same as in Fig.~\ref{comp-lat}.
\label{tab-fitfn2}}
\end{table}
Using eq.~(\ref{proof}), we calculate Taylor expansions of
$V_{\rm C+L}(r)-V_S^{\rm (R)}(r;\mu_f)$ in 
$r$, which are listed in Tab.~\ref{tab-fitfn2}
for $\mu_f=3\,
\Lambda_{\overline{\rm MS}}^{\mbox{\scriptsize 3-loop}}$.
We estimate errors by the size of the NNNLL corrections and
obtain
\bea
[V_{\rm C+L}(r)-V_S^{\rm (R)}(r;3\,
\Lambda_{\overline{\rm MS}}^{\mbox{\scriptsize 3-loop}})]
/\Lambda_{\overline{\rm MS}}^{\mbox{\scriptsize 3-loop}} 
\approx (1.02 \pm 0.02) \, \rho^2 + (0.58 \pm 0.31) \, \rho^3 .
~~~~~~
(\rho <0.5)
\label{est-fac-scheme}
\eea
Thus, $\delta E_{\rm US}(r;3\,
\Lambda_{\overline{\rm MS}}^{\mbox{\scriptsize 3-loop}})$
in the factorization scheme (second scheme)
is given by the sum of eqs.~(\ref{res-delE-force}) and (\ref{est-fac-scheme}).
Given the above estimate, estimating 
$V_{\rm C+L}(r)-V_S^{\rm (R)}(r;\mu_f)$ for other value of
$\mu_f$ is straightforward, 
since the difference of the integrals [eq.~(\ref{proof})]
can be evaluated by integrating over $q$
along the real axis and the integrand is free from singularities;
convergence is fairly good in this region of $q$, so one may simply use
the prediction up to NNNLL.

By the same token, we can estimate $\delta E_{\rm US}(r)$ in the first scheme 
within the factorization
scheme.
The difference between the first scheme and second scheme is 
perturbatively computable.
In practice, up to NNNLL, we do not find a significant difference between
the first scheme and 
the second scheme.
We consider that we may apply the above estimate
(\ref{est-fac-scheme}) also to the first scheme 
within the factorization
scheme.

\section{Summary and Conclusions}
\label{s6}
\clfn

In this paper, we analyzed the static QCD potential in the distance
region relevant to heavy quarkonium spectroscopy,
$0.5~{\rm GeV}^{-1} (0.1~{\rm fm}) \simlt r \simlt 5~{\rm GeV}^{-1}
(1~{\rm fm})$, using perturbative expansion and OPE as basic
theoretical tools.
The analysis consists of three major steps:
\vspace*{3mm}\\
(I)
Behavior of the QCD potential 
at large orders of perturbative expansion was analyzed.
As for the higher-order terms, we used the estimates 
by large-$\beta_0$ approximation
or by RG equation; 
as for the renormalization
scale $\mu$, we varied it around the minimal-sensitivity
scale [or, more precisely the scale defined by eq.~(\ref{xi})].
Then the perturbative expansion of 
the QCD potential, truncated at ${\cal O}(\alpha_S^N)$,
was separated into a scale-independent
(prescription-independent) part
and scale-dependent (prescription-dependent) part
when $N \gg 1$:
\bea
V_N(r) = V_C(r) + {\cal B}(N,\xi) + {\cal C} \, r + {\cal D}(r,N,\xi)
+ (\mbox{terms that vanish as $N\to\infty$}) .
\label{decgen2}
\eea
Here, 
%we fixed the scale $\mu$ such that
%$N=6\pi\xi /(\beta_0 \alpha_S(\mu))$;
$\xi$ is a parameter for changing the scale
($\xi = 1$ corresponds
to an optimal choice of scale).
$V_C(r)$ is a ``Coulomb'' potential, which includes 
logarithmic corrections
at short-distances;
${\cal B}$ is an $r$-independent
constant;
${\cal C} \, r$ is a linear potential;
% and divergent (finite) as $N \to \infty$
%if $\xi \geq 1/3$ ($\xi <1/3$).
${\cal D}(r)$ behaves as 
$r^{3\xi-1}\times(\mbox{log corr.})$
if $\xi < 1$, whereas it is ${\cal O}(r^2)$ and divergent as
$N\to\infty$ if $\xi \geq 1$.
$V_C(r)$ and ${\cal C} \, r$ correspond to a renormalon-free part
of $V_N(r)$ and are finite and independent of $\xi$;
thus the scale-independent part has a ``Coulomb''+linear form.
On the other hand, ${\cal B}$ and ${\cal D}(r)$ 
correspond, respectively, to the
${\cal O}(\LQ)$ IR renormalon and beyond ${\cal O}(\LQ)$ IR
renormalons (starting from the ${\cal O}(\LQ^3r^2)$
renormalon) contained in $V_N(r)$; they are
dependent on $\xi$ and divergent as $N\to\infty$ if $\xi$ 
is sufficiently
large.
Detailed analytic behaviors of each component have been studied.
\vspace{2mm}\\
(II)
In the framework of OPE of the QCD potential,
(a) we gave explicit renormalization prescriptions for
the Wilson coefficient (singlet potential) $V_S^{\rm (R)}(r;\mu_f)$,
which belong to the class of conventional
factorization schemes with a hard cutoff;
(b) the scale-independent part of (I) was generalized 
and promoted to a 
Wilson coefficient $V_{\rm C+L}(r)$, which is {\it independent of}
the factorization scale $\mu_f$.
Both $V_S^{\rm (R)}(r;\mu_f)$ and $V_{\rm C+L}(r)$
are free from IR renormalons and IR divergences.
Several properties of these Wilson coefficients and 
of the corresponding non-perturbative
contributions have been derived 
(partly already in \cite{Brambilla:1999xf}):
\begin{itemize}
\item
$V_S^{\rm (R)}(r;\mu_f)$ and $V_{\rm C+L}(r)$ 
can be computed systematically using perturbative expansion
and log resummation via RG, and (in principle)
the predictions can be improved to arbitrary precision.
Hence, the corresponding non-perturbative contributions
$\delta E_{\rm US}(r)$ are unambiguously defined.
\item
With the usual hierarchy condition 
$\LQ \ll \mu_f \ll 1/r$,
the difference between $V_S^{\rm (R)}(r;\mu_f)$ and $V_{\rm C+L}(r)$
is ${\cal O}(\mu_f^3r^2)$ and perturbatively computable.
$V_{\rm C+L}(r)$ is closer to $V_{\rm QCD}(r)$ than 
$V_S^{\rm (R)}(r;\mu_f)$. 
%for the standard choices of $\mu_f$.
\item
$\delta E_{\rm US}(r)$ in the factorization scheme is
${\cal O}(\mu_f^4r^3)$ at very short-distances ($\Delta V(r)  \gg \mu_f$),
whereas it is ${\cal O}(\mu_f^3r^2)$ in a semi-short-distance region
($\Delta V(r) \ll \mu_f \ll 1/r$).
\item
$\delta E_{\rm US}(r)$ in the C+L scheme ($\mu_f$-independent scheme)
is ${\cal O}(\LQ^4r^3)$ at very short-distances ($\Delta V(r) \gg \LQ$),
whereas its behavior cannot be predicted model-independently 
in a semi-short-distance region
($\Delta V(r) \sim \LQ$).
\end{itemize}
We conjectured that the region of our interest corresponds to a
semi-short-distance region where $\Delta V(r) \sim \LQ \, (\ll \mu_f)$.
\vspace{2mm}\\
(III)
We computed $V_S^{\rm (R)}(r;\mu_f)$ and $V_{\rm C+L}(r)$ numerically for
$n_l=0$ at our
current best knowledge (NNNLL with 
certain estimates of $\bar{a}_3$)\footnote{
As far as US logs are concerned, we observe that their effects are
small, hence, we did not resum the US logs 
but included only up to NNNLO
in the analysis.
} 
and compared
them with the lattice computations of $V_{\rm QCD}(r)$ in the
quenched approximation.
We confirmed that the theoretical predictions of (II)
are either correct
or consistent within the present level of uncertainties.\footnote{
We also observed some limitations of the large-$\beta_0$
approximation.
}
We find that  a linear potential in $\delta E_{\rm US}(r)$ reduces
with increasing order, consistently with vanishing in the infinite-order limit;
at NNNLL, it is much smaller than the string tension as
determined by lattice simulations.
Then,
we performed fits of 
$V_{\rm diff}(r;x)\equiv V_{\rm latt}(r;x)-V_{\rm C+L}(r)$
and determined simultaneously  $\delta E_{\rm US}(r)$ and 
$x = r_0 \Lambda_{\overline{\rm MS}}^{\mbox{\scriptsize 3-loop}} $
(relation between Sommer scale and $\Lambda_{\overline{\rm MS}}$).
A sensitivity to $x$ originates from mixing of a Coulombic
term into $V_{\rm diff}(r;x)$ when $x$ differs from its true value.
Both the QCD potential and QCD force were used for
the fits.
The latter resulted in tighter bounds due to
a smaller correlation of lattice errors.
We obtained
\bea
r_0 \Lambda_{\overline{\rm MS}}^{\mbox{\scriptsize 3-loop}} 
=0.574\pm 0.042 \, ,
\eea
in excellent agreement with the determination via
Schr\"odinger functional method, eq.~(\ref{r0Lambda}) \cite{Capitani:1998mq}.
We also obtained
\bea
0.2 \, \rho + 0.7 \, \rho^2 < 
\delta E_{\rm US}(r) /
\Lambda_{\overline{\rm MS}}^{\mbox{\scriptsize 3-loop}}
< 2.0\, \rho + 0.4\, \rho^2 ,
~~~~~ &
(\mbox{C+L scheme})
\eea
where $\rho=r
\Lambda_{\overline{\rm MS}}^{\mbox{\scriptsize 3-loop}}$.
[See also bounds on the coefficients of quadratic polynomial in
Figs.~\ref{bound-A1A2},\ref{bound-A1A2-force}(i).]
In the factorization scheme, we obtain, for instance, 
\bea
0.2 \, \rho + 1.7 \, \rho^2 < 
\delta E_{\rm US}(r; \mu_f) /
\Lambda_{\overline{\rm MS}}^{\mbox{\scriptsize 3-loop}}
< 2.0\, \rho + 1.4\, \rho^2 \, .
~~~
\left(
\begin{array}{cc}
\mbox{factorization scheme}\\
\mu_f=3
\Lambda_{\overline{\rm MS}}^{\mbox{\scriptsize 3-loop}}
\end{array}
\right)
\eea
Estimating $\delta E_{\rm US}(r;\mu_f)$ for other
$\mu_f$ is easy.
In the factorization scheme, the obtained bounds are consistent with
${\cal O}(\mu_f^3r^2)$, rather than ${\cal O}(\mu_f^4r^3)$.
In the C+L scheme, the obtained bound is consistent with
${\cal O}(\LQ^3r^2)$ (vanishing linear potential) 
at 95\% CL, but existence of a small linear term
is more favored;
furthermore, 
$\delta E_{\rm US}(r) \sim {\cal O}(\LQ^4r^3)$ or $\delta E_{\rm US}(r) = 0$
is disfavored.
Consequently, we find that $\mu_f \gg \Delta V(r)$% 
%(provided $\mu_f \gg \LQ$)
, in accord with our
conjecture.
Also $\Delta V(r) \sim \LQ$ is more likely than
$\Delta V(r) \gg \LQ$.\footnote{
Although we have not considered the possibility $\Delta V(r) \ll \LQ$
in this paper (since it seems to lie outside the applicable range of
our analysis), it may be worth examining this possibility in detail
in view of our present results.
}
\vspace{5mm}

The analysis (I) provides a reasoning within perturbative QCD,
why we observed
agreement between the recent perturbative computations of the QCD
potential with phenomenological
potentials or lattice results:
in the large-order limit, $V_N(r)$ does approach a
``Coulomb''+linear form.
It is quite intriguing that we can separate the renormalon-free
part and renormalon-dominant part 
in a natural way.
Furthermore, in this analysis, the coefficient of the 
linear potential can be computed analytically up to NNLL.

In (II) we defined renormalization prescriptions 
for $V_S^{\rm (R)}(r;\mu_f)$, in which all the
known IR renormalons are subtracted.\footnote{
We ignored the instanton-induced renormalon singularities,
which are known to give very small contributions.
From a general argument,
there should also exist contributions to IR renormalons,
which cannot be written in the form of eq.~(\ref{int-renormalon});
we neglected them too.
}
Moreover, introduction of a 
factorization-scale independent
scheme for the Wilson coefficient, $V_{\rm C+L}(r)$, is new.
$V_{\rm C+L}(r)$ has some appealing theoretical features: it 
has no intrinsic uncertainties,
is systematically computable, and is closer to $V_{\rm QCD}(r)$
than $V_S^{\rm (R)}(r;\mu_f)$.
In any case, since the differences between 
the different schemes are perturbatively
computable with good accuracy, the C+L scheme would serve as a 
useful reference.

In the numerical analysis (III), 
the only assumption we made in the fits is that we can approximate
$\delta E_{\rm US}(r)$ by a quadratic polynomial of $r$
at $r\Lambda_{\overline{\rm MS}}^{\mbox{\scriptsize 3-loop}} <0.5$.
This
should be reasonable since $\delta E_{\rm US}(r) \to 0$
as $r \to 0$ according to OPE.
(An $r$-independent constant is irrelevant in our analysis.)
We checked validity of the assumption by including cubic term and
by varying the number of data points used for the fits.

In Fig.~\ref{comp-lat}
we have seen apparent convergence of $V_{\rm C+L}(r)$
towards the lattice data, up to our current best knowledge.
However, a closer examination of $\delta E_{\rm US}(r)$ 
in the C+L scheme revealed that $\delta E_{\rm US}(r)=0$ is
disfavored.
In fact,
we have improved both quantitatively and
conceptually (in the sense that we removed all the
renormalons from the singlet potential)
the bounds on $\delta E_{\rm US}(r)$,
as compared to the pioneering study by
Pineda \cite{Pineda:2002se}.

The OPE analysis of the QCD potential provided, as a byproduct,
a new method for determining
$x = r_0 \Lambda_{\overline{\rm MS}}^{\mbox{\scriptsize 3-loop}} $.
Our current result gives an error comparable in size to the error of
the conventional result using the Schr\"odinger functional
method \cite{Capitani:1998mq}.
The mechanism for the sensitivity is fairly clear,
as well as sources of errors are understood well.
The present status is that the errors of
the lattice data contribute more significantly than
the errors of $V_{\rm C+L}(r)$.
Hence, information on the correlation of the lattice errors
(in particular, the covariant matrix) is highly demanded
in order to reduce the error.

Finally let us comment on the applicable range
of perturbative expansion and OPE of $V_{\rm QCD}(r)$.
We saw in Fig.~\ref{comp-lat} that the current best
perturbative prediction of the
Wilson coefficient $V_{\rm C+L}(r)$ follows
the lattice data up to 
$r  \simlt r_0
\approx 0.5$~fm.
We consider that the distance, at which string breaking occurs,
serves as a measure of the distance
where the perturbative expansion breaks down
(in the theory with $n_l>0$).
It is around 1~fm according to the recent lattice simulation \cite{Bali:2004pb}.
It is clear that the string breaking phenomenon is 
non-perturbative and that
the present perturbative computation of
the QCD potential lacks ingredients necessary for
description of this phenomenon.
In the context of heavy quarkonium phenomenology, string breaking
corresponds to the decay $\Upsilon(4S) \to B\bar{B}$.
Since empirically
the root-mean-square radius of $\Upsilon(4S)$ is around 1~fm,
it is consistent with the lattice results.
We also know empirically that phenomenological
potentials are approximated well by 
a Coulomb+linear form at $r \simlt 1$~fm.
It means that, if we separate heavy quark and antiquark,
we have a
sensitivity to the linear potential at distances before
string breaking takes place.
Thus, we consider $r \simlt 1$~fm [corresponding to
heavy quarkonium states
below $\Upsilon(4S)$] to be the range in which
perturbative expansion may make sense.
(Certainly more terms of the perturbative expansion
need to be included in order
to have an accurate
prediction of $V_{\rm C+L}(r)$
as $r$ approaches 1~fm.)
Already
in this range the QCD potential exhibits
a linear behavior in addition to the ``Coulomb'' part.
Our analysis indicates that
the qualitative argument presented at the end
of Sec.~\ref{s2.4} may be valid in this very range.
Ultimately, it depends on whether the linear term
in $\delta E_{\rm US}(r)$ is truly vanishing or not.\footnote{
Note that the linear potential we are
concerned here has, a priori, nothing to do with the
linear potential at $r \gg \LQ^{-1}$ usually associated
with confinement (in the theory with $n_l=0$).
A priori, we see no reason that both linear potentials 
should have a common slope.
Nevertheless, empirically these two linear potentials
seem to have a common slope.
}

\section*{Acknowledgements}
The author is grateful to 
A.~Pineda, T.~Onogi and T.~Moroi for fruitful discussion.
He also thanks R.~Sommer for suggestion to use the
QCD force in the analysis.

\newpage

\appendix
\clfn
\section*{Appendices}

\section{Basic Formulas}
\label{appA}

In this Appendix we present detailed formulas useful for computing
$\alpha_V^{\rm PT}(q)$ up to NNNLO
as well as $V_N(r)$ for a finite but large $N$.

The perturbative expansion of the
$V$-scheme coupling in momentum space
$\alpha_V^{\rm PT}(q)$ is defined
by eqs.~(\ref{Wilson-loop})--(\ref{alfVRG}).
The polynomials in eq.~(\ref{alfVPT}) up to NNNLO
are given by
\bea
&&
P_0(\ell) = a_0 ,
\label{P0}
\\ &&
P_1(\ell) = a_1 + 2\,a_0\,\beta_0 \, \ell ,
\\ &&
P_2(\ell) = a_2 + (4\,a_1\,\beta_0 + 2\,a_0\,\beta_1 )\,\ell +
4\,a_0\,{\beta_0}^2 \,\ell^2 ,
\\ &&
P_3(\ell) =a_3  + ( 6\,a_2\,\beta_0+ 
   4\,a_1\,\beta_1 + 
   2\,a_0\,\beta_2)\,\ell + 
   (12\,a_1\,{\beta_0}^2 
+ 10 \,a_0\,\beta_0\,\beta_1)\,\ell^2
+ 8\,a_0\,{\beta_0}^3 \,\ell^3 ,
\eea
where 
\bea
\ell = \log(\mu/q) .
\eea

The coefficients of the beta function $\beta_n$, defined by
eq.~(\ref{RGeq}), are given explicitly by 
\bea
&&
\beta_0 = 11 - \frac{2}{3} n_l ,
\\&&
\beta_1 = 102 - \frac{38}{3} n_l,
\\&&
\beta_2 = \frac{2857}{2} - \frac{5033}{18}n_l
+ \frac{325}{54} n_l^2 
~~~~~\mbox{\cite{Tarasov:1980au,Larin:1993tp}},
\\&&
\beta_3 =
\frac{149753}{6} + 
  3564\,\zeta_3
\nonumber\\&&~~~~~
 + 
  \left( - \frac{1078361}{162}   - 
     \frac{6508\,\zeta_3}{27} \right) \,{n_l}  + 
   \left( \frac{50065}{162} + \frac{6472\,\zeta_3}{81} \right) \,{{n_l}}^2
   + \frac{1093\,{{n_l}}^3}{729} 
~~~~~\mbox{\cite{vanRitbergen:1997va}},
\eea
where $\zeta_3=\zeta(3)=1.2020...$ denotes the Riemann zeta function 
$\zeta(z)=\sum_{n=1}^\infty {1}/{n^z}$ evaluated
at $z=3$.

Presently, $a_n$ are known up to $n=2$.
\bea
&&
a_0=1,
\label{a0}
\\&&
a_1 = \frac{31}{3} - \frac{10}{9} n_l  
~~~~~\mbox{\cite{Appelquist:es,Fischler:1977yf}},
~~~~~
\\ &&
a_2 = {\frac{4343}{18}} + 36\,{{\pi }^2} +   66\,{\zeta_3} - 
  {\frac{9\,{{\pi }^4}}{4}} - 
   \left( {\frac{1229}{27}} + {\frac{52\,{\zeta_3}}{3}} \right)  \,{n_l}
+ {\frac{100}{81}} \,{n_l^2} 
~~~~~\mbox{\cite{Peter:1996ig,Schroder:1998vy}}.
\label{a2}
\eea
It is known that
$a_3$ is IR divergent; the coefficient of 
the divergence and associated logarithm 
have been computed \cite{Brambilla:1999qa,Kniehl:1999ud}:
\bea
a_3 = {72 \pi^2}\left[ \,
\frac{1}{\epsilon} + 4 \left\{ 2\ell +\log(4\pi)-\gamma_E
\right\} \right]
+ \bar{a}_3 ,
\label{a3}
\eea
where the IR divergence is regularized by dimensional regularization
($D=4-2\epsilon$).
$\bar{a}_3$ is just a constant independent of $\epsilon$, $\mu$,
$r$.
So far, only some estimates of its size are known: e.g.\
$\bar{a}_3(\mbox{large-}\beta_0)=(\frac{5}{3}\beta_0)^3\approx 6162$,
$\bar{a}_3(\mbox{Pineda})\approx 18688 $ \cite{Pineda:2001zq,Pineda:2002se},
$\bar{a}_3(\mbox{Chishtie-Elias})\approx 20032$ 
\cite{Chishtie:2001mf} for $n_l =0$.

Physical logarithm associated with the IR divergence can be
extracted as follows 
\cite{Brambilla:1999qa,Brambilla:1999xf}.
Instead of defining the ultrasoft contribution $\delta E_{\rm US}(r)$
in eq.~(\ref{deltaEUS}) as a non-perturbative quantity,
it can be computed in a double
expansion in $\alpha_S$
and $\log \alpha_S$  within pNRQCD:\footnote{
It can also be obtained from the difference between the
resummation of diagrams in Fig.~\ref{Coulomb-resum} and its expansion
in $\alpha_S$ before loop integration.
}
\bea
&&
\Bigl[ \delta E_{\rm US}(r) \Bigr] \!\!
_{\mbox{\hbox{ \raise-3pt\hbox to 0pt{\scriptsize double}\raise-11pt\hbox{\scriptsize exp.} } } }
%_{\begin{array}{ll}\scriptsize \mbox{double}\\ \mbox{\scriptsize exp.}\end{array}} = 
=
\frac{C_FC_A^3\alpha_S(\mu)^4}{24\,\pi \,r}
\nonumber \\ &&
~~~~~~~~~~~~~~~~~~~~~~~~~
\times
\left[ \,
\frac{1}{\epsilon} + 8 \log (\mu\, r) -
2 \log \Bigl({C_A \alpha_S(\mu)} \Bigr) + 
\frac{5}{3} + 2\gamma_E + 4\log (4\pi)
\right] + {\cal O}(\alpha_S^5) ,
\nonumber\\
\label{deltaEUS-pert}
\eea
where $\gamma_E = 0.5772...$ denotes the Euler constant.
Upon Fourier transform, $1/\epsilon$ and $\log \mu$ terms
of eqs.~(\ref{a3}) and (\ref{deltaEUS-pert})
cancel each other.\footnote{
Note that the expansion of $V_{\rm QCD}(r)$ in $\alpha_S$, obtained
from $\alpha_V^{\rm PT}(q)$, coincides with the expansion
of the bare singlet potential $V_S(r)$ in $\alpha_S$;
cf.\ Sec.~\ref{s2.4}.
Hence, the sum of 
%$\delta E_{\rm US}(r)$
$
\bigl[ \delta E_{\rm US}(r) \bigr]_{\rm d.e.}
$,
as given by eq.~(\ref{deltaEUS-pert}), and $V_S(r)$ represents
the expansion of $V_{\rm QCD}(r)$ in $\alpha_S$ and $\log \alpha_S$.
}
The remaining $\log (C_A\alpha_S)$ is the physical logarithm.
The argument of the logarithm in eq.~(\ref{deltaEUS-pert}),
$C_A \alpha_S = 2\,r \, \Delta V(r)$, represents the
ratio of the IR regulator $2\Delta V(r)$ and $1/r$; cf.\ Sec.~\ref{s2.4}.
If we perform OPE in conventional factorization schemes, we
introduce the factorization scale $\mu_f$, and the IR regulator
$2\Delta V(r)$ will be replaced by $\mu_f$. 
Thus, one finds
the ultrasoft logarithm 
\bea
V_S^{(\rm R)}(r;\mu_f) \Bigr|_{\mbox{\scriptsize US-log}} =
- \frac{C_FC_A^3\alpha_S(\mu)^4}{24\,\pi \,r}
\times 2\log (\mu_f r) .
\label{LOUSlog}
\eea
It is easy to verify that the $\alpha_S(\mu)^4 \, \log (\mu_f/q)$ term
of eq.~(\ref{NNLLUSlog}) generates eq.~(\ref{LOUSlog}) after
Fourier transform.
(Oppositely, using RG equation
with respect to evolution of $\mu_f$ within pNRQCD,
one can resum US logs as given in eq.~(\ref{NNLLUSlog}) 
\cite{Pineda:2000gz}.)

One may verify explicitly
that $\alpha_V^{\rm PT}(q)$ up to NNNLO, defined via
eqs.~(\ref{P0})--(\ref{a3}),
and 
%$\delta E_{\rm US}(r)$ 
$
\bigl[ \delta E_{\rm US}(r) \bigr]_{\rm d.e.}
$
in eq.~(\ref{deltaEUS-pert})
are separately consistent with RG equations with respect to
evolution of $\mu$:
\bea
\left[
\mu^2 \frac{\partial}{\partial \mu^2} + \beta (\alpha_S(\mu)) 
\frac{\partial}{\partial \alpha_S(\mu)} \right]
X = 0
~~~~~~~~;~~~~~~~~~
X = \alpha_V^{\rm PT}(q) ~~\mbox{or}~~ 
%\delta E_{\rm US}(r).
\Bigl[ \delta E_{\rm US}(r) \Bigr] \!\!
_{\mbox{\hbox{ \raise-3pt\hbox to 0pt{\scriptsize double}\raise-11pt\hbox{\scriptsize exp.} } } }
\label{RGeqX}
\eea
This should be so, as long as $V_{\rm QCD}(r)$ and
$\alpha_V^{\rm PT}(q)$
are defined from the Wilson loop via 
eqs.~(\ref{Wilson-loop})--(\ref{alfVPT}), 
since the Wilson loop is independent of $\mu$.\footnote{
There exists a definition of the singlet potential through
threshold expansion of diagrams
contributing to a quark-antiquark Green function \cite{Kniehl:2002br};
in this definition, $\mu$-independence of the potential
is not preserved.
}
To verify eq.~(\ref{RGeqX}), one should note that the beta function 
in general dimension (in $\overline{\rm MS}$ scheme)
has a form
\bea
\left[ \beta(\alpha_S)\right]_{\epsilon \neq 0} = - \epsilon \, \alpha_S
+ \left[ \beta(\alpha_S)\right]_{\epsilon = 0} .
\label{betafn-gen-dim}
\eea
\medbreak

In the rest of this Appendix, we present formulas useful for
computing $V_N(r)$ for a large (but finite) $N$.
The expansion of $\alpha_S(q)^n$ in terms of $\alpha_S(\mu)$ 
can be
obtained by iterative operation of a derivative operator as
\bea
&&
\alpha_S(q)^n = \exp \left[
-2 \ell \, \beta(x) \frac{\partial}{\partial x} \right]
\, x^n \Biggr|_{x \to \alpha_S(\mu)}
\nonumber \\ && ~~~~~~~~~
= \sum_{k=0}^\infty \frac{(-2 \ell)^k}{k!} \, 
\left[ \beta(x) \frac{\partial}{\partial x} \right]^k
\, x^n \Biggr|_{x \to \alpha_S(\mu)} ,
\label{evolution}
\eea
where $\beta(\alpha_S)$ is defined in eq.~(\ref{RGeq})
[and eq.~(\ref{betafn-gen-dim}) if necessary].

The $V$-scheme coupling in position space,
$\bar{\alpha}_V(1/r)$, is defined by
$V_{\rm QCD}(r) = - C_F \,\bar{\alpha}_V(1/r)/r$.
The series expansion of $\bar{\alpha}_V(1/r)$ in terms of
the $\overline{\rm MS}$ coupling renormalized at 
$\mu = \exp(-\gamma_E)/r$
is obtained as follows \cite{Jezabek:1998wk}.
Using the coefficients $g_m$ defined by
\bea
\sum_{m=0}^\infty g_m \, u^m = 
\exp \left[ \sum_{k=2}^\infty \frac{\zeta(k)\, u^k}{k}
\, \left\{ 2^k - 1 - (-1)^k \right\} \right] ,
\eea
we may write
\bea
\bar{\alpha}_V(1/r) = 
\sum_{m=0}^\infty g_m 
\left[ -\beta(x) \frac{\partial}{\partial x} \right]^m
\sum_{n=0}^\infty \frac{a_n}{(4\pi)^n} \, x^{n+1}
\Biggr|_{x \to \alpha_S( \exp(-\gamma_E)/r)} .
\label{expand-alfVbar}
\eea
To obtain the expansion of $\bar{\alpha}_V(1/r)$ in terms of
$\alpha_S(\mu)$, we first compute eq.~(\ref{expand-alfVbar}),
then substitute the expansions of 
$\alpha_S( \exp(-\gamma_E)/r)^n$ in terms of $\alpha_S(\mu)$
computed using eq.~(\ref{evolution}).
By truncating the series at an appropriate order in
$\alpha_S(\mu)$, the result reduces 
to a polynomial of $\log (\mu\,r)$.

\section{Integral Representation of
\boldmath $[\alpha_{S}(q)]_\infty$ at NLL}
\label{appB}

\begin{figure}
\begin{center}
\psfrag{Cs}{$C_s$} 
\psfrag{s}{$s$} 
\psfrag{0}{\hspace{-1mm}\raise0mm\hbox{\footnotesize 0}}
\includegraphics[width=5cm]{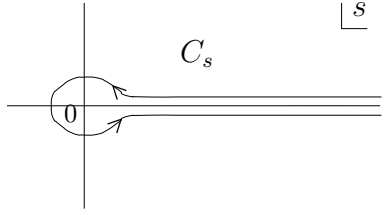} 
\end{center}
\caption{\small
Integral contour $C_s$.
\label{pathCs}
}
\end{figure}

$[\alpha_{S}(q)]_\infty$ for the 2-loop running coupling constant 
can be expressed in a 
one-parameter integral form.
After integrating the RG equation, $\alpha_{S}(q)$ is given
implicitly by the relation
\bea
\log \left( {q}/{\Lambda_{\overline{\rm MS}}^{\mbox{\scriptsize 2-loop}} }
\right)
= -\frac{2\pi}{\beta_0 \alpha_S} + \frac{\delta}{2}\,
\log \left( \frac{4\pi}{\beta_0\alpha_S} + \delta \right) .
\eea
Hence, using Cauchy's theorem, one may write
\bea
\alpha_{S}(q)& =& \frac{i}{\beta_0}\int_{C_s}
ds \, (-s)^{-1} \, 
\left[
\log \left( {q}/{\Lambda_{\overline{\rm MS}}^{\mbox{\scriptsize 2-loop}} }
\right)
+ \frac{\delta}{2}\, \log \left({-2es} \right) + s
\right]^{-1} 
\rule[-8mm]{0mm}{6mm}
\label{alf2L-ints}
\\
&=&
\frac{i}{\beta_0}\int_{C_s}
ds \, (-s)^{-1} \, 
\int_0^\infty dx\, \exp
\left[ 
-x \left\{  \log \left( {q}/{\Lambda_{\overline{\rm MS}}^{\mbox{\scriptsize 2-loop}} }
\right)
+ \frac{\delta}{2}\, \log \left({-2es} \right) + s
\right\}
\right]
\rule[-8mm]{0mm}{6mm}
\nonumber\\
&=&
\frac{2\pi}{\beta_0}\int_0^\infty dx\, 
\left( {q}/{\Lambda_{\overline{\rm MS}}^{\mbox{\scriptsize 2-loop}} }
\right)^{-x}
\left( \frac{x}{2e} \right)^{x\,\delta  /2}
\frac{1}{\Gamma ( 1 + x\,\delta  /2)}
\, .
\label{alf2L-intx}
\eea
The integral contour $C_s$ is shown in Fig.~\ref{pathCs}.
In the last equality, we rescaled $s \to s/x$ and used the
formula
$1/\Gamma(z) = i(2\pi)^{-1}\int_{C_s} ds  \, (-s)^{-z} \, e^{-s}$.

The truncated series expansion of $\alpha_S(q)$ in
$\alpha_\mu \equiv \alpha_S(\mu)$ can be obtained as follows.
One rewrites 
$\Lambda_{\overline{\rm MS}}^{\mbox{\scriptsize 2-loop}}$
in terms of $\alpha_\mu$ in
eq.~(\ref{alf2L-ints}).
After changing variables as 
$s = ( \frac{4\pi}{\beta_0\alpha_\mu} + \delta ) u$, we have
\bea
\alpha_S(q) = \frac{i \, \alpha_\mu}{2\pi} \int_{C_u} du \, (-u)^{-1} \,
\left[
1 + \frac{u}{\alpha_\mu\delta + {4\pi}/{\beta_0}} + \frac{\beta_0\alpha_\mu}{2\pi}
\left\{ \log \Bigl(\frac{q}{\mu}\Bigr) + \frac{\delta}{2} \, \log (-eu) \right\}
\right]^{-1} .
\eea
Expanding the integrand in $\alpha_\mu$ and integrating at each
order of the expansion,
one obtains the series expansion of $\alpha_S(q)$ in
$\alpha_\mu$.
It is then straightforward to truncate at order ${\alpha_\mu}^N$.
Sending $N \to \infty$, we obtain
\bea
&&
\lim_{N \to \infty}
\left[
\alpha_\mu
\left[
1 + \frac{u}{\alpha_\mu\delta + {4\pi}/{\beta_0}} + \frac{\beta_0\alpha_\mu}{2\pi}
\left\{ \log \Bigl(\frac{q}{\mu}\Bigr) + \frac{\delta}{2} \, \log (-eu) \right\}
\right]^{-1} 
\right]_N
\nonumber\\
&&
~~~~~~~~~~~~
=
\left[
\log \left( {q}/{\Lambda_{\overline{\rm MS}}^{\mbox{\scriptsize 2-loop}} }
\right)
+ \frac{\delta}{2}\, \log \left({-2es} \right) + s
\right]^{-1} 
\nonumber\\
&& ~~~~~~~~~~~~~~~~~
\times
\left( 1 - 
\exp
\left[ 
-3\xi \left\{  \log \left( {q}/{\Lambda_{\overline{\rm MS}}^{\mbox{\scriptsize 2-loop}} }
\right)
+ \frac{\delta}{2}\, \log \left({-2es} \right) + s
\right\}
\right]
\right) ,
\eea
where we re-expressed the truncated series in terms of $s$ and
$\Lambda_{\overline{\rm MS}}^{\mbox{\scriptsize 2-loop}}$.
Similarly to eq.~(\ref{alf2L-intx}), we find
\bea
[\alpha_S(q)]_\infty =
\frac{2\pi}{\beta_0}\int^{3\xi}_0 dx\, 
\left( {q}/{\Lambda_{\overline{\rm MS}}^{\mbox{\scriptsize 2-loop}} }
\right)^{-x}
\left( \frac{x}{2e} \right)^{x\,\delta  /2}
\frac{1}{\Gamma ( 1 + x\,\delta  /2)} .
\label{alfinf2L-intx}
\eea
Thus, the one-parameter integral forms given in
eq.~(\ref{one-param-int}) are obtained.

\section{Analytic Formula of the Linear Potential in Case (c)}
\label{appC}

We present the analytic formula for the 
coefficient of the linear potential, defined by eq.~(\ref{genform3}),
in case (c).
The integral eq.~(\ref{genform3}) can be reduced to a one-parameter
integral form by change of variables, e.g.\ from $q$ to $z = 1/\alpha_S$.
Then one readily sees that the integral
can be expressed in terms of the confluent
hypergeometric function except for the coefficient of $a_2$,
while the coefficient of $a_2$ can be expressed in terms of generalized
confluent hypergeometic functions.

For convenience, we first define some auxiliary parameters.
The two solutions to the quadratic equation
\bea
\left. \frac{4\pi \,\beta (\alpha_S)}{\alpha_S^2} \right|_{\rm case\,(c)} =
- \sum_{n=0}^2 \beta_n\, 
\Bigl( \frac{\alpha_S}{4\pi} \Bigr)^{n}
= 0
\eea
[cf.\ eq.~(\ref{RGeq})] are denoted as\footnote{
We assume that the two solutions are complex conjugate of each other.
This is the case when the number of active
quark flavors is less than 6.
}
$\alpha_S = 1/\omega$ and $1/\omega^*$.
Then we define 
\bea
p = \frac{2\pi}{\beta_0}\, \frac{\omega^2}{\omega-\omega^*} .
\eea
We also write $b_0=\beta_0/(4\pi)$.

The coefficient of the linear potential
in case (c)
is given by
\bea
{\cal C}^{\rm (c)} = \frac{2\pi C_F}{\beta_0} \, 
\left( \Lambda^{\rm 3-loop}_{\overline{\rm MS}} \right)^2 \,
\left[
a_0 \, {\cal R}_0 +
\frac{a_1}{4\pi} \, {\cal R}_1 +
\frac{a_2}{(4\pi)^2} \, {\cal R}_2 
\right] ,
\eea
where
\bea
&&
{\cal R}_0= 2 \, {\rm Re}
\left[ \, F(p+{\textstyle \frac{1}{2}}, p^*) + \omega \,
F(p,p^*) \, \right] ,
\\ &&
{\cal R}_1= 2 \, {\rm Re}
\left[ \, F(p,p^*) \, \right] ,
\\ &&
{\cal R}_2= 2 \, {\rm Re}
\left[ \, G \, \right] - 
{b_0}^\delta \,
(-\omega)^{2p-1} \,(-\omega^*)^{2p^*-1} .
\eea
The function $F$ is defined by
\bea
F(x,y)&=&
{b_0}^{x + y+\delta } \,
\frac{
( \omega - \omega^* )^{x + y -1} \, 
e^{-i\pi (x-y) - \delta/2}}{\Gamma(1-2x)}\,
%\nonumber \\
%&&
%\times
W_{y-x,\, x+y-1/2} \Bigl( 
{\textstyle 
\frac{\omega^*-\omega}{b_0}
} \Bigr) ,
\eea
in terms of the Whittaker function,
which is related to the confluent hypergeometric
function $_1F_1$ as
\bea
W_{\kappa,\mu}(z) &=& 
\frac{\Gamma(-2\mu)}{\Gamma(\frac{1}{2}-\mu-\kappa)} \,
z^{\mu+1/2} \, e^{-z/2} \,
_1F_1(\mu-\kappa+{\textstyle \frac{1}{2}},2\mu+1;z) 
\nonumber\\
&& + 
\frac{\Gamma(2\mu)}{\Gamma(\frac{1}{2}+\mu-\kappa)} \,
z^{-\mu+1/2} \, e^{-z/2} \,
_1F_1(-\mu-\kappa+{\textstyle \frac{1}{2}},-2\mu+1;z) .
\eea
On the other hand, $G$ is defined by
\bea
G & = & 
{b_0}^{\delta } \,
e^{(\omega/b_0)+i\pi (2p^*-1) } \,
\nonumber\\
&&
\times
\biggl\{ \, \frac{(\omega^*-\omega)^{-2-\delta}}
{(1-2p)\, B(1-2p,1-2p^*)} \,
\Gamma_1\Bigl(1,2p-1,2+\delta;
{\textstyle \frac{\omega}{\omega^*-\omega},
\frac{\omega-\omega^*}{b_0}
}\Bigr)
\nonumber\\
&&~~~
 - \frac{{b_0}^{-2-\delta }}{\pi} \,
\sin (2\pi p)\, \Gamma(-2-\delta)\,
\Xi_2 \Bigl( 1,1-2p^*,3+\delta;
{\textstyle -\frac{\omega}{b_0},
\frac{\omega^*-\omega}{b_0}
}\Bigr)
\biggr\} .
\eea
$\Gamma_1$ and $\Xi_2$ represent Appell
confluent hypergeometric functions \cite{Erdelyi}
defined by the double series
\bea
&&
\Gamma_1(\alpha,\beta,\beta';x,y)
= \sum_{m,n=0}^\infty 
\frac{(\alpha)_m(\beta)_{n-m}(\beta')_{m-n}}
{m! \, n!} \,
x^m y^n ,
\\
&&
\Xi_2(\alpha,\beta,\gamma;x,y) 
= \sum_{m,n=0}^\infty 
\frac{(\alpha)_m(\beta)_n}
{(\gamma)_{m+n} \, m! \, n!} \,
x^m y^n ,
\eea
where $(a)_n \equiv \Gamma(a+n)/\Gamma(a)$ is the
Pochhammer symbol.

\section{Numerical Evaluation
of \boldmath $V_S^{\rm (R)}(r;\mu_f)$ and $V_{\rm C+L}(r)$}
\label{appD}

In this Appendix, we present a method for 
accurate numerical evaluations of
$V_S^{\rm (R)}(r;\mu_f)$ and $V_{\rm C+L}(r)$.
The former is defined in Sec.~\ref{s4.1} to be
\bea
V_S^{({\rm R})}\!(r;\mu_f\!) = 
 - \frac{2C_F}{\pi} \!\!
\int_{\mu_f}^\infty \!\!\!\! dq \, \frac{\sin (qr)}{qr} \, 
\alpha_{V_S}^{({\rm R})}(q;\mu_f) 
\label{defVSR2}
\eea
with
\bea
\alpha_{V_S}^{({\rm R})}(q;\mu_f)
&=&
\alpha_V^{\rm PT}\! (q) + \delta\alpha_V\! (q;\mu_f) 
\\
&=& 
\alpha_S(q) \, \sum_{n=0}^{N} a^{V_s}_n \,
\biggl( \frac{\alpha_S(q)}{4\pi} \biggr)^n 
.
\eea
$N=0,1,2$, and 3 correspond to $V_S^{\rm (R)}(r;\mu_f)$
up to LL, NLL, NNLL and NNNLL, respectively.
(We do not resum US logs but include only up to NNNLO.)
$a^{V_s}_n=P_n(0)=a_n$ for $n\leq2$, whose explicit forms are given in
eqs.~(\ref{a0})--(\ref{a2}); 
$a^{V_s}_3=\bar{a}_3+144\pi^2 \log(\mu_f/q)$ in the first scheme
[corresponding to eq.~(\ref{delalfV-MSbar})],
while
$a^{V_s}_3=\bar{a}_3 - 120 \pi^2 + 144\pi^2
\left\{\gamma_E+\log[3\alpha_S(q)] \right\}$ in the
second scheme
[corresponding to eq.~(\ref{delalfV-2nd-scheme})].

We deform the integral path of eq.~(\ref{defVSR2}) into
upper half plane:
\bea
V_S^{({\rm R})}\!(r;\mu_f\!) = 
 - \frac{2C_F}{\pi} \, {\rm Im}
\int_{0}^\infty \!\!\!\! dk \, 
\left[ i\,\frac{\exp (iqr)}{qr} \, 
\alpha_{V_S}^{({\rm R})}(q;\mu_f) 
\right]_{q=\mu_f+i\,k} .
\label{defVSR3}
\eea
In order to evaluate the integral numerically,
we first solve the RG equation (\ref{RGeq}) with a
given input value (e.g.\ $\alpha_S(Q)=0.2$) and find
the value of $Q/\Lambda_{\overline{\rm MS}}$
and the value of $\alpha_S(\mu_f)$ for a given 
$\mu_f/\Lambda_{\overline{\rm MS}}$.
Then we solve the RG equation (\ref{RGeq})
along the integral path $q=\mu_f+i\,k$~ 
($0 < k < \infty$) in the complex
plane,
with $\alpha_S(\mu_f)$ as the initial value.
In solving the RG equation, we take the sum 
for $n \leq 0,1,2$ and 3 on the right-hand-side of 
eq.~(\ref{RGeq}), respectively, corresponding to 
$V_S^{\rm (R)}(r;\mu_f)$
up to LL, NLL, NNLL and NNNLL.

The singlet potential in the $\mu_f$-independent
scheme $V_{\rm C+L}(r)$ is defined in Secs.~\ref{s4.1} and \ref{s4.2}.
It is easier to evaluate the 
difference $V_S^{\rm (R)}(r;\mu_f)-V_{\rm C+L}(r)$ accurately, using 
eq.~(\ref{proof}), than  
to directly evaluate $V_{\rm C+L}(r)$:
\bea
&&
V_S^{({\rm R})}(r;\mu_f)  - V_{\rm C+L}(r) 
= 
\frac{2C_F}{\pi} \, {\rm Im}
\int_{C_3}\! dq \, \frac{e^{iqr}-[1+iqr+\frac{1}{2}(iqr)^2]}{qr} \, 
\alpha_{V_S}^{({\rm R})}(q) \,
+ \,{\rm const}.
\nonumber\\
\label{diffVS-VCL}
\eea
Here, we choose the second scheme for
$\alpha_{V_S}^{({\rm R})}(q)$.
The integral path $C_3$ is shown in Fig.~\ref{pathC3},
e.g.\ $q=k+ik^2(k-\mu_f)^2$ for $0\leq k\leq \mu_f$.
We solve the RG equation for $\alpha_S(q)$
along this path similarly to
above.
We may ignore the $r$-independent constant
on the right-hand-side of eq.~(\ref{diffVS-VCL}).
Then, subtracting eq.~(\ref{diffVS-VCL}) from
$V_S^{\rm (R)}(r;\mu_f)$ computed
in the second scheme, we obtain $V_{\rm C+L}(r)$.

\end{document}